\documentclass[11pt]{article}
\usepackage{slashbox}
\usepackage[pdftex]{graphicx,color}    
\usepackage{epstopdf}
\usepackage{amsmath}
\usepackage{amsmath,float,amsfonts,amsbsy,amssymb}
\usepackage{hyperref}
\newcommand{\be}{\text{e}}
\newcommand{\st}{\text{st}}
\newcommand{\tr}{\text{Tr}}
\newcommand{\str}{\text{STr}}
\newcommand{\VEV}[1]{\langle #1 \rangle}  
\newcommand{\bolS}{\text{\bf S}}
\newcommand{\vac}{|\text{vac}\rangle}
\newcommand{\Vac}{|\text{vac}\rangle}

\begin{document}

\begin{titlepage}

\def\slash#1{{\rlap{$#1$} \thinspace/}}

\begin{flushright} 
\end{flushright} 


\begin{Large}
\vspace{-1cm}
\begin{center}

{\bf Topological Many-Body States in Quantum Antiferromagnets \\via Fuzzy Super-Geometry}

\end{center}
\end{Large}
\vspace{1cm}

\begin{center}
{\bf  Kazuki Hasebe and Keisuke Totsuka}   \\ 

\vspace{0.8cm} 
\it{Kagawa National College of Technology, Takuma, 
Mitoyo, Kagawa 769-1192, Japan} \\ 
\vspace{0.1cm}
\it{Yukawa Institute for Theoretical Physics, Kyoto University, 
Oiwake-cho, Kyoto 606-8502, Japan} \\ 

\vspace{0.8cm} 
{\sf
hasebe@dg.kagawa-nct.ac.jp} \\ 
{\sf 
totsuka@yukawa.kyoto-u.ac.jp} 
\vspace{0.4cm} 

{\today} 

\end{center}

\vspace{0.4cm}


\begin{abstract}
\noindent

\baselineskip=16pt
 
  Recent vigorous investigations of topological order   
have not only discovered new topological states of matter  but also 
 shed new light to  ``already known'' topological states. One established example with topological order is the valence bond solid (VBS) states in quantum antiferromagnets. 
 The VBS states are disordered spin liquids with no spontaneous symmetry breaking but most typically manifest topological order known as hidden string order on 1D chain.  Interestingly, the VBS models are  based on mathematics analogous to fuzzy geometry. 
We review applications of the mathematics of fuzzy super-geometry in the construction of supersymmetric versions of VBS (SVBS) states, and  give a pedagogical introduction of SVBS models and their properties [arXiv:0809.4885, 1105.3529, 1210.0299].  
As concrete examples, we present detail analysis of supersymmetric versions of $SU(2)$ and $SO(5)$ VBS states, $i.e.$ $UOSp(N|2)$ and $UOSp(N|4)$ SVBS states whose mathematics are closely related to fuzzy two- and four-superspheres.   
 The SVBS states are physically  interpreted as  hole-doped VBS states with superconducting property that  interpolate  various VBS states depending on value of a hole-doping parameter. 
The parent Hamiltonians for SVBS states are explicitly constructed, and  
 their gapped excitations are derived within the single-mode approximation on 1D SVBS chains.   
Prominent features of the SVBS chains are discussed in detail, such as a generalized  string order parameter and entanglement spectra. It is realized that the entanglement spectra are at least doubly degenerate regardless of the parity of bulk (super)spins. 
Stability of topological phase with supersymmetry is discussed with emphasis on its relation to particular edge (super)spin states.

\end{abstract}



\end{titlepage}

\newpage

\tableofcontents

\section{Introduction}
\label{sec:introduction}
Strongly correlated systems such as cuprate superconductors, quantum Hall systems, 
and quantum anti-ferromagnets (QAFM) have been offering arenas for unexpected emergent 
phenomena brought about by strong many-body correlation. 
In particular, the study of QAFM bears the longest history since Heisenberg introduced 
the celebrated quantum-spin model \cite{Heisenberg1926,Heisenberg1928} 
and Bethe \cite{Bethe-31} found the first non-trivial exact solution to the quantum many-body problem,  
and still is providing us with attractive topics in modern physics.  
Generally, in the presence of many-body interaction, it is very hard  to obtain exact many-body 
ground-state wave functions and, even if possible, it is rare that we are able to write down 
them in compact and physically meaningful forms.  
Fortunately, in the above mentioned three cases (superconductivity (SC), quantum Hall effects (QHE), 
and QAFM), the paradigmatic ground-state wave functions have been known and greatly contributed  
to our understanding of exotic physics of these systems: the Bardeen-Cooper-Schrieffer (BCS) 
state \cite{BCS1957} for SC, the Laughlin wave function \cite{Laughlin1983} for QHE, 
and for QAFM the valence bond solid (VBS) states \cite{Afflecketal1987,Afflecketal1988}, 
which are the exact ground states of a certain class of quantum-spin models called 
the VBS models.   
In the present paper, we give a pedagogical review of the VBS models 
and their supersymmetric (SUSY) extension, i.e. the supersymmetric valence bond solid (SVBS) models \cite{Arovasetal2009,Hasebe-Totsuka2011,Hasebe-Totsuka2012}, 
with particular emphasis on their relation to fuzzy geometry.   
The VBS models had been originally introduced by 
Affleck, Kennedy, Lieb and Tasaki (AKLT) \cite{Afflecketal1987, Afflecketal1988} 
as a class of ``exactly solvable'' models that exhibit the properties conjectured by Haldane \cite{Haldane1983I,Haldane1983II}, 
namely the qualitative difference in excitation gaps and spin correlations in one-dimensional (1D) QAFM 
between half-odd-integer-spin and integer-spin cases.     

\begin{figure}[!t]
\centering
\includegraphics[width=13cm]{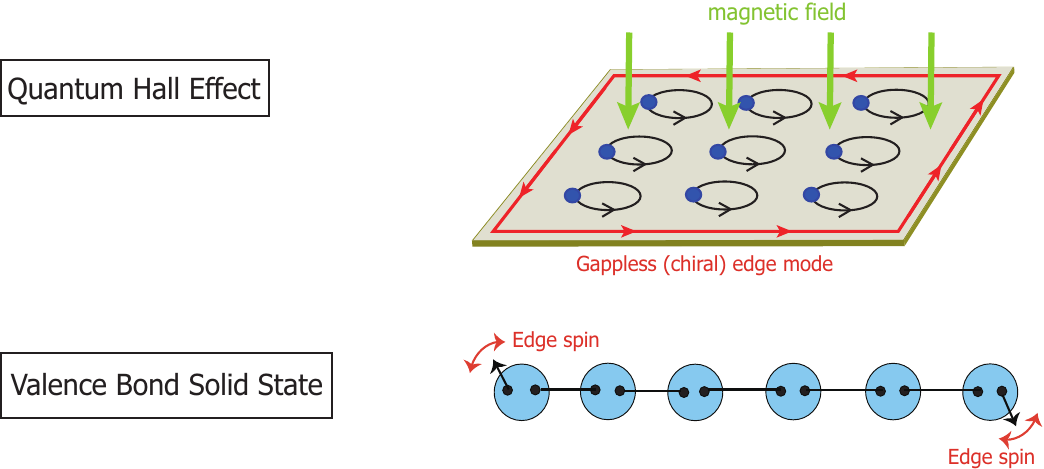}
\caption{(Color online) 
Physical analogies between QHE and VBS state. 
The bulk excitation on QHE is gapful while the edge mode is a gapless (chiral) mode. 
Meanwhile, the bulk excitation on the VBS state is gapful while the motion of edge spins is a freely rotating gapless mode.   
 \label{QHEVBS.fig} }
\end{figure}

As other quantum-disordered paramagnets, the VBS states have a finite excitation gap and 
exponentially decaying spin correlations.      
However, the VBS states are not  ``mere'' disordered non-magnetic spin states; the spin-$S$ 
VBS states necessarily have gapless modes at their edges, or more specifically, 
emergent spin-$S/2$ edge spins \cite{Hagiwara-K-A-H-R-90,Kennedy-90}.  
This might remind the readers of the (chiral) gapless modes at the edge of the QHE systems 
where excitations are gapful in the bulk (see Fig.~\ref{QHEVBS.fig}).   
Similar features are found in topological insulators as well 
\cite{QiandZhangPhysToday,Hasan-Kane-10,Qi-Zhang-11} 
and considered as a hallmark of the topological state of matter.          
Furthermore,  in 1D, the VBS states are known to exhibit a non-local order 
called the hidden string order \cite{NijsRommelse-1989,Tasaki1991}.   
What is prominent in the VBS states is that they most typically exhibit a certain kind of 
topological order {\em even} in 1D, 
while QHE 
needs at least 2D to work.  
This is a great advantage of the investigation of the VBS states 
since in 1D  most calculations of physical quantities can be carried out exactly 
by using the matrix product state (MPS) representation 
 \cite{Fannes-etal1992,Klumperetal1991, Klumperetal1992,Totsuka-Suzuki1995,Perez-Garciaetal2007} 
combined with the transfer-matrix method.    

Since the topological character of a system is believed to be encoded in the many-body ground-state 
wave functions rather than in its Hamiltonian, 
the relatively simple structure of the VBS wave function is suitable 
to investigate its topological properties by using such modern means as entanglement entropy 
\cite{Nielsen-Chuang,Amico-RMP-08}
or entanglement spectrum \cite{LiHaldane2008}.  
For the above reasons, the VBS states, or more generally, the MPS as a class of model states 
that satisfy the so-called area-law constraint \cite{Hastings2007} 
have attracted renewed 
attention as a ``theoretical laboratory'' in the recent study of topological states of matter.  
In fact, the MPS representation is now regarded as a natural and efficient way 
to describe quantum entangled many-body states, and for a given (1D) Hamiltonian  
the density matrix renormalization group (DMRG) \cite{White-92} 
provides a powerful tool to find the optimized variational wave function in the form of MPS  
(see Refs.~\cite{OstlundandRommer1995,Verstraete-M-C-08,Schollwoeck-11} for more details 
about the MPS method and DMRG).   The key idea there is that for generic short-range 
interacting systems, only a small part of the entire Hilbert space is important and, 
depending on the problems in question, there are variety of ways to parametrize 
this physically relevant subspace.  
Although the Wigner's $3j$-symbol may give a convenient description of $SU(2)$-invariant MPS states \cite{Dukelskyrtal1995,Romanal1998,Fledderjohannetal2011}, 
we adopt in the present paper the Schwinger-particle formalism (see, for instance, 
chapters 7 and 19 of \cite{auerbach1994book}) to emphasize the analogies 
to the lowest Landau level physics.
The list of possible applications includes a convenient description 
of gapped quantum ground states 
\cite{Verstraete-Cirac2006,Hastings-06,Hastings2007} as well as 
its application to efficient simulations of dynamics 
\cite{Vidal2003PRL,Vidal2004PRL,White-F-04,Daley-K-S-S-V-04,Vidal2007PRL,OrusVidal2008PRB}  
and variational calculations \cite{Schadschneider-Z-95,Kolezhuketal1997,Weichselbaum-V-S-C-D-09,Porras-V-C-06}.  
Due to their simple structure, the entanglement entropy of the VBS states comes only from 
the bond(s) cut by the entanglement bipartition \cite{Fan-K-R-04,Katsura2007,Katsuraetal2008,Xuetal2008}.  

In some `anisotropic' MPSs, it is known that a generalized quantum phase transitions (QPT) occur 
as the parameters contained in the MPSs are varied 
\cite{Niggemannetal1997,Niggemannetal2000,Klumperetal1993,Wolfetal2006}.  
However, a remark is in order about the interpretation of this kind of 
`quantum phase transitions' in MPS.  Normally, these QPTs in MPS are characterized by 
the divergence of {\em spatial} correlation lengths \cite{Wolfetal2006}.  
In generic lattice models, on the other hand, the diverging spatial correlation does not necessarily mean  
the vanishing of the excitation gap, while in the Lorentz-invariant systems, these two occur hand in hand. 
In fact, by the structure of MPS, the block (with size $L$) entanglement entropy never exceeds 
the $L$-independent value $2\ln D$ \cite{Vidaletal2003} (with $D$ being the dimension of the MPS matrix) 
while in quantum critical ground systems 
whose low-energy physics is well described by conformal field theories, the block  
entanglement entropy is proportional to $\ln L$ \cite{Vidaletal2003,Calabrese-C-04}.  
Therefore, one should take the QPTs in MPS mentioned above in a wider sense.  
This kind of phase transitions in the VBS-type of states in 2D have been discussed 
in \cite{Niggemannetal1997, Niggemannetal2000}.

Finally, we would like to mention the recent application of MPS to the classification of 
the gapped topological phases in 1D.  
As is well-known, there is no true topological phase characterized by long-range entanglement 
\cite{Chenetal2011}.  However, if certain symmetries (e.g. time reversal) are imposed, 
topological phases characterized by short-range entanglement are possible.  
Since this kind of topological phases are stable {\em only} in the presence of certain 
protecting symmetries, they are called {\em symmetry-protected topological} (SPT) 
\cite{GuWen2009}.   
As has been mentioned above, an appropriate choice of MPS faithfully represents 
any given gapped (short-range entangled) ground state.  Therefore, the problem of 
classifying all possible gapped topological phases reduces to classifying all possible 
MPSs by using group cohomology \cite{Chenetal2011,Schuchetal2011}.  
This program has been carried out for such elementary symmetries as 
time-reversal and link-parity in Refs.~\cite{Pollmannetal2010, Pollmannetal2012} 
(for the fermionic systems, see Refs.~\cite{Turner-P-B-11,Fidkowski-K-11}) 
and for the Lie-group symmetries in Ref.~\cite{DuivenvoordenandQuella2012}.  
Topological quantum phase transitions among these SPT phases have been discussed in, e.g.,  \cite{GuWen2009,Zangetal2010,Zhengetal2011}.   

In a sense, the second important keyword of this paper, 
fuzzy (super)geometry obtained by replacing the ordinary (commuting) coordinates 
with non-commutative ones, is again closely related to Heisenberg,   
who made a pioneering contribution in physics when he had built 
quantum mechanics on the basis of non-commuting phase-space variables. Snyder first substantiated Heisenberg's idea of non-commutative coordinates in his paper ``quantized space-time'' \cite{Snyder1947}\footnote{See Ref.\cite{BalachandranQuresh2006} for Heisenberg's contribution to the original idea of non-commutative geometry and related historical backgrounds.}.   
In fact, the VBS models have many interesting connections to QHE and fuzzy geometry. 
To explain the interesting relationship among them, let us begin with an analogy between 
the VBS states and the Laughlin wave functions of fractional quantum Hall effects (FQHE).  
Soon after the proposal of AKLT \cite{Afflecketal1987, Afflecketal1988}, 
Arovas, Auerbach and Haldane \cite{Arovasetal1988} 
realized that the Laughlin wave function and the VBS states (generalized to higher-spin cases) 
have analogous mathematical structures upon identifying the odd integer $m$, which 
characterizes the filling fraction $\nu=1/m$, and the spin quantum number $S$.  
Specifically, the VBS state is transformed to the Laughlin wave function 
of the electron system on a two-sphere, $i.e.$ the Laughlin-Haldane wave function \cite{haldane1983}  
by using the coherent-state representation of the VBS states and 
assigning appropriate correspondences between their physical quantities.    
In such translation, the $\it{external}$ symmetry (or, $SU(2)$-rotation of the spatial coordinates) 
of the Laughlin-Haldane wave function for QHE on a two-sphere 
is translated into the $\it{internal}$ symmetry 
(or, spin-$SU(2)$ symmetry) of the VBS states for the integer-spin chains.  
This analogy can be generalized and one can readily see that 
the Laughlin-Haldane wave functions with some external symmetry are 
generally transformed to give the VBS states with the identical $\it{internal}$ symmetry.    

In the past decade, there have been remarkable developments in higher-dimensional generalization 
of QHE  (see Refs.~\cite{Hasebe2010review,Karabali-Nair2006} for reviews).    
So far, the set-up \cite{Laughlin1983,haldane1983} of 2D QHE has been generalized 
to higher-dimensional manifolds, 
such as 4D \cite{ZhangHu2001}, 8D \cite{Bernevigetal2003}, $S^{2n}$  \cite{Hasebe-Kimura2004} and $\mathbb{C}P^N$ \cite{Karabali-Nair2002} \cite{Belluccietal-2003}. There also exist $q$-deformed QHE \cite{Gelounetal2006} and QHE on non-compact manifolds \cite{Jellal2005,DaoudJella2008,Hasebe2008,Hasebe2009}.    
As is well-known, QHE is a physical realization of the non-commutative geometry \cite{Bellissard-E-B-94},  
and the non-commutative geometry brings exotic properties to QHE \cite{Girvin1984, Girvin-M-P-86, Ezawaetal2003}.  
Therefore, for each higher-dimensional QHE, one can think of the underlying 
higher-dimensional fuzzy geometry, such as fuzzy two-sphere \cite{berezin1975,Hoppe1982,madore1992}, four-sphere\cite{hep-th/9602115}, $2n$ sphere \cite{hep-th/9711078,Ramgoolam2001,Ho2002,Kimura2002,Kimura2003,AzumaBagnoud2003}, 
$\mathbb{C}P^N$ \cite{Alexanianetal2002,Balachandranetal2002a,Watamuraetal2005}, $q$-deformed sphere \cite{Podles-1987} and fuzzy hyperboloids \cite{HoLi2000,FakhriImaanpur2003,DeBellisetal2010,HasebeNPB2012}. 

\begin{figure}[!t]
\centering
\includegraphics[width=13cm]{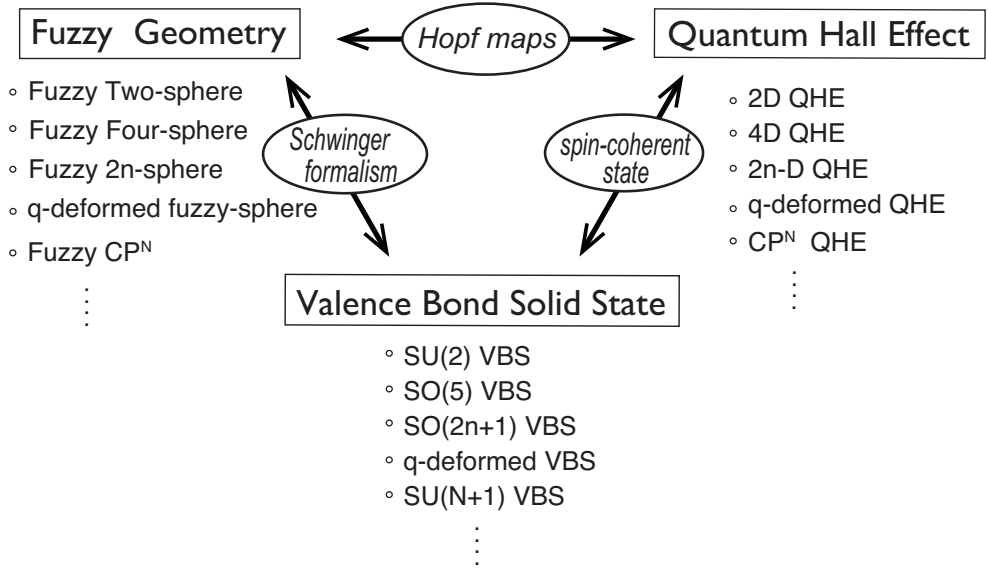}
\caption{
Close relations among fuzzy geometry, QHE, and VBS.  
They are ``transformed'' to each other with appropriate translations. 
 \label{connections.fig} }
\end{figure}

On the other hand, we have already seen that the Bloch spin-coherent state enables us to relate 
the Laughlin-Haldane wave functions to the VBS states.  
In correspondence with  such QHE, a variety of VBS models have been constructed with the symmetries, such as $SO(5)$, $SO(2n+1)$ \cite{Tuetal2008I,Tuetal2008II,Tuetal2009},  $SU(N+1)$ \cite{Greiteretal2007I,Greiteretal2007II,Arovas2008,Rachel-spinon-2009}, $Sp(N)$ \cite{Schuricht-Rachel2008,Rachel2009}, and $q$-deformed $SU(2)$ \cite{Klumperetal1991,Klumperetal1992,Totsuka-Suzuki1994,Arita-Motegi2010,Santosetal12012} [see Fig.~\ref{connections.fig}]. 
The VBS states demonstrate manifest relevance to fuzzy geometry also, 
by adopting the Schwinger operator formalism (see Sec.~\ref{sec:fuzzy-vs-LLL}). 

One of the main goals of this review is to illustrate their relations which have been less emphasized 
in previous literatures. 
As an explicit  demonstration of the correspondence,  we discuss a supersymmetric (SUSY) generalization of the VBS models.  
Since fuzzy super-spheres have already been explored in Refs.~%
\cite{Grosseeetal1995,GrosseReiter1998,Landi2001,Balachandranetal2002,Hasebe2011} 
and the SUSY QHE in Refs.~%
\cite{Hasebe-Kimura2005,Hasebe2005PRL,Hasebe2005PRD,Hasebe2006PRD}\footnote{A variety of super Landau models with super-unitary symmetries have been constructed in  
Refs.\cite{Ivanovetal2004,Ivanovetal2005,Curtright2007,Beylinetal2008,Beylinetal2010,BychkovIvanov2012,Goykhmanetal2012}
 and fuzzy supergeometries have also been investigated in Refs.\cite{Ivanovetal2003,MurraySaemann2008,Lazaroiuetal2009,DeBellisetal2010}.  },       
 we can develop a SUSY version of the VBS models (SVBS models) based on the mutual relations \cite{Arovasetal2009,Hasebe-Totsuka2011,Hasebe-Totsuka2012}. 
  For the inclusion of fermionic degrees of freedom,   
 the SVBS states exhibit particular features not observed in the VBS states with higher dimensional $\it{bosonic}$ groups. For instance, while the VBS states only exhibit a property of quantum magnets, the SVBS states accommodate two distinct sectors, spin sector and charge sector, due to the inclusion of fermionic degrees of freedom.     
Physically, the SVBS states can be interpreted as  hole-doped  VBS states with superconducting property. 
Mathematically, the SVBS states are regarded as a type of ``superfield'' in terminology of SUSY field theory.    
As the superfield unifies a various fields as its components, the SVBS states realize a variety of ordinary VBS states as the coefficients of the expansion with respect to Grassmann quantities.

By investigating topological features of the SVBS states, we address how SUSY affects the stability of topological phase or  entanglement spectrum.  
 As concrete examples, 
 we discuss topological feature of  the  SVBS states with $UOSp(1|2)$ and $UOSp(2|2)$ symmetries, which we call type-I and type-II SVBS states, respectively. To  perform detail analyses of the SVBS states on 1D chain,  we develop  a SUSY version of the MPS formalism (SMPS) \cite{Totsuka-Suzuki1995,Fannes-etal1992,Perez-Garciaetal2007}.  
 MPS formalism is now regarded as an appropriate formalism to treat gapped 1D quantum systems \cite{Verstraete-Cirac2006,Hastings2007}.  Since the MPS formalism naturally incorporates edge states, MPS provides a powerful formalism to discuss relations  between topological order and edge state.  Taking this advantage, we can obtain general  lessons for SUSY  effects in topological phases. 
Specifically, the robustness of the topological phase in the presence of SUSY is discussed in light of modern symmetry-protected topological order argument with emphasis on  edge superspin picture.   We also generalize the lower SUSY analyses to the case of higher SUSY; fuzzy four-superspheres and  $UOSp(1|4)$ SVBS states.     
 
The following is the organization of the present paper.    
In Sec.~\ref{sec:fuzzytwoVBS}, we give a brief introduction to the original $SU(2)$ VBS states 
and explain its relations to fuzzy two-sphere and the 2D QHE 
through the Schwinger-operator representation and the Hopf map.    
In Sec.~\ref{SecFuzzySuper-geometry}, the SUSY extensions of fuzzy two-sphere are presented in detail.  
The corresponding SVBS states and their basic properties are reviewed in Sec.~\ref{secsvbs}.   
In Sec.~\ref{secsmat}, the SMPS formalism is introduced for the analysis of SVBS states.   
Gapped excitations of SVBS states are derived with use of SMPS formalism within single mode approximation.  
We investigate topological properties of the SVBS states in Sec.~\ref{Sectopological}, where the entanglement spectrum and entanglement entropy are derived. (Unpublished results for $UOSp(2|2)$ SVBS chain are also included here.)   The stability of topological phases in the presence of SUSY  is discussed too.    
In Sec.~\ref{SecthigherDfuzzy}, we extend the discussions to the case of fuzzy four-superspheres and the $UOSp(1|4)$ SVBS states.            
 Sec.~\ref{SectSummary} is devoted to summary and discussions.  
In Appendix \ref{Appen:SecUOSp(MN)}, we provide mathematical supplements for the super-Lie group, $UOSp(N|2K)$.   Fuzzy four-superspheres with arbitrary number of SUSY are described in Appendix \ref{SectMoreSUSY}.

\section{Fuzzy Geometry and Valence Bond Solid States}\label{sec:fuzzytwoVBS}
In this section, 
we give a quick introduction to fuzzy two-spheres and the (bosonic) VBS states.  
Their mutual relations will be discussed, too.  

The  original idea of quantization of two-sphere was first introduced by Berezin in '70s \cite{berezin1975}.  
In the beginning of '80s, Hoppe \cite{Hoppe1982} explored 
the algebraic structure and field theories on a quantized sphere, and subsequently in the early '90s,   
Madore \cite{madore1992}  further examined their structures coining the name ``fuzzy sphere''. 
Unlike the ordinary sphere, the fuzzy sphere has a minimum scale of area, 
while respecting the $SU(2)$ $\it{continuum}$ (rotational) symmetry as the ordinary two-sphere. 
This is a remarkable property of the fuzzy sphere, when we consider the field theories 
on it.  In fact, in the other regularized field theories, 
such as lattice field theories, the extrinsic cut-off cures the UV divergence 
at the cost of continuous symmetries and the resulting theories only respect 
the $\it{discrete}$ space-group symmetry of the lattice on which they are defined.  
On the other hand, field theories defined on the fuzzy manifolds contain an intrinsic ``cut-off'' 
coming from the minimum area of the fuzzy sphere, and the non-commutative field theories 
constructed on it were expected to have the innate property which might soften 
the UV divergence to be appropriately renormalized in conventional field theories.   

In the middle of '90s, Grosse et al. \cite{hep-th/9602115} generalized the notion of the fuzzy two-sphere 
to construct four-dimensional fuzzy spheres and supersymmetric  fuzzy spheres, 
$i.e.$ fuzzy superspheres \cite{Grosseeetal1995, GrosseReiter1998}. 
In late 90s, the fuzzy geometry was rediscovered in the context of the ``second revolution 
of string theory'' and attracted a lot of attentions.  Researchers began to recognize that 
the fuzzy geometry is closely related to the geometry that the string theory attempts 
to describe: The geometry of multiple D-branes is naturally described by fuzzy geometry 
(see Refs.~\cite{hep-th/0007170,hep-th/0101126,hep-th/0512054} as reviews) 
and fuzzy manifolds are also known to arise as classical solutions of Matrix theory 
(see e.g. Refs.\cite{hep-th/9711078,hep-th/9910053}).   
Similarly, field theories on fuzzy superspheres provide a set-up for  SUSY field theories 
with UV regularization \cite{Grosseeetal1995,hep-th/9903112,hep-th/9903202}, 
and the fuzzy superspheres were also found to arise as classical solutions 
of supermatrix models \cite{hep-th/0311005,hep-th/0312307}.   
For details and applications of the fuzzy sphere, interested readers may consult  
Refs.~\cite{hep-th/0106048,Azumathesis,BalachandranealReview,Abethesis}.  
Non-commutative geometry and fuzzy physics also found their applications 
in gravity \cite{arXiv:hep-th/0504183,arXiv:hep-th/0506157,hep-th/0606197} 
and  in condensed matter physics \cite{Girvin1984,Girvin-M-P-86,Ezawaetal2003}. 
\subsection{Fuzzy two-spheres and the lowest Landau level physics}\label{subsec:fuzzytwoandlll}
\label{sec:fuzzy-vs-LLL}
The fuzzy two-sphere  \cite{berezin1975,Hoppe1982,madore1992} is one of the simplest curved fuzzy manifolds \footnote{The fuzzy spheres also play a crucial role in studies of string theory (see  \cite{Taylor1998,TAzuma2004,YAbe2005} for reviews.)}.  
The coordinates of fuzzy two-sphere, $X_i$ $(i=1,2,3)$, are regarded as the operators that are constituted  of the $SU(2)$ generators  
\begin{equation}
X_i=\frac{2R}{d} J_i. 
\label{defXi}
\end{equation}
Here, $R$ denotes the radius of the sphere and  $d$ is the dimension of the $SU(2)$ irreducible representation:  
\begin{equation}
d=2j+1,  
\end{equation}
with  the $SU(2)$ Casimir index $j$ $(j=0, \frac{1}{2}, 1, \frac{3}{2}, 2,\cdots)$, and   
 $J_i$ $(i=1,2,3)$ are the $SU(2)$ matrices of the corresponding representation that satisfy 
\begin{equation}
[J_i, J_j]=i\epsilon_{ijk}J_k, 
\label{eqn:SU2-comm-rel}
\end{equation}
and 
\begin{equation}
J_iJ_i=j(j+1)\bold{1}_d, 
\end{equation}
where $\bold{1}_d$ denotes  $d\times d$ unit matrix. 
The `coordinates' $X_i$ defined in \eqref{defXi} satisfy 
\begin{equation}
[X_i, X_j]=i \frac{2R}{d} \epsilon_{ijk}X_k, \quad 
X_iX_i=R^2\left(1-\frac{1}{d^2}\right)\bold{1}_d .  
\label{eqn:coord-comm-rel}
\end{equation}
Since $J_3$ takes the eigenvalues $j, j-1, j-2, \cdots, -j$, the eigenvalues of $X_3$ are given by   
\begin{equation}
X_3=\left\{ 
R \biggl(1-\frac{1}{d}\biggr),~ R \biggl(1-\frac{3}{d}\biggr),
~ R \biggl(1-\frac{5}{d}\biggr),~ \cdots,~  R \biggl(-1+\frac{1}{d}\biggr) 
\right\} \; .
\label{distrix3}
\end{equation}
Each ``latitude'' of $X_3$ corresponds to a patch of width $2R/d$, 
{\it i.e.} the minimum state, on the fuzzy sphere\footnote{%
Here two different viewpoints are possible.  
One assumes, as is done here, $R$ to be fixed and, in the large-$d$ (or, large-$j$) limit, the width of 
the patch $2R/d$ becomes zero to recover the continuous space.  
The other fixes the width $2R/d$ and the radius of the fuzzy two-sphere $R \sim d$ diverges 
in the large-$d$ limit. The former is reminiscent of the scaling limit of lattice field theories upon 
identifying $R$ with the physical mass.}.    
If the limit of large $SU(2)$ representation $d\rightarrow \infty$ (with fixed radius $R$) is taken, 
the patches become invisible and the discrete nature of the fuzzy sphere is smeared off.  
In fact, 
one readily sees $[X_i, X_j]=0$ and $X_iX_i|_{d \rightarrow \infty} =R^2$  
and  the discrete spectrum \eqref{distrix3} of $X_3$ becomes continuum raging between 
$-R$ and $R$.  
In this sense, the fuzzy sphere reduces to the ordinary (commutative) sphere with radius $R$ 
in the limit 
$d \rightarrow \infty$. 

As a realization of the fuzzy two-sphere, a convenient way is to utilize the Schwinger operator formalism \cite{watamuras1997,BalachandranealReview} and introduce the following operator
\begin{equation}
\Phi=(a,~b)^t 
\label{eqn:def-Schwinger}
\end{equation}
whose components satisfy the following ordinary bosonic commutation relations:  
\begin{equation}
[a,a^{\dagger}]=[b,b^{\dagger}]=1,~~ 
[a,b]=[a,b^{\dagger}]=0. 
\end{equation}
By sandwiching the Pauli matrices with the Schwinger operators $\Phi$, 
one may simply represent the coordinates $X_i$ of the fuzzy two-sphere as:  
\begin{equation}
X_i=\frac{R}{d} \Phi^{\dagger}\sigma_i \Phi, 
\label{Schwingersf2}
\end{equation}
where $\sigma_i$ $(i=1,2,3)$ are the Pauli matrices:  
\begin{equation}
\sigma_1=\begin{pmatrix}
0 & 1 \\
1 & 0 
\end{pmatrix},~~\sigma_2=
\begin{pmatrix}
0 & -i \\
i & 0 
\end{pmatrix},~~
\sigma_3=
\begin{pmatrix}
1 & 0 \\
0 & -1 
\end{pmatrix}. 
\end{equation}
This is the well-known construction of the $SU(2)$ angular momentum operators 
introduced by Schwinger \cite{Schwinger1965}.   
In fact, one can easily check that $J_i = (1/2)\Phi^{\dagger}\sigma_i \Phi$ satisfy 
the standard $SU(2)$ commutation relations \eqref{eqn:SU2-comm-rel} and that 
the Schwinger-operator representation \eqref{Schwingersf2} coincides 
with the original definition \eqref{defXi}.  
The square of the radius is given by 
\begin{equation}
X_iX_i=\left( \frac{R}{d} \right)^2 {n}({n}+2)\bold{1}_{n+1},   
\label{fuzzytwosphereradius}
\end{equation}
where ${n}$ denotes the eigenvalues of the number operator for  the Schwinger bosons,  
$\hat{n}=\Phi^{\dagger}\Phi=a^{\dagger}a+b^{\dagger}b$. 
Comparing this with eq.\eqref{eqn:coord-comm-rel}, one sees $j=n/2$ and 
\begin{equation}
d(n)=n+1 .  
\label{su2repre}
\end{equation}
Since we utilize the Schwinger operator, the  irreducible representation is given by the fully symmetric representation of $SU(2)$:   
\begin{equation}
|n_1,n_2\rangle=
\frac{1}{\sqrt{n_1!~n_2!}}
{a^{\dagger}}^{n_1}{b^{\dagger}}^{n_2}|\text{vac}\rangle, 
\label{su2irredu}
\end{equation}
where $n_1$ and $n_2$ are non-negative integers that satisfy $n_1+n_2=n$.  
Physically, $|n_1,n_2\rangle$ stand for a finite number of basis states 
constituting fuzzy two-sphere which are the eigenstates  of $X_3$ with the eigenvalues:      
\begin{equation}
X_3=\frac{R}{d}(n_1-n_2)=\frac{R}{d}(n-2k),  
\label{X3fuzzysphere}
\end{equation}
where $k\equiv n_2=0,1,2,\cdots,n$. 
 The dimension of the space spanned by the basis states (\ref{su2irredu}) is given by $d(n)$.  
From the expression \eqref{X3fuzzysphere}, it is not obvious that the definition \eqref{Schwingersf2} 
respects the $SU(2)$ rotational symmetry of the fuzzy two-sphere.  
However this is an artifact of the particular choice of the $X_3$-diagonal basis states (\ref{su2irredu}).   
Any other complete sets of the basis states of the fuzzy two-sphere can be obtained by 
applying $SU(2)$ transformations to the basis set (\ref{su2irredu}). 
In this sense, whole Hilbert space of fuzzy two-sphere is ``symmetric'' with respect to 
the $SU(2)$ transformation, 
and hence the fuzzy two-sphere is $SU(2)$ (rotationally) symmetric like the ordinary (continuum) sphere.     

Another  description of fuzzy two-sphere is to utilize the  coherent state\footnote{%
The coherent state here is usually referred to as the Bloch spin coherent state 
in literature.}  formalism, which will be useful in understanding the connection with the VBS states. 
The coherent state $\phi$, or more precisely the Hopf spinor, 
that is labeled by a point on two-sphere $(x_1,x_2,x_3)$ $(x_ix_i=1)$,  is defined as 
a 2-component complex vector satisfying\footnote{%
The usual coherent `state' $|\phi\rangle$ (for $S=1/2$) defined as
$|\phi\rangle \equiv \Phi^{\dagger}{\cdot}\phi|0\rangle$ 
satisfies 
\[ (2x_i{\cdot}J_i)|\phi\rangle = |\phi\rangle \; . \] 
Expressing the above in the basis $a^{\dagger}|0\rangle$ and $b^{\dagger}|0\rangle$, 
we obtain eq.\eqref{su2spincoherenteq}.} 
\begin{equation}
(x_i \cdot \sigma_i) \phi (\{x_i\})= \phi(\{x_i\})  . 
\label{su2spincoherenteq}
\end{equation}
In general, the (normalized) coherent state is represented in terms of the Euler angles as 
\begin{equation}
\phi(\{x_i\}) = 
\begin{pmatrix}
u \\
v
\end{pmatrix}=\frac{1}{\sqrt{2(1+x_3)}}
\begin{pmatrix}
1+x_3 \\
x_1+ix_2
\end{pmatrix}e^{-\frac{i}{2}\chi}
=\begin{pmatrix}
\cos\frac{\theta}{2}\\
\sin\frac{\theta}{2}e^{i\varphi}
\end{pmatrix}e^{-\frac{i}{2}\chi}, 
\end{equation}
where $(x_1,x_2,x_3)=(\sin\theta\cos\varphi,\sin\theta\sin\varphi,\cos\theta)$ 
($0\leq \varphi <2\pi$, $0\leq \theta \leq \pi$, $0\leq \chi < 2\pi$), 
and $e^{-\frac{i}{2}\chi}$ denotes an arbitrary $U(1)$ phase factor.  
The relation between the point $(x_1,x_2,x_3)$ on a two-sphere and the two-component 
Hopf spinor $(u,v)$ is given by the so-called 1st Hopf map\footnote{%
For construction of higher dimensional Hopf maps, see Ref.\cite{Hasebe2010review} for instance.}
\begin{equation}
x_i =\phi^{\dagger}\sigma_i\phi \; ,
\label{theexplicitmapof1stHopf}
\end{equation}
which maps the three-sphere $S^3$ onto the two-sphere $S^2$:   
\begin{equation}
S^3~\overset{S^1}\longrightarrow ~S^2.   
\label{hopfmap1st}
\end{equation}
In fact, a space of normalized two-component complex spinors $\phi=(u,v)^t$ 
subject to $\phi^{\dagger}\phi=1$ is isomorphic to $S^{3}$ 
and eq.\eqref{theexplicitmapof1stHopf} gives the mapping onto $S^{2}$:
\begin{equation}
x_ix_i=(\phi^{\dagger}\phi)^2=1. 
\label{twossphereradius}
\end{equation}

The Hopf spinor $\phi$ is regarded as the classical counterpart of the Schwinger operator $\Phi$ 
[eq.\eqref{eqn:def-Schwinger}] that satisfies  
\begin{equation}
 X_i \cdot \Phi^{\dagger}\sigma_i  =\frac{R}{d} ({n}+2)\Phi^{\dagger}. 
\label{coherentschwingereq}
\end{equation}
If we note that $R({n}+2)/d \rightarrow R$, $X_i \rightarrow R x_i$ in the limit $d\rightarrow \infty$, 
the resemblance between (\ref{su2spincoherenteq}) and (\ref{coherentschwingereq}) is clear. 
Multiplying $(R/d)\Phi$ from the right to both sides of Eq.(\ref{coherentschwingereq}) 
and using $\Phi^{\dagger}\Phi=n$, we reproduce Eq.(\ref{fuzzytwosphereradius}).     
 By replacing the Schwinger operator with the Hopf spinor, one sees that 
 the Schwinger operator construction (\ref{Schwingersf2}) of fuzzy coordinates 
is an operator-generalization ($x_i \mapsto X_i$, $\phi \mapsto \Phi$) 
 of the Hopf map \eqref{theexplicitmapof1stHopf} \cite{Grosseetal1996}.

The fully symmetric $SU(2)$ representation in terms of the Hopf spinor $(u,v)$  is obtained 
if  the Schwinger operator in  (\ref{su2irredu}) is replaced with the Hopf spinor 
$(a^{\dagger},b^{\dagger})~\rightarrow~(u,v)$ 
\begin{equation}
\begin{split}
& u_{n_1,n_2}={{\frac{1}{\sqrt{n_1!~n_2!}}}} u^{n_1}v^{n_2}, \\
& J^{+} = u\frac{\partial}{\partial v} \; , \;\;
J^{-} = v\frac{\partial}{\partial u} \; , \;\;
J^{z}= \frac{1}{2}\left(
u\frac{\partial}{\partial u}
- v \frac{\partial}{\partial v}
\right) 
\end{split}
\label{symmetrichopf}
\end{equation}
with $n_1+n_2=n$, $n_1,n_2\geq 0$.  

\begin{figure}[!t]
\centering
\includegraphics[width=14cm]{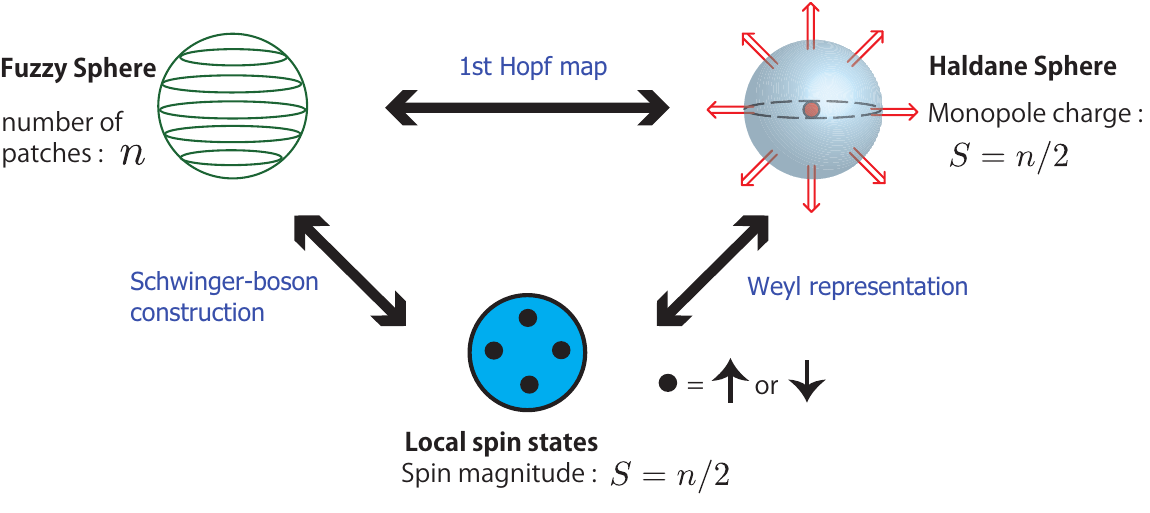}
\caption{(Color online) One-body-level relationship among 
fuzzy two-sphere (upper left), Haldane sphere (upper right) 
and local spin states of the VBS state (lower middle). 
The fuzzy two-sphere consists of a finite number of patches, $i.e.$, the basis states, 
with width $2R/(n+1)$.  
The Haldane sphere is a two-sphere with Dirac monopole at its center. 
The $S=n/2$ is the  monopole charge quantized as half-integer or integer by the Dirac quantization condition. 
In the local spin state of the VBS state (lower center), each blob denotes spin-$1/2$ degrees of freedom, 
and $n$ blobs amount to $S=n/2$  local spin by a large Hund coupling on each site.  
\label{newfuzzysphereanVBsite.fig} }
\end{figure}

So far, $(u,v)$ are just the two auxiliary variables needed to realize the fuzzy two-sphere. 
However, if we regard them as the physical coordinates of the {\em usual} two-dimensional 
sphere [via the Hopf map \eqref{hopfmap1st}], 
we may think that $u_{n_1,n_2}$ represents the wave functions of a certain kind 
of quantum-mechanical systems in two dimensions.   
In fact, the wave functions $u_{n_1,n_2}$ (\ref{symmetrichopf}) coincide with those of 
the lowest eigenstates of the Landau Hamiltonian on a two-sphere (with radius $R$) 
in the Dirac monopole background ({\em i.e.} the so-called Haldane's sphere \cite{haldane1983}): 
\begin{equation}
H=\frac{1}{2MR^2}\Lambda_i\Lambda_i, 
\end{equation}
where $\Lambda_i$ $(i=1,2,3)$ are the covariant angular momenta  
\begin{equation}
\Lambda_i=-i\epsilon_{ijk}x_j(\partial_k+iA_k)
\end{equation} 
in the presence of the gauge field $A_i$ generated by the Dirac monopole at the origin:
\begin{equation}
A_i=\frac{n}{2}\epsilon_{ij3}\frac{x_j}{R(R+x_3)}.
\end{equation}
In the fuzzy geometry side, the monopole charge $n/2$ corresponds to the quantized radius of the fuzzy two-sphere (\ref{fuzzytwosphereradius})  (see also Fig.~\ref{newfuzzysphereanVBsite.fig}).  
Thus, the lowest Landau level eigenfunctions (on a two-sphere) 
can be ``derived'' from the fuzzy two-sphere 
by switching from the Schwinger operator formalism to the coherent state formalism (or, 
the Weyl representation).  
This is the key observation of correspondence between the fuzzy geometry 
and the lowest Landau level physics \cite{Hasebe2010review}.

\subsection{Valence bond solid states}\label{subsec:valenvebond}
\label{sec:VBS-introduction}
In order to translate the above features to those of QAFM, we first express the spin $1/2$ state 
in terms of the Schwinger bosons:   
\begin{equation}
|\!\uparrow\rangle=a^{\dagger}|\text{vac}\rangle,~~~~~|\!\downarrow\rangle=b^{\dagger}|\text{vac}\rangle.
\label{su2updoenstates}
\end{equation}
The fully symmetric representation (\ref{su2irredu}) constructed out of $n$ Schwinger operators 
automatically realizes the local spin defined on each site with the spin magnitude   
\begin{equation}
S={\overbrace{(\frac{1}{2}\otimes \frac{1}{2}
\otimes \cdots \otimes \frac{1}{2})}^{n}}_{\text{fully~symm.}}=\frac{n}{2}. 
\label{eqn:symmetric-rep}
\end{equation}
Physically, this bosonic construction realizes the ferromagnetic (Hund) coupling 
among $n$ spin-1/2s to yield the maximal spin, $S=n/2$, at each site.   
We have already seen that the spin wave function, written in terms of the Hopf spinor 
$(u,v)$, coincides with those of the non-interacting electrons moving on a two-sphere 
in the presence of the monopole magnetic field.  

\begin{figure}[!t]
\centering
\includegraphics[width=15cm]{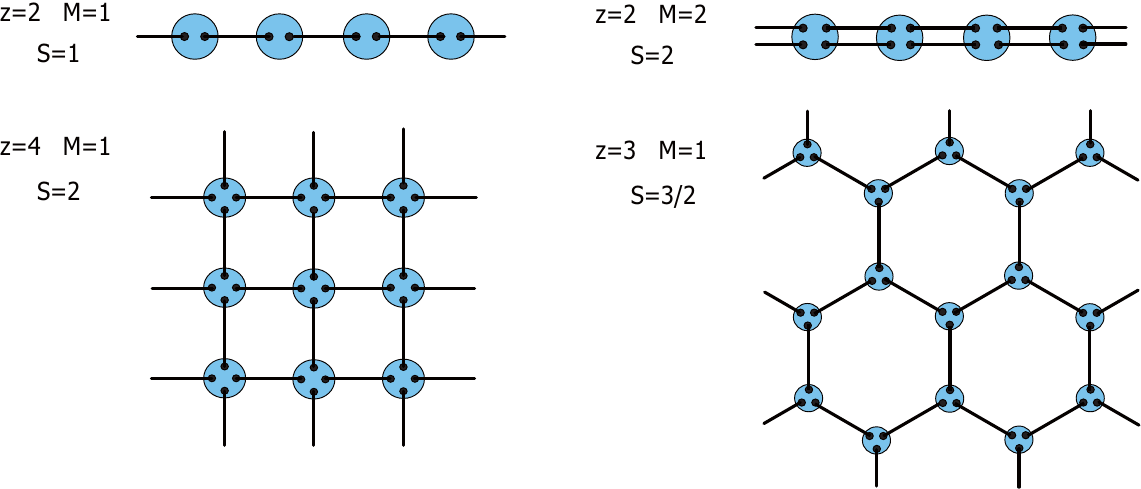}
\caption{(Color online) 
 VBS states on 1D and 2D lattices.  Filled circles denote auxiliary spin-1/2 objects 
 which are finally symmetrized to form $S=Mz$ physical spins at each site.  
 Solid lines stand for singlet valence bonds between the spin-1/2s.   
 \label{examplesAKLT.fig} }
\end{figure}

Now we demonstrate that a similar analogy exists even when we switch on strong interactions.  
Specifically, the exact ground-state wave functions of a class of interacting spin models called 
the valence-bond solid (VBS) model \cite{Afflecketal1987,Afflecketal1988,Arovasetal1988} 
closely resemble the Laughlin wave functions on a two-sphere.  
The ground states of the VBS models (dubbed the VBS states) on a lattice with coordination 
number $z$ (see Fig.~\ref{examplesAKLT.fig}) are constructed as follows.  
As the first step of the construction, 
we prepare $n$ ($=z$) local $S=1/2$ spins ({\em auxiliary spins}) on each vertex (site) of the lattice.   
Next, we connect every pair of two spin-1/2s on the nearest neighbor sites 
by spin-singlet bonds\footnote{%
Since the spin-singlet state of the two spin-1/2s maximizes the entanglement entropy,  
the pair is sometimes called {\em maximally entangled} in modern literatures.} 
called the {\em valence bonds} (VB) 
(see Fig.~\ref{examplesAKLT.fig}).  
Last, we project the entire $2^{n}$-dimensional Hilbert space of $n$ spin-1/2s onto the subspace 
with the desired value of the ({\em physical}) spin $S$ ($S\leq n/2$) 
on each site to obtain the spin-$S$ many-body  
singlet state.  Depending on which irreducible representation we use to realize the physical spin, 
we obtain different states even for the same lattice structure (some examples may be found in, 
{\em e.g.}, Ref.~\cite{Tuetal2009}).   
This kind of construction ({\em valene-bond construction}) applies 
to any lattice in any dimensions (see Fig.~\ref{examplesAKLT.fig} for typical examples 
in 1D and 2D) and the state obtained thereby is called 
the VBS state or, in a modern terminology, the projected entangled-pair state (PEPS) 
\cite{Verstraete-W-G-C-06,Garcia-V-C-W-08}.  

It is also possible to construct the VBS states for other symmetry groups 
(e.g., $SU(N)$ \cite{Greiteretal2007I,Greiteretal2007II} 
and $SO(N)$ \cite{Tuetal2008I,Tuetal2008II}) with due 
extension of the notion of the singlet valence bond.  
In these examples, the VBS states thus constructed are constrained by specific symmetry groups 
and normally they do not have any tunable parameter\footnote{%
However, the parent Hamiltonians ({\em i.e.} the VBS models) may have tunable parameters.  
For instance, the parent Hamiltonian of the spin-2 VBS state (in 1D) allows one 
free parameter to tune even after we fix the overall energy scale.}.  
However, if we use the matrix-product representation \cite{Fannes-etal1992} in 1D or,  
in general, tensor-product representation 
(or, equivalently, the vertex-state representation \cite{Niggemannetal1997,Niggemannetal2000}) 
in higher dimensions to express the VBS states 
and consider their ``anisotropic'' extensions of the tensors, 
it is possible to obtain parameter-dependent states.   
These generalized VBS states are known to have interesting properties.  
See, for instance, Refs.~\cite{Fannes-etal1992,Klumperetal1992,Klumperetal1993} for 1D 
and Refs.~\cite{Niggemannetal1997,Niggemannetal2000} for 2D honeycomb- and square lattices.  

Now let us come back to the SU(2) VBS states.  
If we represent the up and down degrees of freedom of auxiliary spin-1/2 by the Schwinger bosons $a^{\dagger}$ 
and $b^{\dagger}$ (see eq.\eqref{su2updoenstates}), the singlet valence bond 
on the bond $\langle i j \rangle$ reads
\begin{equation}
(a_i^{\dagger}b_j^{\dagger}-b_i^{\dagger}a_j^{\dagger})  .
\end{equation}
Physically, the spin singlet bond denotes the state with no specific spin polarization made of
two spin 1/2 states, and hence the valence bond represents a non-magnetic spin pairing 
between the two neighboring sites. 

Of course, 
we can promote the power of the singlet bond from one to an arbitrary integer $M$
\begin{equation}
(a_i^{\dagger}b_j^{\dagger}-b_i^{\dagger}a_j^{\dagger})^M 
\end{equation}
to represent $M$ valence bonds between  the sites $i$ and $j$.   
Thus, one sees that the original valence-bond construction \cite{Afflecketal1987,Afflecketal1988} 
of the VBS states is equivalent to the following representation 
in terms of the Schwinger bosons \cite{Arovasetal1988}
\begin{equation}
|\text{VBS}\rangle = 
\prod_{\langle ij\rangle\in \text{N.N.}}
(a_i^{\dagger}b_j^{\dagger}-b_i^{\dagger}a_j^{\dagger})^M|\text{vac}\rangle, 
\label{generalsu2vbs}
\end{equation}
where $\langle ij\rangle\in \text{N.N.}$  implies that the product is taken over all nearest-neighbor 
bonds $\langle ij\rangle$ and $|\text{vac}\rangle$ denotes the vacuum of the Schwinger bosons.  
From $[S^{a}_{i}+S^{a}_{j},a_i^{\dagger}b_j^{\dagger}-b_i^{\dagger}a_j^{\dagger}]=0$ ($a=x,y,z$), 
it is obvious that the state eq.\eqref{generalsu2vbs} is spin-singlet. 
As $z$ bonds emanate from each site of the lattice (for instance, $z=2$ is for 1D chain  and  $z=2D$ for 
$D$-dimensional hypercubic lattice), we have $Mz$ Schwinger bosons per site 
in the VBS state (\ref{generalsu2vbs}) 
\begin{equation}
a^{\dagger}_ia_i+b^{\dagger}_ib_i=zM, 
\end{equation}
and hence the local spin quantum number $S_i=\frac{1}{2}(a^{\dagger}_ia_i+b^{\dagger}_ib_i)$ is given by 
\begin{equation}
S_i=\frac{1}{2}{zM}. 
\end{equation}
In particular, for the 1D (i.e., $z=2$) $M=1$ VBS state, we have 
\begin{equation}
S_i=1 
\end{equation}
and the local Hilbert space is spanned by the following three basis states: 
\begin{align}
&|1\rangle=\frac{1}{\sqrt{2}}{a_i^{\dagger}}^2|\text{vac}\rangle,\quad 
|0\rangle={a_i^{\dagger}}b_i^{\dagger}|\text{vac}\rangle, \quad 
|{-1}\rangle=\frac{1}{\sqrt{2}}{b_i^{\dagger}}^2|\text{vac}\rangle.
\end{align}

We were a bit sloppy in writing down eq.\eqref{generalsu2vbs}.  
When considering a finite open chain (with length $L$), 
we should be careful in dealing with the edges of the system, while,  
for the VBS states on a circle, the expression \eqref{generalsu2vbs} is correct without 
any modification.   
In fact, in \eqref{generalsu2vbs}, the number of the Schwinger bosons at the sites 1 and $L$ 
is $M$ (half of that of the other sites) and we have to add the {\em extra} edge degrees of 
freedom represented by the $M$-th order homogeneous polynomials in 
$a^{\dagger}$ and $b^{\dagger}$ to recover the correct spin-$M$s at the edges.   
The edge polynomials for $M=1$ are given by  
\begin{equation}
f_{\uparrow}(a^{\dagger},b^{\dagger})=a^{\dagger}, ~~
f_{\downarrow}(a^{\dagger},b^{\dagger})=b^{\dagger} .
\end{equation}
This representation naturally incorporates the physical emergent edge 
spins \cite{Hagiwara-K-A-H-R-90,Kennedy-90} localized {\em around} the two edges.
For general $M$, the precise form of the VBS states on an open chain reads as
\begin{equation}
|\text{VBS}\rangle = (a^{\dagger}_{1})^{p}(b^{\dagger}_{1})^{M-p}
\prod_{j=1}^{L-1}(a_j^{\dagger}b_{j+1}^{\dagger}-b_{j}^{\dagger}a_{j+1}^{\dagger})^M
(a^{\dagger}_{L})^{M-q}(b^{\dagger}_{L})^{q}  |\text{vac}\rangle \;\; 
(0 \leq p,q \leq M) \; .
\label{generalsu2vbs-2}
\end{equation}

\begin{figure}[!t]
\centering
\includegraphics[width=12cm]{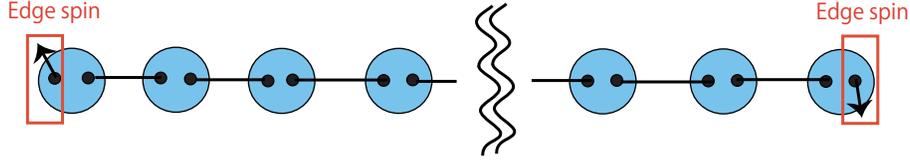}
\caption{(Color online) For the $S=1$ VBS state on a finite {\em open} chain, 
there exist spin-1/2 degrees of freedom at each edge.   
By construction, the VBS state is the ground state of the VBS Hamiltonian 
regardless of the spin states at the edges. 
   \label{AKLTedgeex.fig} }
\end{figure}

Since the VBS Hamiltonian is defined as the projection operator for 
the Hilbert space of a pair of neighboring spins (see Sec.~\ref{sec:parent-Ham} for the detail), 
the state of the form \eqref{generalsu2vbs} is the ground state {\em regardless} of 
the edge polynomials.  Therefore, there appear $(M+1)\times (M+1)$ degenerate ground states 
for the spin-$S$($=M$) VBS state 
on a finite open chain (see Fig.~\ref{AKLTedgeex.fig}) \cite{Afflecketal1988}. 
From similar considerations, it is obvious that the higher-dimensional VBS ground states 
(e.g., those in Fig.~\ref{examplesAKLT.fig}) have degeneracy exponentially large in 
the size of the boundary.  
In the following, unless otherwise stated, we implicitly assume that the VBS state is defined in 1D 
and will focus on the spin-1 case. 

Let us look at some key features of the VBS states in more detail. 
Classically, AFM on a {\em bipartite} lattice assumes the N\'{e}el-ordered state, 
where any pair of neighboring spins point the opposite directions.  
To be specific, in the N\'{e}el state of (classical) spin-1 AFM, the $z$-component $S_z$ 
of local spins at the site $i$ 
takes either $+1$ or $-1$ depending on to which sublattice the site belongs. 
(Without loss of generality, we may assume that ordered spins are parallel to the $z$-axis.) 
The spin configuration described by the VBS state is totally different from that of 
the classical N\'{e}el state described above. 
First of all, we note that the VBS state is $SU(2)$-invariant (being a product of singlet bonds 
between pairs of adjacent sites; see eq.\eqref{generalsu2vbs} and Fig.~\ref{examplesAKLT.fig}) 
and  is  non-degenerate,\footnote{%
This is true for a periodic chain and the bulk in a infinite chain. 
On a finite open chain, the ground state may not be spin-singlet and hence may show 
degeneracy corresponding to the edge states.} 
while the classical ground state, $i.e.$ N\'{e}el  state, is infinitely degenerate 
with respect to the $SU(2)$ rotational symmetry.  
This implies that though the magnitude of the local spin is $S_i=1$, 
its expectation value is zero $\langle \bold{S}_i\rangle_{\text{VBS}}=\mathbf{0}$ 
in the bulk\footnote{%
On a finite {\em open} chain, $\langle \bold{S}_i\rangle_{\text{VBS}}$ may take non-zero finite 
value near the two boundaries and decays exponentially to zero toward the center of the system.  
In this sense, magnetism revives near the boundaries and this is the manifestation of 
the emergent edge states.} and the system is non-magnetic.     
In this sense,  the VBS state is purely quantum-mechanical and does not have 
any classical counterpart.    

For better understanding of the VBS state, let us expand the spin-1 VBS state in the $S^{z}$-basis: 
\begin{align}
|\text{VBS}\rangle
&=\prod_{i} (a_i^{\dagger}b_{i+1}^{\dagger}-b_i^{\dagger}a_{i+1}^{\dagger})|\text{vac}\rangle \nonumber\\
&
=
|\cdots 000 \cdots  \rangle + 
|\cdots  00- \cdots \rangle + |\cdots  0-+ \cdots  \rangle+|\cdots  0-0 \cdots  \rangle\nonumber\\
& + |\cdots 0+0\cdots \rangle+|\cdots -+-\cdots \rangle+|\cdots -0+\cdots \rangle 
+ |\cdots -+0\cdots \rangle\nonumber\\
& + |\cdots +00\cdots \rangle+|\cdots +0-\cdots \rangle+|\cdots +-+\cdots \rangle
 + |\cdots +-0\cdots \rangle\nonumber\\
&+|\cdots 0+0\cdots \rangle + |\cdots 0+-\cdots \rangle + |\cdots 00+\cdots \rangle+|\cdots 000 \cdots\rangle
\; , 
\label{expVBSeq}
\end{align}
where the coefficient in front of each term on the right-hand side is omitted for simplicity.  
What is remarkable with the above VBS state is that all the states appearing on the right-hand side 
have a very special feature; the states $S^{z}=+1$ and $-1$ appear {\em in an alternating manner} 
with intervening $S^{z}=0$ states.  Namely, the ground state exhibits an analogue of 
the classical N\'{e}el order called the {\em string order} \cite{NijsRommelse-1989,Tasaki1991} 
albeit ``diluted'' by randomly inserted zeros (see Sec.~\ref{sec:string-ordinary-VBS} for 
further detail).  
Unlike in the case of the classical N\'{e}el order, by the SU(2) symmetry of the state, 
the existence of the string order does not rely on the particular choice of the quantization axis 
($z$-axis here).  

For the sake of later discussions, we introduce here a concise representation  
of the VBS state and point out remarkable 
similarity \cite{Arovasetal1988}
to the Laughlin-Haldane wave function \cite{haldane1983} for FQHE on a two-sphere. 
First, we note that by using the $SU(2)\simeq USp(2)$ invariant matrix given 
by the 2$\times$2 antisymmetric matrix  (see Appendix \ref{Appen:SecUOSp(MN)}) 
\begin{equation}
\mathcal{R}_{2}= i\sigma_2=\begin{pmatrix}
0 & 1 \\
-1 & 0 
\end{pmatrix}, 
\end{equation}
the VBS states (\ref{generalsu2vbs}) on generic lattices can be rewritten compactly as 
\begin{equation}
|\text{VBS}\rangle = \prod_{\langle ij\rangle \in \text{N.N.}} 
(\Phi_i^{\dagger}\mathcal{R}_2 \Phi_j^*)^M|\text{vac}\rangle, 
\label{eqn:VBS-by-Phi-R-Phi}
\end{equation}
where $\Phi_i$ denotes the Schwinger operator on the site $i$
\begin{equation}
\Phi_i^*\equiv (a_i^{\dagger},b_i^{\dagger})^t. 
\end{equation}

Then, we rewrite the VBS state \eqref{eqn:VBS-by-Phi-R-Phi} 
into the form of the wave function on a two-sphere.  
Specifically, by replacing the Schwinger operator with the Hopf spinor 
\begin{equation}
\Phi^{\ast} =(a^{\dagger},b^{\dagger})^t ~\rightarrow ~\phi=(u,v)^t, 
\end{equation}
we obtain the coherent-state (or, Weyl) representation of the VBS state\footnote{%
Precisely, if we use the coherent-state basis
\[ |\boldsymbol{\Omega_{i}\rangle} = \frac{1}{\sqrt{(Mz)!}}
(u_i a_{i}^{\dagger}+v_i b_{i}^{\dagger})^{Mz}
|\text{vac}\rangle\]
for the local spin-$Mz/2$ states and expand the VBS state in these basis, we obtain 
the `wave function' 
$\langle \{ \boldsymbol{\Omega}_{j} \} |\text{VBS}\rangle 
\propto \Phi_{\text{VBS}}(\{u^{\ast}_i,v^{\ast}_i\})$.}: 
\begin{equation}
\Phi_{\text{VBS}}(\{u_i,v_i\})
=\prod_{\langle ij\rangle \in \text{N.N.}}(\phi_i^t~\mathcal{R}_{2}~\phi_j)^M  
=\prod_{\langle ij\rangle \in \text{N.N.}}(u_iv_j-v_iu_j)^M,
\label{coherentrepLHstates} 
\end{equation}
Now the formal similarity to the Laughlin-Haldane wave function of QHE 
\begin{equation}
\Phi_{\text{LH}}= \prod_{i<j}(\phi_i^t~ \mathcal{R}_{2}~\phi_j)^m  = \prod_{i<j}(u_iv_j-v_iu_j)^m 
\label{LaughlinHaldanewavefunc}
\end{equation}
on a two-sphere \cite{Arovasetal1988} is clear.\footnote{%
By the stereographic projection from a two-sphere to a complex plane:
\begin{equation}
\phi=\begin{pmatrix}
u\\
v
\end{pmatrix} ~\longrightarrow ~ z={v/u}, 
\end{equation}
$\Phi_{\text{LH}}$ (\ref{LaughlinHaldanewavefunc}) is, in the thermodynamic limit, reduced to the celebrated Laughlin wave function, 
\begin{equation}
\Phi_{\text{L}}=\prod_{i<j}(z_i-z_j)^m e^{-\sum_i z_i z_i^*}. 
\label{celebLaugh}
\end{equation}
}  
Though the physical interpretation of the quantities appearing in these wave functions 
are different (Table \ref{correspondenceII}),   
mathematical similarities between the VBS model and QHE may be manifest from the above constructions.  Similarities between topological properties of VBS and QHE have also been discussed 
in Refs.\cite{Girvin-Arovas1989,Hatsugai1992}.

\hspace{-0.5cm}
\begin{table}
\renewcommand{\arraystretch}{1}
\begin{center}
\begin{tabular}{|c||c|c|}
\hline    & QHE &  QAFM  \\
\hline
\hline   Many-body state  & Laughlin-Haldane wave function  
 &  VBS state 
\\
& $\Phi_{\text{LH}}=\prod_{ i<j }^N(u_iv_j-v_i u_j)^m$  &  $|\Phi\rangle=\prod_{\langle ij\rangle}^z(a_i^{\dagger}b_j^{\dagger}-b_i^{\dagger}a_j^{\dagger})^M\vac$  
\\
\hline  Power  & $m$: inverse of filling factor &  $M$: number of VBs between neighboring sites    
\\ \hline 
 Charge & $S=m{N}/{2}$: monopole charge & $S={Mz}/{2}$: local spin magnitude   
\\ \hline
\end{tabular}
\end{center}
\caption{Correspondences between physical quantities of many-body wavefunctions of QHE and QAFM. 
}
\label{correspondenceII}
\end{table}

\section{Fuzzy Two-supersphere}\label{SecFuzzySuper-geometry}
Next, we proceed to the SUSY version of fuzzy two-sphere\footnote{%
Fuzzy superspheres are also realized as a classical solution of supermatrix models \cite{hep-th/0311005,hep-th/0312307}.} and generalized VBS states. 
We mostly focus on the cases of the SUSY numbers $\mathcal{N}=1$ and $2$. 

\subsection{$\mathcal{N}=1$}

First, we introduce  $\mathcal{N}=1$ SUSY algebra, $UOSp(1|2)$ \cite{ParisRittenberg1976,Scheunertetal1977,Marcu1980}, which contains the $SU(2)$ algebra as its maximal bosonic subalgebra. The $UOSp(1|2)$ algebra consists of the five generators three of which are $SU(2)$ (bosonic) generators $L_i$ $(i=1,2,3)$ and the remaining two are  $SU(2)$
fermionic spinor $L_{\alpha}$ $(\alpha=\theta_1,\theta_2)$, which amount to satisfy    
\begin{equation}
[L_i,L_j]=i\epsilon_{ijk}L_k,~~~
[L_i,L_{\alpha}]=\frac{1}{2}(\sigma_{i})_{\beta\alpha}L_{\beta},~~~\{L_{\alpha},L_{\beta}\}=\frac{1}{2}(i\sigma_2\sigma_i)_{\alpha\beta}L_i.\label{osp3} 
\end{equation}
The $UOSp(1|2)$ Casimir is constructed as  
\begin{equation}
\mathcal{K}=L_iL_i+\epsilon_{\alpha\beta}L_{\alpha}L_{\beta}, 
\end{equation}
and its eigenvalues are given by $j(j+1/2)$:  
$j$ is  referred to as the superspin taking non-negative integer or half-integer values, $j=0,1,\frac{1}{2},1,\frac{3}{2},\cdots$. 
The $UOSp(1|2)$ irreducible representation with the superspin index $j$ consists of the $SU(2)$ $j$ and $j-1/2$ spin representations, and hence the dimension of the $UOSp(1|2)$  representation of superspin $j$ is given by    
\begin{equation}
(2j+1)+(2j)=4j+1.  
\end{equation}
The fundamental representation $(j=1/2)$ matrices of the $UOSp(1|2)$ generators are expressed by the following 3$\times$3 matrices: 
\begin{align}
l_i=\frac{1}{2}
\begin{pmatrix}
\sigma_i & 0 \\
0 &  0 
\end{pmatrix},~~~~l_{\alpha}=
\frac{1}{2}
\begin{pmatrix}
0 & \tau_{\alpha}\\
-(i\sigma_2\tau_{\alpha}) & 0
\end{pmatrix},\label{superPaulimat}
\end{align}
where $\tau_{\alpha}$ ($\alpha=1,2$) are 
\begin{equation}
\tau_{1}=
\begin{pmatrix}
1 \\
0\end{pmatrix},~~~  \tau_{2}=
\begin{pmatrix}
0 \\
1\end{pmatrix}.  
\end{equation}
Eq.(\ref{superPaulimat}) may be regarded as a SUSY extension of the Pauli matrices.  
They are ``Hermitian'' in the sense 
\begin{equation}
l_{i}^{\ddagger}=l_i,~~~l_{\alpha}^{\ddagger}=\epsilon_{\alpha\beta}l_{\beta}, \label{adjointofosp12generators}
\end{equation}
where $\ddagger$ signifies the super-adjoint defined by 
\begin{align}
\begin{pmatrix}
A & B \\
C & D
\end{pmatrix}^{\ddagger}\equiv  \begin{pmatrix}
A^{\dagger} & C^{\dagger} \\
-B^{\dagger} & D^{\dagger}
\end{pmatrix}. 
\label{eqn:def-superadjoint}
\end{align}

Similar to the case of fuzzy sphere (\ref{Schwingersf2}), the coordinates of fuzzy two-supersphere \cite{Grosseeetal1995,GrosseReiter1998} are constructed by the graded version of the Schwinger operator formalism \cite{Balachandranetal2002,Hasebe2011,Hasebe-Kimura2005}.   
The graded Schwinger operator consists of two bosonic components, $a$ and $b$, and one fermion component $f$,  
\begin{equation}
\Psi=
(a, 
b, f)^t,   
\end{equation}
with satisfying   
\begin{align}
&[a,a^{\dagger}]=[b,b^{\dagger}]=\{f,f^{\dagger}\}=1,\nonumber\\
&[a,b]=[a,f]=[b,f]=[a,b^{\dagger}]=[a,f^{\dagger}]=[b,f^{\dagger}]=0. 
\end{align}

\begin{figure}[!t]
\centering
\includegraphics[width=12cm]{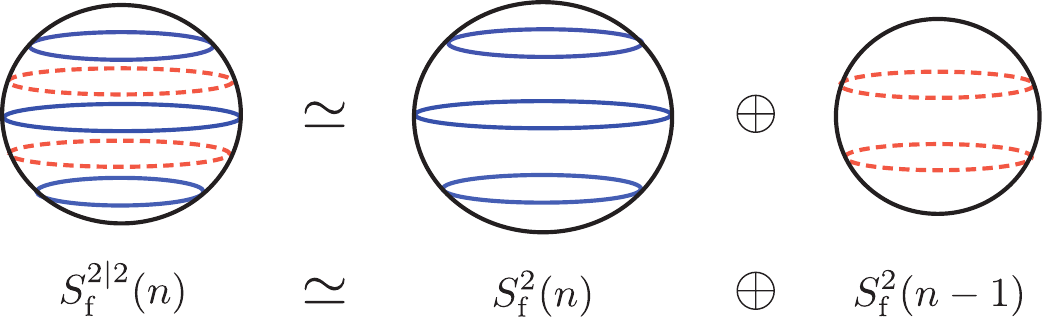}
\caption{(Color online) $\mathcal{N}=1$ fuzzy supersphere is a ``compound'' 
of two fuzzy two-spheres with radii, 
$\frac{R}{d}n$ and $\frac{R}{d}(n-1)$. This figure corresponds to $n=2$. \label{SupersphereDecomp} }
\end{figure}

With use of $\Psi$, the coordinates of fuzzy two-supersphere,  $S_{\mathrm{f}}^{2|2}$ (2 on the left to $|$ denotes the bosonic degrees of freedom, while 2 on the right to $|$ does the fermionic degrees of freedom), are constructed as 
\begin{equation}
X_i= \frac{2R}{d}\Psi^{\dagger}l_i\Psi,~~~~\Theta_{\alpha}=\frac{2R}{d}\Psi^{\dagger}l_{\alpha}\Psi,  
\label{SchRepF2sphere}
\end{equation}
which satisfy 
\begin{align}
[X_i,X_j]=i\frac{2R}{d}\epsilon_{ijk}X_k,~~~~~[X_i,\Theta_{\alpha}]=\frac{R}{d}(\sigma_i)_{\beta\alpha}\Theta_{\beta},~~~~~\{\Theta_{\alpha},\Theta_{\beta}\}=\frac{R}{d}(i\sigma_2\sigma_i)_{\alpha\beta}X_i, 
\end{align}
where $d=n+1$ with     
${n}=\Psi^{\dagger}\Psi={n}_{\mathrm{B}}+{n}_{\mathrm{F}}=a^{\dagger}a+b^{\dagger}b+f^{\dagger}f.$ 
Square of the radius of fuzzy supersphere is given by the $UOSp(1|2)$ Casimir  
\begin{equation}
X_i X_i +\epsilon_{\alpha\beta}\Theta_{\alpha}\Theta_{\beta}= 
\biggl(\frac{R}{d}\biggr)^2 n(n+1)\bold{1}_{2n+1}. 
\label{casimirofF2sphere}
\end{equation}
Notice the zero-point energy in (\ref{casimirofF2sphere}) reflects the difference between the bosonic and fermionic degrees of freedom.   
The basis states on fuzzy supersphere consist of the graded fully symmetric representation specified by the superspin $j=n/2$: 
\begin{subequations}
\begin{align}
&|n_1,n_2\rangle= \frac{1}{\sqrt{n_1!~n_2!}}
 {a^{\dagger}}^{n_1}{b^{\dagger}}^{n_2}\vac,\label{bosonicosp12}\\
&|m_1,m_2) = \frac{1}{\sqrt{m_1!~m_2!}} 
{a^{\dagger}}^{m_1} {b^{\dagger}}^{m_2}f^{\dagger}\vac, \label{fermionicosp12}  
\end{align}\label{UOSp12irredu}
\end{subequations}
where $n_1, n_2 , m_1$ and $m_2$ are non-negative integers that satisfy $n_1+n_2=m_1+m_2+1=n$.  
$|m_1,m_2)$ is the fermionic counterpart of $|n_1,n_2\rangle$, and thus $|n_1,n_2\rangle$ and $|m_1,m_2)$ exhibit ${N}=1$ SUSY. 
The bosonic and fermionic basis states are the eigenstates of the fermion parity $(-1)^{{n}_\mathrm{F}}$ 
with the eigenvalues $+1$ and $-1$, respectively. 
Their degrees of freedom  are also respectively given by  
\begin{equation}
d_\mathrm{B}=d(n)=n+1, ~~~d_\mathrm{F}=d(n-1)=n,
\label{usp21bosferdof}
\end{equation}
and the total degrees of freedom is 
\begin{equation}
d_\mathrm{T} = d_\mathrm{B} + d_\mathrm{F} = 2n+1. 
\label{usp21totdof}
\end{equation}
The eigenvalues of $X_3$ for (\ref{UOSp12irredu}) read as    
\begin{equation}
X_3=\frac{R}{d}(n-{k}), 
\label{X3fuzzysupersphere}
\end{equation}
where $k=0,1,2,\cdots,2n$.  Even $k$ ($k= 2n_2$) correspond to the bosonic states (\ref{bosonicosp12}), while odd $k$ ($k=2m_2+1$) the fermionic states (\ref{fermionicosp12}). Compare the $X_3$ eigenvalues  of fuzzy supersphere (\ref{X3fuzzysupersphere}) with those of the fuzzy (bosonic) sphere (\ref{X3fuzzysphere}): 
The degrees of freedom of fuzzy supersphere for even $k$ are accounted for by those of fuzzy sphere with radius $n$, while those for odd $k$ are by fuzzy sphere with radius $n-1$. 
In this sense, the fuzzy two-supersphere $S_{\mathrm{f}}^{2|2}(n)$ 
of radius ${n}$ can be regarded as a ``compound'' of  two fuzzy  spheres with different radii, $\frac{R}{d}n$ and $\frac{R}{d}(n-1)$ [Fig.~\ref{SupersphereDecomp}]: 
\begin{equation}
S_{\mathrm{f}}^{2|2}(n)\simeq S_{\mathrm{f}}^{2}(n)\oplus S_{\mathrm{f}}^{2}(n-1). 
\label{compoundfuzzysupersph}
\end{equation}
The first $S_{\mathrm{f}}^{2}(n)$ accounts for the bosonic degrees of freedom (\ref{bosonicosp12}), while the second $S_{\mathrm{f}}^{2}(n-1)$ does for the fermionic degrees of freedom (\ref{fermionicosp12}).

Similar to the 1st Hopf map, we can construct a graded version of the 1st Hopf map  \cite{LandiMarmo1987,Bartocci1990} 
\begin{equation}
\psi ~~\rightarrow~~ x_i =2\psi^{\ddagger}L_i\psi,~~\theta_{\alpha}=2\psi^{\ddagger}L_{\alpha}\psi, 
\label{1stgradedHopf}
\end{equation}
where $l_i$ and $l_{\alpha}$ are the $UOSp(1|2)$ matrices of fundamental representation (\ref{superPaulimat}), and  $\psi$ denotes a normalized $UOSp(1|2)$ superspinor 
\begin{equation}
 \psi=\begin{pmatrix}
u\\
v\\
\eta\end{pmatrix} 
\label{spinholeeq}
\end{equation}
with 
\begin{equation}
\psi^{\ddagger}\psi=u^*u+v^*v-\eta^*\eta=1.    
\label{noramlizationpsipseudo}
\end{equation}
Here, the super-adjoint of the superspinor is defined by 
$\psi^{\ddagger}\equiv (u^*,v^*,-\eta^*)$ [see also eq.\eqref{eqn:def-superadjoint}] 
and $*$ represents the pseudo-conjugation.\footnote{%
The pseudo-conjugation is defined as  
$(\eta^*)^*=-\eta$ and $(\eta_1\eta_2)^* =\eta_1^*\eta_2^*$ for Grassmann odd quantities. 
See Ref.\cite{BookFrappat} for instance. } 
The first two components of $\psi$ are Grassmann-even and the third component is Grassmann-odd.  
 $\psi$ subject to the normalization (\ref{noramlizationpsipseudo}) can be regarded as a coordinate of the  manifold $S^{3|2}$. 
From (\ref{1stgradedHopf}), we find that $x_i$ and $\theta_{\alpha}$ satisfy the condition of $S^{2|2}$:  
\begin{equation}
x_ix_i +\epsilon_{\alpha\beta}\theta_{\alpha}\theta_{\beta}=(\psi^{\ddagger}\psi)^2=1. 
\label{normalization2supersphere}
\end{equation}
Consequently, the map (\ref{1stgradedHopf}) represents 
\begin{equation}
S^{3|2}\overset{S^1}\longrightarrow S^{2|2}.   
\label{absgraded1sthopf}
\end{equation}
The bosonic part of (\ref{absgraded1sthopf})  exactly corresponds to the 1st Hopf map. 
Note that $\psi$ satisfies the super-coherent state equation 
\begin{equation}
x_i\cdot l_i\psi+\epsilon_{\alpha\beta}l_{\alpha}\psi \cdot \theta_{\beta}=\frac{1}{2}\psi,  
\end{equation}
and $\psi$ is referred to as the super-coherent state or spin-hole coherent state \cite{auerbach1994book} in literature.  
 $x_i$ are Grassmann even but not usual c-numbers, since the square of $x_i$ is not a c-number,  $x_ix_i=1-\epsilon_{\alpha\beta}\theta_{\alpha}\theta_{\beta}$, as seen from (\ref{normalization2supersphere}).  
Instead of $x_i$, one can introduce $c$-number quantities $\{y_i\}$ as  
\begin{equation}
y_i=\frac{1}{\sqrt{1-\epsilon_{\alpha\beta}\theta_{\alpha} \theta_{\beta}}} ~ x_i, 
\end{equation}
which  denote coordinates on $S^2$, the ``body'' of $S^{2|2}$, as confirmed from $y_iy_i=1$.   
With use of the coordinates on $S^{2|2}$, $\psi$ can be expressed as \cite{Hasebe-Kimura2005}  
\begin{align}
\psi&=\frac{1}{\sqrt{2(1+y_3)(1+\theta_1\theta_2)}}\begin{pmatrix}
1+y_3\\
y_1+iy_2\\
(1+y_3)\theta_1+(y_1+iy_2)\theta_2
\end{pmatrix}e^{i\chi} 
\nonumber\\
&=\frac{1}{\sqrt{2(1+x_3)}}
\begin{pmatrix}
(1+x_3)(1-\frac{1}{2(1+x_3)} \theta_1\theta_2)\\
(x_1+ix_2)(1+\frac{1}{2(1+x_3)} \theta_1\theta_2)\\
(1+x_3)\theta_1+(x_1+ix_2)\theta_2
\end{pmatrix}e^{i\chi}, 
\label{exphopfspinor} 
\end{align}
where $e^{i\chi}$ stands for the arbitrary $U(1)$ phase factor.  
The last expression on the right-hand side manifests the graded Hopf fibration, $S^{3|2}\sim S^{2|2}\otimes S^1$ (here, $\sim$ denotes local equivalence): The $S^1(\simeq U(1))$-fibre, $e^{i\chi}$, is canceled in the graded 
Hopf map (\ref{1stgradedHopf}), and the other quantities, $x_i$ and $\theta_{\alpha}$ in (\ref{exphopfspinor}), correspond to the coordinates on $S^{2|2}$.

\subsection{$\mathcal{N}=2$}

As the geometric structure of $S_{\mathrm{f}}^{2|2}$ is determined by  
 the $UOSp(1|2)$ algebra,   $\mathcal{N}=2$ fuzzy supersphere $S_{\mathrm{f}}^{2|4}$ is formulated by the $UOSp(2|2)$ algebra.  
The $UOSp(2|2)$ algebra contains the bosonic subalgebra, $usp(2)\oplus o(2)\simeq su(2)\oplus u(1)$, and is  isomorphic to $SU(2|1)$. The dimension is given by 
\begin{equation}
\dim[uosp(2|2)]=\dim[su(2|1)]=4|4=8. 
\end{equation}
Denoting the four bosonic generators as $L_i$ $(i=1,2,3)$ and $\Gamma$ and the four fermionic generators as $L_{\alpha}$ and $L_{\alpha}'$ $(\alpha=\theta_1,\theta_2)$, we can express the  $UOSp(2|2)$ algebra as 
\begin{align}
&[L_i,L_j]=i\epsilon_{ijk}L_k,~~~~
[L_i,L_{\alpha\sigma}]=\frac{1}{2}(\sigma_i)_{\beta\alpha}L_{\beta\sigma},\nonumber\\
&[\Gamma,L_i]=0,~~~~~~~~~~~~~[\Gamma,L_{\alpha\sigma}]=\frac{1}{2}\epsilon_{\tau\sigma}L_{\alpha\tau}, \nonumber\\
&\{L_{\alpha\sigma},L_{\beta\tau}\}=\frac{1}{2}\delta_{\sigma\tau}(i\sigma_2\sigma_i)_{\alpha\beta}L_i+\frac{1}{2}\epsilon_{\sigma\tau}\epsilon_{\alpha\beta}\Gamma,
\label{osp22algebranew}
\end{align}
where $L_{\alpha\sigma}=(L_{\alpha},L'_{\alpha})$.  
 $L_i$ and $L_{\alpha}$ form the $UOSp(1|2)$ subalgebra.  
There are two sets of fermionic generators, $L_{\alpha}$ and  $L_{\alpha}'$, which bring $\mathcal{N}=2$ SUSY. 
The $UOSp(2|2)$ algebra has two Casimirs, quadratic 
\begin{equation}
\mathcal{K}=L_iL_i+\epsilon_{\alpha\beta}L_{\alpha}L_{\beta}+\epsilon_{\alpha\beta}L'_{\alpha}L'_{\beta}+\Gamma\Gamma 
\label{osp22casimir}
\end{equation}
and cubic ones \cite{Scheunertetal1977}. 

\begin{figure}[!t]
\centering
\includegraphics[width=15cm]{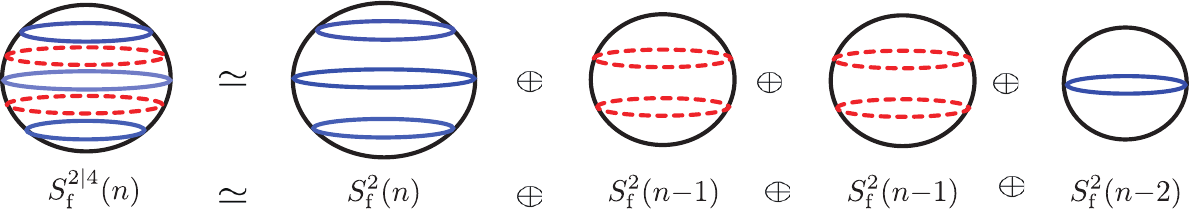}
\caption{(Color online) $S_{\mathrm{f}}^{2|4}$ is a ``compound'' made 
of four fuzzy two-spheres that are considered 
as $\mathcal{N}=2$ superpartners. 
The above picture corresponds to $n=2$. \label{SupersphereDecomp2} }
\end{figure}

To specify a fuzzy manifold, an appropriate choice of  irreducible representation  is crucial as well.   
The irreducible representation of $UOSp(2|2)$ is classified into two categories; typical representation and atypical representation \cite{Scheunertetal1977,Marcu1980}. 
Since the quadratic Casimir  (\ref{osp22casimir}) is identically zero for atypical representation,  we adopt  typical representation for the construction of  $S_{\mathrm{f}}^{2|4}$ \cite{Hasebe2011}.   
The $UOSp(2|2)$ matrices for typical representation for minimal dimension are represented by the following $4\times 4$ matrices:   
\begin{align}
&l_i=\frac{1}{2}
\begin{pmatrix}
\sigma_i & 0 _2 \\
0_2 & 0_2
\end{pmatrix},~~~l_{\alpha}=\frac{1}{2}
\begin{pmatrix}
0_2 & \tau_{\alpha} & 0 \\
-(i\sigma_2\tau_{\alpha})^t & 0 & 0  \\
0 & 0 & 0 
\end{pmatrix},\nonumber\\
&l'_{\alpha}=\frac{1}{2}
\begin{pmatrix}
0_2& 0  & \tau_{\alpha}  \\
0 & 0 & 0  \\
-(i\sigma_2\tau_{\alpha})^t & 0 & 0 
\end{pmatrix},~~~\gamma=\frac{1}{2}\begin{pmatrix}
0_2 & 0_2 \\
0_2 & i\sigma^2
\end{pmatrix}.
\label{osp22typicalmatrices}
\end{align}
Applying the Schwinger construction, we introduce the coordinates of $S_{\mathrm{f}}^{2|4}$ as  
\begin{equation}
X_i=\frac{2R}{d}\Psi^{\dagger}L_i\Psi,~~~~\Theta_{\alpha}=\frac{2R}{d}\Psi^{\dagger}L_{\alpha}\Psi,~~~~
\Theta'_{\alpha}=\frac{2R}{d}\Psi^{\dagger}L'_{\alpha}\Psi,~~~~G=\frac{2R}{d}\Psi^{\dagger}\Gamma\Psi. 
\label{Schwingerconstrfuzzytwosphere}
\end{equation}
where   $\Psi$ signifies the four-component $UOSp(2|2)$ Schwinger operator  
\begin{equation}
\Psi=
(a,
b, 
f, 
g)^t,    
\end{equation}
and $d=n+1$ with $n=\Psi^{\dagger}\Psi=a^{\dagger}a+b^{\dagger}b+f^{\dagger}f+g^{\dagger}g$. 
Here $a$ and $b$ are bosonic operators, while $f$ and $g$ are fermionic operators  that satisfy   
\begin{align}
&[a,a^{\dagger}]=[b,b^{\dagger}]=\{f,f^{\dagger}\}=\{g,g^{\dagger}\}=1,\nonumber\\
&[a,b]=[a,f]=[a,g]=\cdots=\{f,g\}=\{f,g^{\dagger}\}=0. 
\end{align}
It is straightforward to evaluate the square of the radius of $S^{2|4}_{\mathrm{f}}$: 
\begin{equation}
X_iX_i+\epsilon_{\alpha\beta}\Theta_{\alpha}\Theta_{\beta}+\epsilon_{\alpha\beta}\Theta'_{\alpha}\Theta'_{\beta}+GG=\biggl(\frac{R}{d}\biggr)^2 n^2.   
\label{radiusoftyposp22ori}
\end{equation}
With a given $n$, the basis states of $S_{\mathrm{f}}^{2|4}$ are constituted of the graded fully symmetric representation of the $UOSp(2|2)$:  
\begin{subequations}
\begin{align}
&|n_1,n_2\rangle= \frac{1}{\sqrt{n_1!~n_2!}}
{a^{\dagger}}^{n_1}{b^{\dagger}}^{n_2}\vac,  \label{grosp22b1}\\
&|m_1,m_2)= \frac{1}{\sqrt{m_1!~m_2!}} 
 {a^{\dagger}}^{m_1}{b^{\dagger}}^{m_2}f^{\dagger}\vac,  \label{grosp22b2} \\
&|m'_1,m'_2)=\frac{1}{\sqrt{m'_1!~m'_2!}}
{a^{\dagger}}^{m'_1}{b^{\dagger}}^{m'_2}g^{\dagger}\vac, \label{grosp22f1} \\
&|l_1,l_2\rangle= \frac{1}{\sqrt{l_1!~l_2!}}
{a^{\dagger}}^{l_1}{b^{\dagger}}^{l_2}f^{\dagger}g^{\dagger}\vac, \label{grosp22f2} 
\end{align} \label{fullysymmosp22rep}
\end{subequations}
where $n_1$, $n_2$, $m_1$, $m_2$, $m'_1$, $m'_2$, $l_1$ and $l_2$ denote non-negative integers that satisfy $n_1+n_2=m_1+m_2+1=m_1'+m_2'+1=l_1+l_2+2=n$. 
 The dimension of bosonic basis states,  $|n_1,n_2\rangle$ and $|l_1,l_2\rangle$, and that of fermionic basis states,  $|m_1,m_2\rangle$ and $|m'_1,m'_2\rangle$, are found to be equal:  
\begin{align}
&d_{\mathrm{B}}=d(n)+d(n-2)=2n,\nonumber\\
&d_{\mathrm{F}}=2\times d(n-1)=2n,  
\end{align}
and the total degrees of freedom amount to 
\begin{equation}
d_{\mathrm{T}}=d_{\mathrm{B}}+d_{\mathrm{F}}=4n.
\end{equation}
Square of the radius of $S_{\mathrm{f}}^{2|4}$  (\ref{radiusoftyposp22ori}) does not have the zero-point ``energy'', since the bosonic and fermionic degrees of freedom are canceled exactly\footnote{ In the cases of fuzzy superspheres based on the $UOSp(2|\mathcal{N})$ algebra for $\mathcal{N}>2$, the dimension of fermionic basis states is larger than that of the bosonic basis states, and the zero-point ``energy'' becomes minus.  In other words, the radius of fuzzy supersphere would become minus for the $UOSp(2|\mathcal{N})$ representations with sufficiently small dimensions. Thus, we only deal with the fuzzy superspheres for $\mathcal{N}=1,2$.}.  
The four sets of the basis states (\ref{fullysymmosp22rep}) forms the $\mathcal{N}=2$ SUSY 
multiplet, and  suggest the following geometrical structure of $S_{\mathrm{f}}^{2|4}$,  
\begin{equation}
S_{\mathrm{f}}^{2|4}(n) 
\simeq 
S_{\mathrm{f}}^{2}(n)\oplus S_{\mathrm{f}}^2(n-1)\oplus S_{\mathrm{f}}^2(n-1)\oplus S_{\mathrm{f}}^2(n-2). 
\end{equation}
The latitudes for the the basis states (\ref{fullysymmosp22rep}) are given by 
\begin{equation}
X_3=\frac{R}{d}(n-k),
\label{x3eigenfuzzyfour}
\end{equation}
with $k=0,1,2,\cdots,2n$. The even $k$s ($k=2n_2$, $k=2l_2+2$) correspond to the bosonic basis states, (\ref{grosp22b1}) and (\ref{grosp22b2}), while odd $k$s ($k=2m_2+1$, $k=2m_2'+1$) the fermionic basis states, (\ref{grosp22f1}) and (\ref{grosp22f2}) [Fig.~\ref{SupersphereDecomp2}].  
Except for non-degenerate states at the north and south poles $X_3=\pm n$,  all the other eigenvalues of $X_3$ (\ref{x3eigenfuzzyfour}) are doubly-degenerate. 

\section{Supersymmetric Valence Bond Solid States}\label{secsvbs}
In this section, we review the basic properties of the SVBS states 
\cite{Arovasetal2009,Hasebe-Totsuka2011} and discuss its intriguing connection 
to the SUSY QH wave function and BCS function.  
\subsection{Construction of SVBS states}\label{subsec:SVBS}

\subsubsection{$\mathcal{N}=1$ }

 Here, we consider the SVBS states with $UOSp(1|2)$ SUSY ($\mathcal{N}=1$),    
which we shall call the type-I SVBS states.  
We apply the mathematical procedure of the constructions of the VBS states described in Sec.\ref{subsec:valenvebond}.   The first we prepare is the $UOSp(1|2)$ invariant matrix 
(see Appendix \ref{Appen:SecUOSp(MN)}) 
\begin{equation}
\mathcal{R}_{1|2}=
\begin{pmatrix}
0 & 1 & 0 \\
-1 & 0 & 0 \\
0 & 0 & -1 
\end{pmatrix}. 
\end{equation}
With the parameter dependent Schwinger operator
\begin{equation}
\Psi(r)=(a,b,\sqrt{r}f)^t, 
\label{paraSchwingergrade}
\end{equation}
the type-I SVBS state is constructed as 
\begin{equation}
|\text{SVBS-I}\rangle=\prod_{\langle ij\rangle} 
(\Psi^{\dagger}_i(r)~\mathcal{R}_{1|2}~\Psi_j^*(r))^M|\text{vac} \rangle=\prod_{\langle ij\rangle} 
(a_i^{\dagger}b_j^{\dagger}-b_i^{\dagger}a_j^{\dagger}
-r f_i^{\dagger}f_j^{\dagger})^M|\text{vac} \rangle.
\label{SVBSstateI}
\end{equation}
  The operators $a_i$, $b_i$ and $f_i$ 
are components of the graded Schwinger operator defined on each site $i$ and satisfy  the commutation relations,  
$[a_i,a^{\dagger}_j]=[b_i,b^{\dagger}_j]=\delta_{ij}$ 
and 
$\{f_i,f^{\dagger}_j\}=\delta_{ij}$.   
Physically, the three fundamental states, $a^{\dagger}|\text{vac}\rangle$,  $b^{\dagger}|\text{vac}\rangle$and  $f^{\dagger}|\text{vac}\rangle$, are interpreted as spin $\uparrow$, $\downarrow$ and 
spinless hole states (see Table \ref{physicalintSch})\footnote{%
In Ref.\cite{2010Duffetal}, such $UOSp(1|2)$ triplet is dubbed the superqubit.}.  
Since the fermions always appear in pairs of the form 
$f_i^{\dagger}f_j^{\dagger}$ ($i$, $j$ are adjacent), 
the SVBS states can be regarded as  {\em hole-pair doped} VBS states, and 
$r$ stands for a hole doping parameter.  
\hspace{-0.5cm}
\begin{table}
\renewcommand{\arraystretch}{1}
\begin{center}
\begin{tabular}{|c||c|c|}
\hline  Schwinger operator  & $SU(2)$ quantum number & Spin state  \\
\hline
\hline   $a^{\dagger}$  & 1/2  
 &  $|\!\uparrow\rangle =a^{\dagger}\vac$ 
\\
\hline  $b^{\dagger}$  &  $-1/2$ &   $|\!\downarrow\rangle =b^{\dagger}\vac$     
\\ \hline  $f^{\dagger}$    & 0 &  $| h\rangle =f^{\dagger}\vac$  
\\ \hline
\end{tabular}
\end{center}
\caption{The physical interpretation of the local states made by the Schwinger operators. $f^{\dagger}$ denotes the hole degrees of freedom.  }
\label{physicalintSch}
\end{table}
 
Here, some comments are added. 
The $UOSp(1|2)$ specific feature does not enter the local Hilbert space on each site.   
For instance, (\ref{UOSp12irredu}) can also be regarded as an irreducible representation of $SU(2|1)$.   Meanwhile, in the construction of the type-I SVBS states (\ref{SVBSstateI}), the $UOSp(1|2)$ invariant matrix was utilized, and then 
the $UOSp(1|2)$ structure explicitly enters in the many-body states.    
 This implies that (super)spin interaction between adjacent sites reduces the $SU(2|1)$ symmetry  on each site to the lower symmetry $UOSp(1|2)$ in many-body physics.

In the type-I SVB states (\ref{SVBSstateI}),  
the total particle number of the Schwinger particles at each site is given by   
\begin{equation}
z M=a^{\dagger}_ia_i+b^{\dagger}_ib_i+f^{\dagger}_if_i \; . 
\end{equation}
Since the fermion number $f^{\dagger}f$ takes either 0 or 1, the following 
two eigenvalues are possible for the local spin quantum number 
$S_i=\frac{1}{2}(a^{\dagger}_ia_i+b^{\dagger}_ib_i)$:  
\begin{equation}
S_i=M,~ M-\frac{1}{2}.
\end{equation}
In particular, for $M=1$, each site can take two spin values
\begin{equation}
S_i=1,~ \frac{1}{2}, 
\label{M1typeISVBS}
\end{equation}
and the local Hilbert space is spanned by the five ($4M{+}1$, in general) basis states 
\begin{equation}
\begin{split}
&|1\rangle=\frac{1}{\sqrt{2}}{a_i^{\dagger}}^2|\text{vac}\rangle,
\;\; |0\rangle={a_i^{\dagger}b_i^{\dagger}}|\text{vac}\rangle, 
\;\; |{-}1\rangle=\frac{1}{\sqrt{2}}{b_i^{\dagger}}^2|\text{vac}\rangle, \\
&\quad\quad\quad |\!\uparrow\rangle=
 a_i^{\dagger}f_i^{\dagger}|\text{vac}\rangle, 
\;\; |\!\downarrow \rangle =b_i^{\dagger}f_i^{\dagger}|\text{vac}\rangle \; .
\end{split}
\end{equation}
These constitute  $\mathcal{N}{=}1$ SUSY multiplet with the $UOSp(1|2)$ superspin $\mathcal{S}=1$.   
Similarly, the edge states consist of $\mathcal{N}=1$ SUSY multiplet 
with the $UOSp(1|2)$ superspin $\mathcal{S}=1/2$:   
\begin{equation}
|\!\uparrow\rangle\!\rangle=a^{\dagger}|\text{vac}\rangle, \;\;
|\!\downarrow\rangle\!\rangle=b^{\dagger}|\text{vac}\rangle, \;\;
|0\rangle\!\rangle=f^{\dagger}|\text{vac}\rangle.
\end{equation}
As we will see in Sec.\ref{secsmat}, 
the ground state of a finite {\em open} chain is nine-fold degenerate (corresponding 
to $3\times 3$ matrix-components of the $M=1$ type-I SVBS states). 

\begin{figure}[!t]
\centering
\includegraphics[width=10cm]{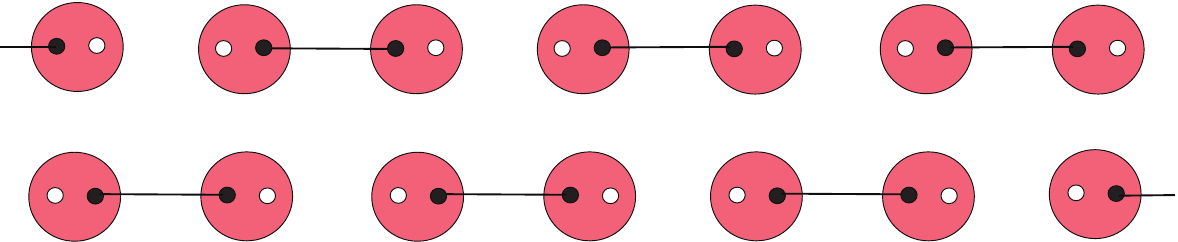}
\caption{(Color online) Two MG dimer states related by translation. \label{dimer.fig} }
\end{figure}

Intriguingly, the $M=1$ type-I SVBS chain interpolates  two VBS states 
in the two extreme limits of the hole doping: 
In the limit $r\rightarrow 0$,  $|\text{SVBS-I}\rangle$  reproduces the original spin-1 VBS state $|\text{VBS}\rangle$, 
\begin{equation}
|\text{SVBS-I}\rangle \rightarrow |\text{VBS}\rangle=\prod_{i}
(a^{\dagger}_i b_{i+1}^{\dagger}-b_i^{\dagger}a_{i+1})|\text{vac} \rangle, 
\label{r0limitsvbsI}
\end{equation}
while in the limit $r\rightarrow \infty$,   $|\text{SVBS-I}\rangle$  reduces 
to the Majumdar-Ghosh (MG) dimer state\cite{Majumdar-G-69,Majumdar-70}  
$|\text{MG}\rangle$,  
\begin{equation}
|\text{SVBS-I}\rangle  \rightarrow 
\left\{
\prod_i  f^{\dagger}_i 
\right\}
|\text{MG}\rangle, 
\label{rinfinitylimitsvbsI}
\end{equation}
where $|\text{MG}\rangle$ is either of the two dimerized states of 
the MG model\footnote{%
The open boundary condition has been implicitly assumed here; 
if the periodic boundary condition had been used, the two states would have been summed 
up with a minus sign due to the anti-commutating property of the holes.
} [Fig.~\ref{dimer.fig}]:
\begin{equation}
|\text{MG}\rangle=
\begin{cases}
|\Phi_{\text{A}}\rangle=\prod_{i:\text{even}}(a^{\dagger}_i b^{\dagger}_{i+1} - b^{\dagger}_i a^{\dagger}_{i+1})|\text{vac}\rangle, 
\\
|\Phi_{\text{B}}\rangle=\prod_{i:\text{odd}}(a^{\dagger}_i b^{\dagger}_{i+1} - b^{\dagger}_i a^{\dagger}_{i+1})|\text{vac}\rangle
\; .
\end{cases}\label{SchRepMG}
\end{equation}
For larger $M$, $|\text{MG}\rangle$ should be replaced with the inhomogeneous VBS 
states \cite{Arovasetal1988} where the number of valence bonds alternates from bond 
to bond. 

\begin{figure}[h]
\centering
\includegraphics[width=7cm]{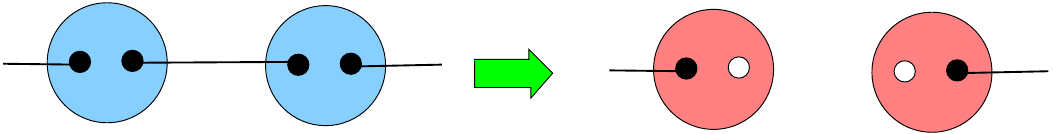}
\caption{(Color online) 
When the exponential factor (\ref{expofac}) acts to the VBS state, 
the factor breaks the VB between $i$ and $i+1$ site and creates  
hole-pair instead. \label{bondbreaking.fig} (Figure and Caption 
are taken from Ref.\cite{Arovasetal2009}).}
\end{figure}

\begin{figure}[!t]
\centering
\includegraphics[width=7cm]{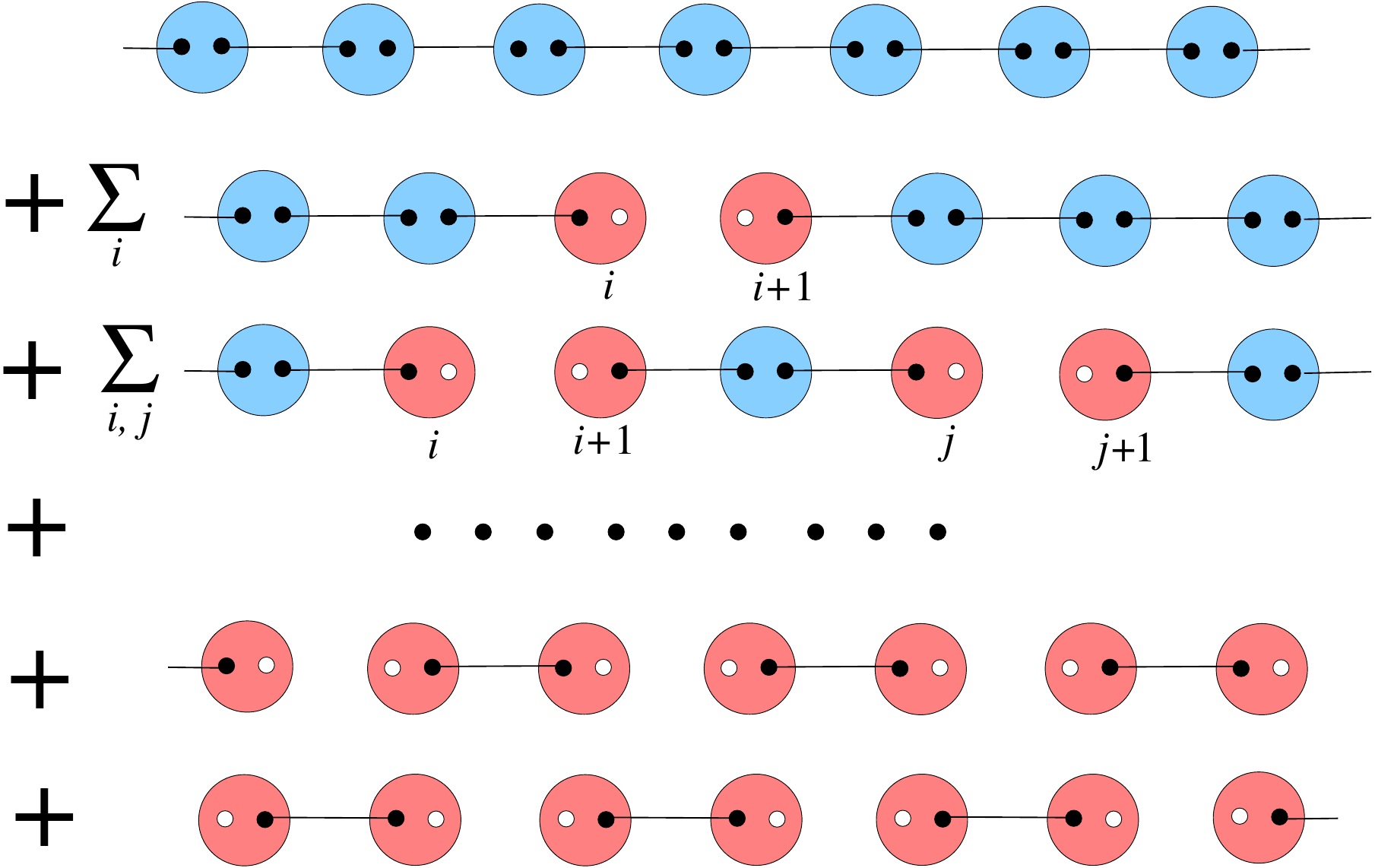}
\caption{(Color online) The type-I SVBS is a superposed state of hole-pair doped VBS states. 
With finite hole-doping parameter $r$, all of the hole-pair doped VBS states are superposed to form the SVBS state, and the SVBS state exhibits the SC property. At $r=0$, the SVBS state is reduced to the original VBS state (depicted as the first chain), while $r\rightarrow \infty$, the SVBS state is reduced to the MG dimer state (depicted as the last two chains). \label{expSUSY.fig} 
(Figure and Caption are taken from Ref.\cite{Hasebe-Totsuka2011}). }
\end{figure}

In the super-coherent state (\ref{spinholeeq}) representation\footnote{%
In the following discussion, the explicit form of the superspinor is not important. 
Only the Grassmann property of the components matters.}, 
the SVBS state is expressed as \cite{Arovasetal2009}
\begin{equation}
\Psi_{\text{SVBS-I}}= \prod_{\langle ij\rangle} (u_iv_{j}-v_iu_{j}-r\eta_i\eta_{j})^M. 
\label{supercoheresvbs}
\end{equation}
From the Grassmann (odd) property of $\eta$, $\Psi_{\text{SVBS-I}}$ (\ref{supercoheresvbs}) 
can be rewritten as  
\begin{equation}
\Psi_{\text{SVBS-I}}=\exp\biggl(-{Mr\sum_{\langle ij \rangle} 
\frac{\eta_i\eta_j}{u_iv_j-v_iu_j}} \biggr) \cdot \Phi_{\text{VBS}}(\{u_i,v_i\}) \; , 
\label{exptimesVBS}
\end{equation}
where $\Phi_{\text{VBS}}$ is the coherent state 
representation of the original VBS state (\ref{coherentrepLHstates}). We can deduce a nice physical interpretation of the SVBS states from this expression. 
Since the SVBS states are written as a ``product'' of the exponential and the original VBS 
wave function $\Phi_{\text{VBS}}$, 
all physics inherent in the SVBS must be included in this exponential factor:
\begin{equation}
\exp\biggl(-{Mr\sum_{\langle ij \rangle} 
\frac{\eta_i\eta_j}{u_iv_j-v_iu_j}} \biggr)
= \prod_{\langle ij \rangle} \left\{
1-{Mr 
\frac{\eta_i\eta_j}{u_iv_j-v_iu_j}}
\right\}  \; .
\end{equation}
Since every time when the factor 
\begin{equation}
\frac{\eta_i\eta_j}{u_iv_j-v_iu_j}
\label{expofac}
\end{equation}
acts to the VBS wave function, the VB between the adjacent sites $i$ and $j$ is replaced with a hole-pair 
(see Fig.~\ref{bondbreaking.fig}): 
\begin{equation}
u_iv_j-v_iu_j ~~\xrightarrow{\frac{\eta_i\eta_j}{u_iv_j-v_iu_j}}~~ \eta_i\eta_j \; ,
\end{equation}
one sees that the SVBS wave function \eqref{exptimesVBS} may be expanded as
\begin{align}
\Psi_{\text{SVBS-I}}&=\Phi_{\text{VBS}}-{Mr\sum_i  
\frac{\eta_i\eta_{i+1}}{u_iv_{i+1}-v_iu_{i+1}}} \cdot \Phi_{\text{VBS}} +\frac{1}{2}(Mr)^2 \biggl(\sum_{i} 
\frac{\eta_i\eta_j}{u_iv_{i+1}-v_iu_{i+1}}\biggr)^2+\cdots\nonumber\\
&+(-Mr)^{L/2}\prod_i \eta_i\cdot (\prod_{i:\text{even}}-\prod_{i:\text{odd}})
\frac{1}{u_iv_{i+1}-v_iu_{i+1}}\cdot \Phi_{\text{VBS}}.
\label{expansionofsvbs}
\end{align}
Thus, the SVBS chain is expressed as the superposition of many-body states on the right-hand side (r.h.s.) of (\ref{expansionofsvbs}). The first term on the r.h.s. is the original VBS chain (which is consistent with (\ref{r0limitsvbsI})).  
The second term is the VBS chain with one hole-pair doped, and the third term is the VBS chain with two hole-pairs doped.  
In general, $n$th term represents the VBS chain with $(n-1)$ hole pairs doped.   As the last term, partially dimerized chains,  $i.e.$, the VB chains whose  half sites are occupied with hole-pairs, are realized    
(see Fig.~\ref{expSUSY.fig})\footnote{For general $M$ and arbitrary lattice coordination number $z$, the last term of the expansion realizes a resonating valence bond (RVB) state \cite{Anderson1987}. For instance, on 2D square lattice, $M=2$ SVBS state with $\mathcal{N}=3$ SUSY gives 
the Rokhsar-Kivelson RVB state \cite{Rokhsar-Kivelson1988} as the last term.}. 
For $M=1$, the last term gives rise to the Majumdar-Ghosh dimer states, 
\begin{equation}
(\prod_{i:\text{even}}-\prod_{i:\text{odd}})\frac{1}{u_iv_{i+1}-v_iu_{i+1}}\cdot 
\Phi_{\text{VBS}}=(\prod_{i:\text{odd}}-\prod_{i:\text{even}})(u_iv_{i+1}-v_iu_{i+1})
=  -\Phi_{\text{A}} + \Phi_{\text{B}}, 
\end{equation}
where $\Phi_\text{A}$ and $\Phi_\text{B}$ are the coherent state representation of 
$|\Phi_\text{A}\rangle$ and $|\Phi_\text{B}\rangle$ (\ref{SchRepMG}). 
Now, the physical meaning of the SVBS states is transparent: The SVBS states signify a superposed state by all possible hole-pair doped VBS states, which can be viewed as a generalization of the resonating valence bond state \cite{Anderson1987}  [see Sec.\ref{subsec:scproperty} for more details].   

As in the original correspondence between VBS and QHE, 
the super-coherent state representation of SVBS (\ref{supercoheresvbs}) shows striking analogies to  
the SUSY Laughlin-Haldane wave function \cite{Hasebe2005PRL},  
\begin{equation}
\Phi_{\text{SLH}}=\prod_{ i<j}^N 
(\psi_i(r)^t ~\mathcal{R}_{1|2}~\psi_j(r))^M=\prod_{i<j}^N (u_iv_j-v_iu_j-r\eta_i\eta_j)^m. 
\label{susylaughhaldanefunc}
\end{equation}
Also for one-particle mechanics, there exist apparent relations between the SVBS and the SUSY Landau problem. 
Indeed,   the super-coherent state  representation of 
the basis states (\ref{UOSp12irredu}),  
\begin{subequations}
\begin{align}
& \varphi^{(\mathrm{B})}_{n_1,n_2}=\frac{1}{\sqrt{n_1!n_2!}}u^{n_1}v^{n_2},\\
& \varphi^{(\mathrm{F})}_{m_1,m_2}=\frac{1}{\sqrt{m_1!m_2!}}u^{m_1}v^{m_2}\eta,
\end{align}
\end{subequations}
gives the  lowest Landau level eigenstates of a SUSY Landau Hamiltonian \cite{Hasebe-Kimura2005}. 
Here,  $n_1$, $n_2$, $m_1$ and $m_2$ are  non-negative integers that satisfy $n_1+n_2=m_1+m_2+1=I$. 
By the stereographic projection $z_i=v_i/u_i$ and $\xi_i=\eta_i/u_i$, 
the SUSY Laughlin-Haldane wave function $\Phi_{\text{SLH}}$ (\ref{susylaughhaldanefunc}) is 
transformed to the  SUSY Laughlin wave function defined by 
\begin{equation}
\Phi_{\text{SL}}
\equiv \prod_{i<j}^N(z_i-z_j-r\xi_i\xi_j)^m e^{-\sum_i (z_iz_i^*+\xi_i\xi_i^*)},  
\end{equation}
By expanding the polynomial part $(z_i-z_j-r\xi_i\xi_j)^m$ 
in the Grassmann odd quantities, we have 
\begin{equation}
\Phi_{\text{SL}}=\Phi_{\text{L}}-mr\sum_{i<j}\frac{\xi_i\xi_j}{z_i-z_j} \Phi_{\text{L}}
+\cdots +\frac{1}{(N/2)!}m^{\frac{N}{2}}(-r)^N\xi_1,\xi_2\cdots \xi_{N} \cdot 
  \text{Pf} \left( \frac{1}{z_i-z_j} \right)    \Phi_{\text{L}},  
 \label{expansionslinwav}
\end{equation}
where $N$ is the total number of particles (which we assumed to be an even integer) 
and $\Phi_{\text{L}}$ coincides with the Laughlin wave function on a 2D plane 
(up to the Grassmann factor $e^{-\sum_i \xi_i\xi_i^*}$): 
\begin{equation}
\Phi_{\text{L}}=\prod_{i<j}^N(z_i-z_j)^m e^{-\sum_i (z_iz_i^*+\xi_i\xi_i^*)}.
\end{equation}
Interestingly, the Pfaffian wave function 
$\text{Pf} \left( \frac{1}{z_i-z_j} \right)\prod_{i<j}^N(z_i-z_j)^m e^{-\sum_i z_iz_i^*}$ 
for the ground state of the non-Abelian QHE \cite{Moore-Read-91} 
appears in the last term of the expansion (\ref{expansionslinwav}) at $m=2$.

\subsubsection{$\mathcal{N}=2$}
\label{eqn:type-II-intro}

The $\mathcal{N}=2$ SVBS states, which we call the type-II SVBS states, 
are constructed as  $UOSp(2|2)$ invariant VBS states.  
With the $UOSp(2|2)$ invariant matrix 
\begin{equation}
{\cal R}_{2|2} = 
\begin{pmatrix}
0 & 1 & 0 & 0\\
-1 & 0 & 0 & 0\\
0 & 0 & -1 &0 \\
0 & 0 & 0 & -1  
\end{pmatrix}  
\; , 
\end{equation}
and $UOSp(2|2)$ Schwinger operator
\begin{equation}
\Psi(r)\equiv 
(a, b, \sqrt{r}f, \sqrt{r}g)^t, 
\label{uosp22schwingerope}
\end{equation}
 we introduce the type-II SVBS states: 
\begin{equation}
|\text{SVBS-II}\rangle=\prod_{\langle ij\rangle} 
(\Psi^{\dagger}_i(r)~ \mathcal{R}_{2|2}~\Psi_j^{*}(r))^M|\text{vac} \rangle=\prod_{\langle ij\rangle} 
(a_i^{\dagger}b_j^{\dagger}-b_i^{\dagger}a_j^{\dagger}-rf_i^{\dagger}f_j^{\dagger}-rg_i^{\dagger}g_j^{\dagger})^M|\text{vac} \rangle. 
\label{SVBSstateII}
\end{equation}
The inclusion of  two species of holes, $f$ and $g$, allows us to write down 
a wave function more symmetric with respect to the bosonic  
 and fermionic degrees of freedom. 
The new fermionic degrees of freedom $g_i$ are interpreted as another species of  (spinless) hole, and 
 satisfy the standard anti-commutation relations 
$\{g_i,g^{\dagger}_j\}=\delta_{ij}$, $\{f_{i},g_{j}\}=0$ and etc. 
In the type-II VBS states,  
there appear the local sites such as $f^{\dagger}_i g_i^{\dagger}\vac$ with spin-0, 
which are not realized in the type-I SVBS states.

We have two species of fermions, and the total particle number of the Schwinger particles at each site $i$ reads as  
\begin{equation}
zM=a^{\dagger}_ia_i+b^{\dagger}_ib_i+f^{\dagger}_if_i+g^{\dagger}_ig_i.  
\end{equation}
Since the eigenvalues of $n_{f}(i){=}f^{\dagger}_{i}f_{i}$ and 
$n_{g}(i){=}g^{\dagger}_{i}g_{i}$ can take either 0 or 1, 
in the type-II SVBS chain $(z=2)$, the following four eigenvalues are allowed 
for the local spin quantum number $S_i=\frac{1}{2}(a^{\dagger}_ia_i+b^{\dagger}_ib_i)$: 
\begin{equation}
S_i={M},~ M-\frac{1}{2},~M-\frac{1}{2},~ M-1. 
\end{equation}
In particular, for the $M=1$ SVBS chain, the  local spin values are given by 
\begin{equation}
S_i=1,~\frac{1}{2},~\frac{1}{2},~ 0 \; , 
\end{equation}
and the local Hilbert space is spanned by the following nine basis states,   
\begin{align}
&|1\rangle=\frac{1}{\sqrt{2}}{a_i^{\dagger}}^2|\text{vac}\rangle,\quad 
|0\rangle={a_i^{\dagger}}b_i^{\dagger}|\text{vac}\rangle, \quad 
|{-1}\rangle=\frac{1}{\sqrt{2}}{b_i^{\dagger}}^2|\text{vac}\rangle, \nonumber\\
& ~~~~~~~~~|\!\uparrow\rangle=a_i^{\dagger}f_i^{\dagger}|\text{vac}\rangle, \quad 
|\!\downarrow\rangle=b_i^{\dagger}f_i^{\dagger}|\text{vac}\rangle,\nonumber\\
& ~~~~~~~~~|\!\uparrow'\rangle=a_i^{\dagger}g_i^{\dagger}|\text{vac}\rangle, \quad 
|\!\downarrow'\rangle=b_i^{\dagger}g_i^{\dagger}|\text{vac}\rangle,\quad\nonumber\\
& ~~~~~~~~~ ~~~~~~~~~|0'\rangle=g_i^{\dagger}f_i^{\dagger}|\text{vac}\rangle.
\end{align}
The edge states are now given by  
\begin{equation}
|\!\uparrow\rangle\!\rangle=a^{\dagger}|\text{vac}\rangle, \;\;
|\!\downarrow\rangle\!\rangle=b^{\dagger}|\text{vac}\rangle, \;\;
|0\rangle\!\rangle=f^{\dagger}|\text{vac}\rangle, \;\; |0'\rangle\!\rangle=g^{\dagger}|\text{vac}\rangle, 
\end{equation}
and correspondingly, there appear $4\times 4=16$ degenerate ground states 
for the $M=1$ type-II SVBS chain. 

\begin{figure}[!t]
\centering
\includegraphics[width=7cm]{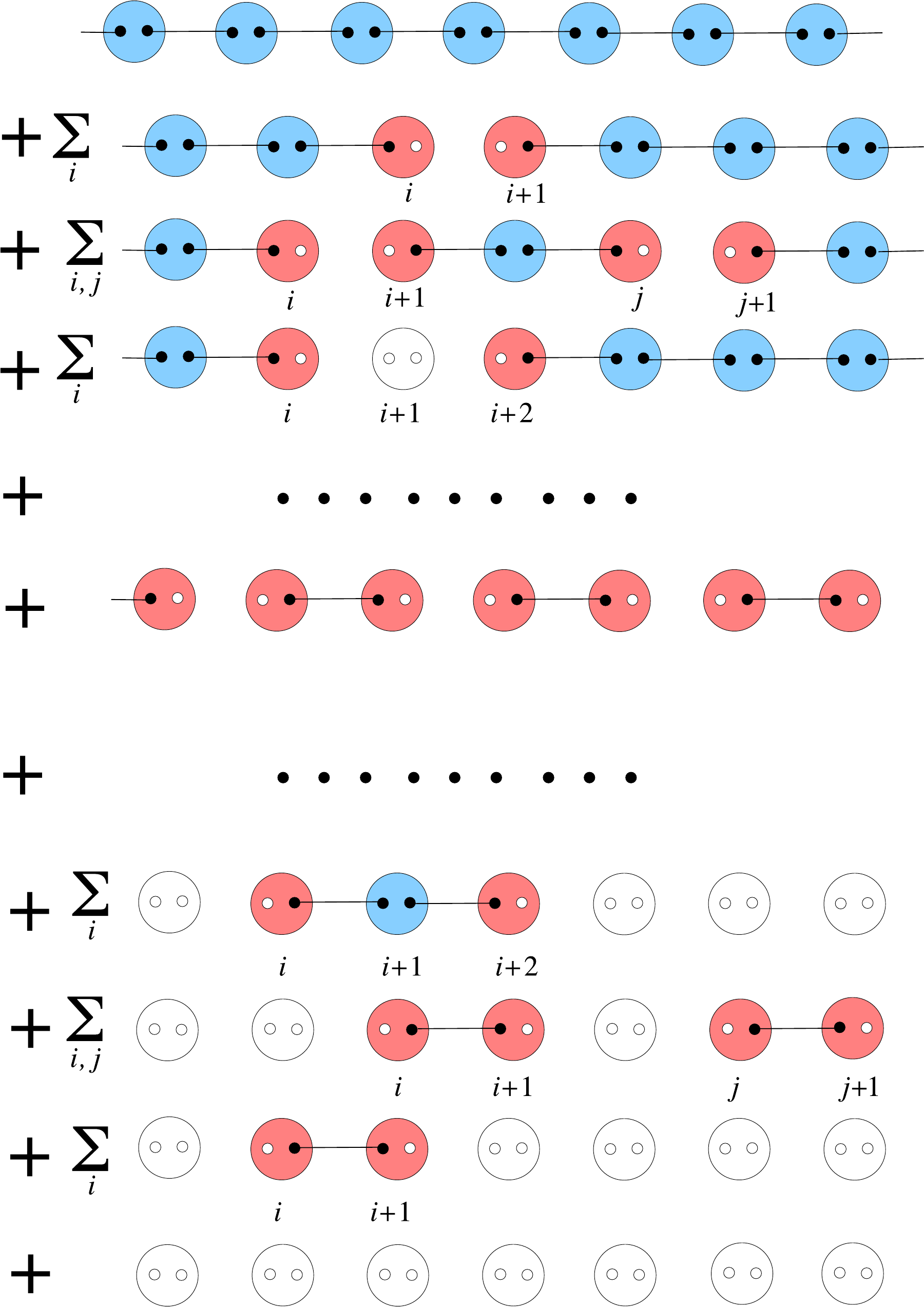}
\caption{(Color online) Like the type-I SVBS chain, the type-II SVBS chain is also expressed as a superposition of the 
hole-pair doped VBS chains. What is different to the type-I SVBS chain is the appearance of 
the spinless sites, depicted by the large white circles with double holes.   
The MG states are realized in the ``middle'' of the sequence. 
The original VBS state and the hole-VBS state are respectively realized 
in the first and last lines.%
 \label{expNewSUSY.fig} [Figure and Caption are taken from Ref.\cite{Hasebe-Totsuka2011}.] }
\end{figure}

The $M=1$ type-II SVBS chain has the following properties.  
As in the type-I SVBS state, the type-II SVBS state reproduces the original VBS state 
for $r\rightarrow 0$: 
\begin{equation}
|\text{SVBS-II}\rangle \rightarrow |\text{VBS}\rangle
=\prod_{i}(a^{\dagger}_i b_{i+1}^{\dagger}-b_i^{\dagger}a^{\dagger}_{i+1})|\text{vac}\rangle 
\; .\label{rzerotypeII}
\end{equation}
On the other hand, when $r\rightarrow \infty$, it reduces to the  
{\em totally uncorrelated} fermionic (F) state filled with holes:
\begin{equation}
\begin{split}
|\text{SVBS-II}\rangle \rightarrow |\text{F-VBS}\rangle
& \equiv 
\pm \prod_{i}f^{\dagger}_i g^{\dagger}_i|\text{vac}\rangle . 
\end{split} \label{rinftypeII}
\end{equation}
Here we have assumed the open boundary condition\footnote{%
If the periodic boundary condition is used, we have zero state for odd-length chains.  
(The sign factor depends on both the parity of the system size and the edge states).}. 
Note that, unlike type-I, type-II SVBS states have no spin degrees of freedom for $r\rightarrow \infty$.    
The super-coherent state representation of $|\text{SVBS-II}\rangle$ is given by 
\begin{align}
\Psi_{\text{SVBS-II}}&= \prod_{<ij>}(u_iv_j-v_iu_j-r\eta_i\eta_j-r\eta'_i\eta'_j)^M\nonumber\\
&=\exp\biggl(-Mr\sum_{<ij>}\frac{\eta_i\eta_j+\eta'_i\eta'_j}{u_iv_j-v_iu_j}  \biggr)\cdot \exp\biggl(-Mr^2\sum_{<ij>}\frac{\eta_i\eta_i \eta'_j\eta'_j}{(u_iv_j-v_iu_j)^2}   \biggr)          
{\cdot}  \Phi_{\text{VBS}}. 
\end{align}
By expanding the exponentials in terms of  $r$, 
the type-II SVBS states can be expressed as a superposition of the hole-pair doped VBS states as shown in Fig.~\ref{expNewSUSY.fig}.  

\subsection{Superconducting Properties}\label{subsec:scproperty}

In both $\Psi_{\text{SVBS-I}}$ and $\Psi_{\text{SVBS-II}}$, the fermions always appear in pairs 
and the wave functions can be expressed by a superposition of fermion pairs. 
We can point out interesting similarities between the SVBS states and  BCS state of SC \cite{Schrieffer1999}: 
\begin{equation}
|\text{BCS}\rangle =\prod_{k\ge 0} \frac{1}{\sqrt{1+|g_k|^2}}(1+g_k c_k^{\dagger}c_{-k}^{\dagger})\Vac,  
\label{BCSstatedef}
\end{equation}
 with electron operator $c_{k}$ and coherence factor $g_k$.    
$|\text{BCS}\rangle$ can be expressed as 
\begin{align}
|\text{BCS}\rangle \propto &\prod_{k\ge 0} (1+g_k c_k^{\dagger}c_{-k}^{\dagger})\vac 
=  \exp(\sum_{k\ge 0} g_k c_k^{\dagger}c_{-k}^{\dagger} )\Vac \nonumber\\
&= \Vac + (\sum_{k\ge 0} g_k c_k^{\dagger}c_{-k}^{\dagger})  \Vac 
+\frac{1}{2}(\sum_{k\ge 0} g_k c_k^{\dagger}c_{-k}^{\dagger})^2 \Vac 
+\cdots+\prod_{k\ge 0} g_k c_k^{\dagger} c_{-k}^{\dagger} \Vac \; .   
\label{bcsexpansion}
\end{align}
This expansion may remind us of the hole-pair expansion of the SVBS state (\ref{expansionofsvbs}). 
Furthermore, in the limit $g_k\rightarrow 0$, the BCS state reduces to the electron vacuum (no fermions), 
and for $g_k\rightarrow\infty$, it coincides with the filled Fermi sphere $|\text{Fermi}\rangle$: 
\begin{subequations}
\begin{align}
&|\text{BCS}\rangle ~~\overset{g_k\rightarrow 0}\longrightarrow ~~ \Vac \; , \\
&|\text{BCS}\rangle~~\overset{g_k\rightarrow \infty}{\longrightarrow} ~~|\text{Fermi}\rangle\equiv \prod_{k\ge 0}c^{\dagger}_kc^{\dagger}_{-k}|\text{vac}\rangle. 
\end{align}
\end{subequations}
These also show apparent similarities  with the asymptotic behaviors of the type-II SVBS chain, (\ref{rzerotypeII}) and (\ref{rinftypeII}), under the correspondence  
\begin{subequations}
\begin{align}
& \Vac~~\leftrightarrow~~|\text{VBS}\rangle, \\
& c_k^{\dagger}c_{-k}^{\dagger}~~\leftrightarrow~~f^{\dagger}_ig^{\dagger}_i. 
\end{align}
\end{subequations}

\begin{figure}[h]
\begin{center}
\includegraphics[scale=0.7]{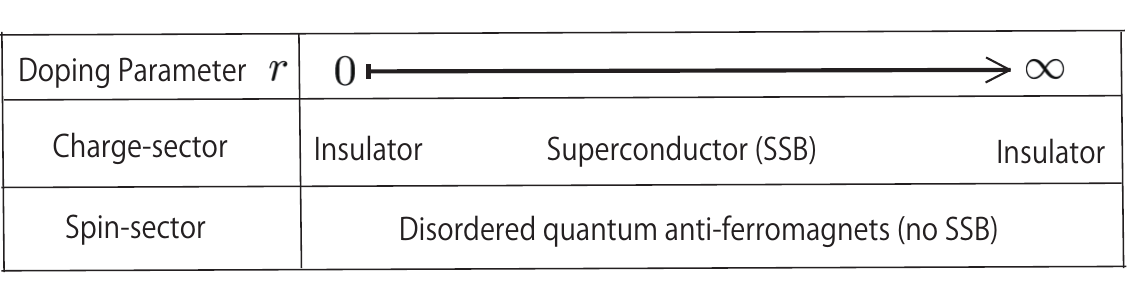}
\caption{ The SVBS states exhibit superconducting property  in the charge sector with finite $r$ in addition to quantum AFM property in the spin sector.    
\label{fig:PhasesSVBS}} 
\end{center}
\end{figure}

 From the analogies to the BCS state, the SVBS states are expected to exhibit SC property in the charge sector by the immersion of hole-pairs to  (insulating) VBS states [see Fig.~\ref{fig:PhasesSVBS}]. This is quite similar to the mechanism of the Anderson's RVB picture of high Tc SC \cite{Anderson1987}:  A finite amount of  hole-doping transforms the insulator of resonating valence bond state to  
  high Tc SC state.  In the following, we explore qualitative arguments for the SC aspect of the SVBS states. 

\subsubsection{$\mathcal{N}=1$}

\begin{figure}[h]
\begin{center}
\includegraphics[scale=0.5]{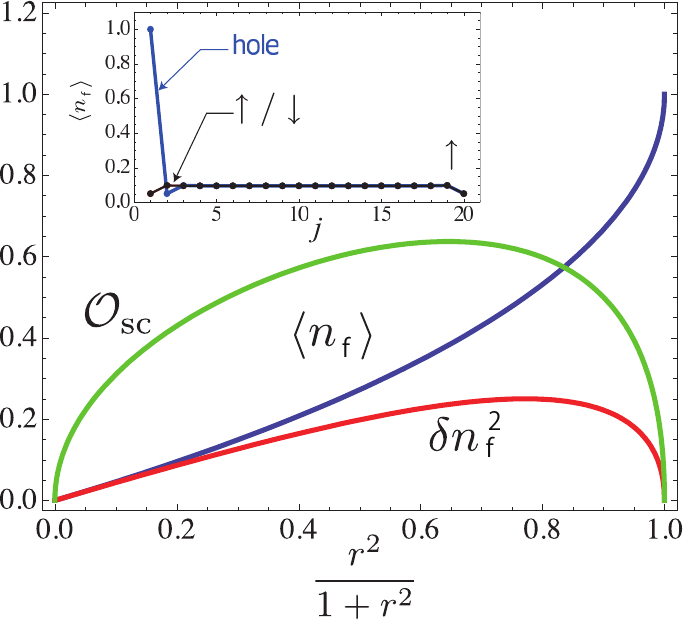}
\caption{(Color online) Plot of 
${\cal O}_{\text{sc}}=\VEV{\Delta}$, 
the hole density $\VEV{n_{\rm f}}$ and 
the hole-number fluctuation 
$\delta n_{\rm f}^2$ 
as a function of $r$.  Here the bulk values are plotted.  
Inset: Profile of the hole density ($r=0.5$) 
for a finite system ($L=20$) with different left edge states 
($\uparrow$, $\downarrow$, and ``hole'').  Only the left edge state 
is changed with the right one fixed to $s_{\text{R}}=\uparrow$. 
The hole density approaches exponentially to the bulk value 
as we move away from the edge. 
\label{fig:SC-orderparam} (Figure and Caption are taken from Ref.\cite{Hasebe-Totsuka2011}.)} 
\end{center}
\end{figure}

Since the naive SC order parameter such as $\langle\, f_k^{\dagger}f_{k+1}^{\dagger}\,\rangle $ vanishes due to the violation of particle number conservation at each site, as the SC order parameter of the type-I VBS state we adopt the following quantity:   
\begin{equation}
{\rm{\Delta}}=(a^{\vphantom{\dagger}}_k\,b^{\vphantom{\dagger}}_{k+1} - b^{\vphantom{\dagger}}_k\,a^{\vphantom{\dagger}}_{k+1})\,f^{\dagger}_k\,f^{\dagger}_{k+1}, 
\label{swaveorder}
\end{equation}
which is calculated as 
\begin{equation}
\langle\, {\rm{\Delta}}\,\rangle 
={2M(M+{\frac{1}{2}})^2\,r\over \left(\sqrt{M(M\!+\!1)(1\!+\!|r|^2) + {1\over 4}}+{1\over 2}(M\!+\!{1\over 2})\right)^{\!2}-
{1\over 4}(M\!+\!{1\over 2})^2}\ . 
\label{swaveorderpara}
\end{equation}
It  takes the maximum value
\begin{equation}
|\Delta_{\text{max}}|=(\sqrt{5}-2)\sqrt{\frac{2M(1+\sqrt{5})}{M+1}}
\end{equation}
at
\begin{equation}
|r|=\biggl(M+\frac{1}{2}\biggr)\sqrt{\frac{1+\sqrt{5}}{2M(M+1)}}.
\end{equation}
In particular for $M=1$, $|r|=\frac{3}{2}\sqrt{\frac{1+\sqrt{5}}{6}}\simeq 1.10$.  
The expectation values for the boson and fermion numbers, $n_{\rm b}(i)=a^{\dagger}_ia_i+b^{\dagger}_ib_i$ and 
$n_{\rm f}(i)=f^{\dagger}_if_i$, 
are respectively calculated as
\begin{align}
&\langle n_{\rm b}\rangle =2M-1 +\frac{2M+1}{\sqrt{4M(M+1)(1+|r|^2)+1}}, \nonumber\\
&\langle n_{\rm f}\rangle = 1-\frac{2M+1}{\sqrt{4M(M+1)(1+|r|^2)+1}}.
\end{align}\label{expectnumber}
As expected, with increase of the hole doping $|r|$, $\langle n_{\rm b}\rangle$ monotonically decreases, while $\langle n_{\rm f}\rangle $ monotonically increases.
The fluctuations for the boson and fermion numbers, $\delta n_{\rm b}^2 =\langle{n^2_{\rm b}\rangle-\langle{n_{\rm b}}\rangle}^2$ and $\delta n_{\rm f}^2=\langle n^2_{\rm f}\rangle-\langle n_{\rm f}\rangle^2$, are given by 
\begin{equation}
\delta n_{\rm b}^2=\delta n_{\rm f}^2=\frac{2M+1}{\sqrt{4M(M+1)(1+|r|^2)+1}}-\frac{(2M+1)^2}{{4M(M+1)(1+|r|^2)+1}}, 
\label{SVBSparticlenumberfluc}
\end{equation}
and their maximum $\delta n_{\rm b}=\delta n_{\rm f}={1\over 2}$ is met at
\begin{equation}
|r|=3\bigg(1+{1\over 4M(M+1)}\bigg)\ .
\end{equation}
The behaviors of such quantities are depicted in Fig.~\ref{fig:SC-orderparam}. 

While the BCS state (\ref{BCSstatedef}) has the particle-hole symmetry
\footnote{ For (\ref{BCSstatedef}), 
the order parameter $\Delta_k=\langle c^{\dagger}_{k}c^{\dagger}_{-k}\rangle$,  the electron number $n_k=c_k^{\dagger}c_k$, 
and its fluctuation $\delta n_k$, are calculated as   
\begin{align}
&\Delta_k=
\frac{g_k^*}{1+|g_k|^2}=\frac{1}{{g_k}+{g_k^*}^{-1}},\nonumber\\ 
&\langle n_k\rangle =\frac{|g_k|^2}{1+|g_k|^2},\nonumber\\
&\delta n_k^2=\langle (n_k-\langle{n_k}\rangle)^2\rangle = \frac{|g_k|^2}{(1+|g_k|^2)^2}=
\frac{1}{(g_k+{g_k^*}^{-1})(g_k^*+g_k^{-1})}.
\end{align}
$\Delta_k$ and $\delta n_k$ are symmetric under $g_k \leftrightarrow 1/g_k^*$ due to the particle-hole symmetry.}, the SVBS state  does not show an exact particle-hole symmetry, $r~\leftrightarrow ~1/r$ [see the SC order parameter (\ref{swaveorderpara}) for instance]. This is because of the unequal properties between boson and fermion operators, $(a_ib_j-b_ia_j) \leftrightarrow f_if_j$.

\subsubsection{$\mathcal{N}=2$ }\label{trancpseudohamilSVBS22}

\begin{figure}[h]
\begin{center}
\includegraphics[scale=0.7]{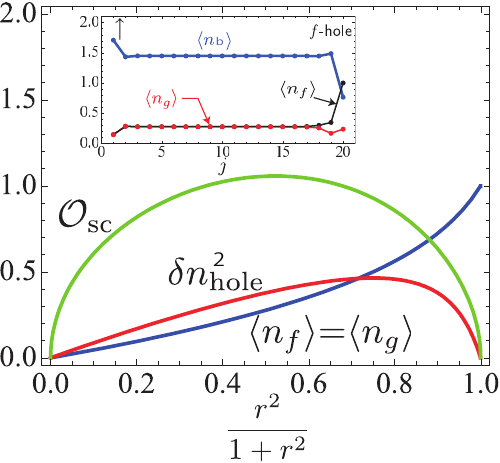}
\caption{(Color online) Plot of 
${\cal O}_{\text{sc}}=\VEV{\Delta_{i}}$, 
the hole density $\VEV{n_{\text{hole}}(i)}=\VEV{f^{\dagger}_{i}f_{i}}$ and 
the hole-number fluctuation 
$\delta n_{\text{hole}}^2$ 
as a function of $r$. 
Inset: Profile of the hole density [$r=0.5$]  
for a finite system ($L=20$).  Only the left edge state 
is changed with the right one fixed to $s_{\text{R}}=\uparrow$. 
(Figure and Caption are taken from Ref.\cite{Hasebe-Totsuka2011}.) 
\label{fig:SC-orderparam-2}}
\end{center}
\end{figure}

We define the order parameter for the type-II SVBS state (\ref{SVBSstateII}) as follows: 
\begin{equation}
\Delta_{i}  \equiv 
(a_{i}b_{i+1} - b_{i}a_{i+1})
(f^{\dagger}_{i}f^{\dagger}_{i+1}
+ g^{\dagger}_{i}g^{\dagger}_{i+1}) .
\end{equation}
In Fig.~\ref{fig:SC-orderparam-2}, we plotted its expectation value 
\begin{equation}
{\cal O}_{\text{sc}}=\VEV{\Delta_{i}},
\end{equation}
 the hole-number $\langle n_{\rm f}\rangle$, and hole fluctuation $\delta n_{\text{hole}}$, 
\begin{equation}
\begin{split}
& \VEV{n_{\rm f}}=\VEV{f_{i}^{\dagger}f_{i}}=
\VEV{g^{\dagger}_{i}g_{i}}=\VEV{n_{g}} \, , \\
&\delta n_{\text{hole}}^2=\VEV{n^{2}_{\text{hole}}} 
- \VEV{n_{\text{hole}}}^{2} \; 
(n_{\text{hole}}\equiv n_{f}+n_{g}) \; .
\end{split}
\end{equation}
The SC order parameter ${\cal O}_{\text{SC}}$    
takes its maximal value  at $|r|\simeq  1.05$ for $M=1$.  Comparing Fig.~\ref{fig:SC-orderparam} and Fig.~\ref{fig:SC-orderparam-2}, one may find that  the behavior of the order parameter of the type-II SVBS chain is more symmetric with respect to $|r|=1$ than that of the type-I SVBS chain. This is because of the almost compensation of the contributions from the equal number of the boson  and  fermion  species in the type-II SVBS states.  

\subsection{Parent Hamiltonians}
\label{sec:parent-Ham}

The VBS state is the exact and unique ground state of a many-body Hamiltonian which we call the parent Hamiltonian \cite{Afflecketal1987,Afflecketal1988}. 
The relation between the VBS state and its parent Hamiltonian is quite unique.  
Usually in quantum mechanics,  Hamiltonian is firstly given, and then we solve the eigenvalue problem of the given Hamiltonian. 
In most cases, particularly in the presence of many-body interaction,  it is formidable to exactly solve the eigenvalue problem, and so we need to rely on some appropriate approximation method.   
Interestingly in VBS models, the procedure is completely $\it{inverse}$: the many-body state (VBS state) is firstly given, and next the parent Hamiltonian is constructed such that its ground state is exactly given by the VBS state.      

\begin{figure}[H]
\begin{center}
\includegraphics[scale=0.8]{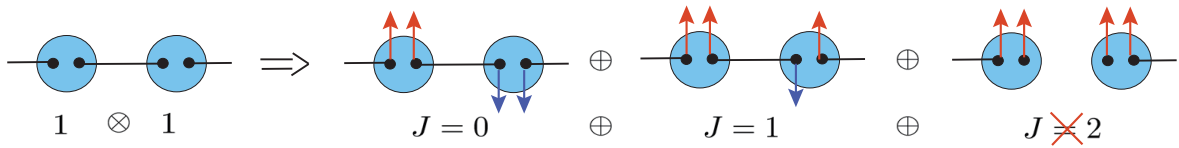}
\caption{(Color online) The bond spin $J=2$ cannot be realized in the $S=1$ VBS chain.  
\label{fig:constparentHam}}
\end{center}
\end{figure}

We briefly review the procedure for the construction of the parent Hamiltonian. Consider the VBS chain with bulk spin $1$.  
The $SU(2)$ decomposition rule of two spin 1 gives the total spin $J=0,1,2$: 
\begin{equation}
1\otimes 1=0\oplus 1\oplus 2. 
\end{equation}
However, the values for the bond spin of the VBS chain does not take $J=2$, since in that case all four 1/2 spins are aligned to a same direction and do not form the spin-singlet bond between  neighboring sites [see Fig.~\ref{fig:constparentHam}].    
Hence, the VBS state satisfies the condition 
\begin{equation}
P_{J=2}(i,i+1)|\text{VBS}\rangle=0,
\end{equation}
where $P_J$ denotes a projection operator to the bond spin $J=2$.  
Since the eigenvalue of the  projection operator is either 0 or 1  for arbitrary adjacent sites, the minimum eigenvalue of the ``many-body operator'', $\sum_i P_{J=2}(i,i+1)$, is zero. This simple fact is the key observation for the construction of the parent Hamiltonian.  We then construct the parent Hamiltonian for the VBS chain  as 
\begin{equation}
H=\sum_{i}V_{J=2}P_{J=2}(i,i+1), 
\label{basicparentha}
\end{equation}
where $V_{J=2}$ denotes a positive coefficient\footnote{$V_{J=2}$ can depend on the lattice site $i$, but here we postulate lattice translation  symmetry and drop the lattice index $i$.}.  It is obvious that the eigenvalues of the parent Hamiltonian (\ref{basicparentha}) are semi-positive definite and its ground-state energy is zero. 
Also remember that the VBS state vanishes by the projection operators in the parent Hamiltonian, and hence  
\begin{equation}
H|\text{VBS}\rangle=0.  
\end{equation}
Therefore, the VBS chain is a zero-energy ground state of the parent Hamiltonian (\ref{basicparentha}). 
Uniqueness of the ground state\footnote{%
On a finite-size chain, the ground state is, by construction, either unique (for a periodic chain) or degenerate 
with respect to the edge states (for an open chain). 
In order to prove the uniqueness in the infinite-size system, one has to define the {\em infinite-size} 
ground state carefully \cite{Afflecketal1988}.
}  
 for the parent Hamiltonian (up to the degeneracy coming from the edge degrees of freedom) 
 can also be proven \cite{Afflecketal1988}.  The projection operator is explicitly derived as  
\begin{align}
P_J(i,i+1)&=\prod_{J'\neq J}\frac{(\bold{S}(i)+  \bold{S}(i+1)  )^2-J'(J'+1) }{J(J+1)-J'(J'+1)}\nonumber\\
&=\prod_{J'\neq J}\frac{2\bold{S}(i)\cdot \bold{S}(i+1)+2S(S+1)-J'(J'+1) }{J(J+1)-J'(J'+1)}.
\end{align}
In the present case $S=1$,  $P_{J=2}(i,i+1)$ is given by  
\begin{align}
P_{J=2}(i,i+1)&= \prod_{J'=0,1}\frac{2\bold{S}(i)\cdot \bold{S}(i+1)+4-J'(J'+1)   }{6-J'(J'+1)}\nonumber\\
&=\frac{1}{2}\biggl(  \bold{S}(i)\cdot \bold{S}(i+1)+\frac{1}{3}(\bold{S}(i)\cdot \bold{S}(i+1))^2+\frac{2}{3}  \biggr).
\end{align}
After all, for the $S=1$ VBS chain the parent Hamiltonian takes the following form:   
\begin{equation}
H=V \sum_{i} \left\{
\bold{S}(i)\cdot \bold{S}(i+1)+\frac{1}{3}(\bold{S}(i)\cdot \bold{S}(i+1))^2+\frac{2}{3}
\right\} , 
\label{basicparentharewritten}
\end{equation}
where  the overall proportional factor $1/2$ has been absorbed in the coefficient $V$. The parameter $V$ simply determines the energy scale of the system, and is not important in the dynamical behaviors of the system. 
The construction of the parent Hamiltonian is rather technical, but the resulting parent Hamiltonian (\ref{basicparentharewritten}) appears to be ``physical'':  The spin-spin interaction represents the Heisenberg AFM, though it contains the quadratic term whose coefficient is one-third of the first Heisenberg AFM term. 
It may be obvious from the above discussion that the VBS state is the exact ground state of the parent Hamiltonian, but   
if the Hamiltonian (\ref{basicparentharewritten}) was firstly given, one might think that it is almost impossible to derive its exact energy spectra and corresponding many-body states.  However at least for the  ground-state, we know the exact form of the wave function and its energy.  Taking this advantage,  we can develop precise discussions about physical quantities relevant to the ground-state, such as entanglement spectrum. For excitations, we need to rely on some approximation technique to extract useful information from the parent Hamiltonian (see Sec.\ref{subsec:crackion}).   
It should also be emphasized that the present construction can be generalized to higher spin VBS states  on arbitrary lattice in any dimensions. Generally, the parent Hamiltonian is given by 
\begin{equation}
H=\sum_{\langle i,j\rangle} \sum_{(z-1)M < J }^{zM}V_J P_{J}(i,j), 
\label{parentHamilgene}
\end{equation}
where $V_J$ denotes a positive coefficient and $P_J(i,j)$ is given by 
\begin{align}
P_J(i,j)&=\prod_{J'\neq J}\frac{(\bold{S}(i)+  \bold{S}(j)  )^2-J'(J'+1) }{J(J+1)-J'(J'+1)}\nonumber\\
&=\prod_{J'\neq J}\frac{2\bold{S}(i)\cdot \bold{S}(j)+2S(S+1)-J'(J'+1) }{J(J+1)-J'(J'+1)}.
\end{align}

In QHE, the parent Hamiltonian for the Laughlin-Haldane wave function is called the Haldane's pseudo-potential Hamiltonian \cite{haldane1983}. As discussed in Sec.\ref{subsec:valenvebond}, the mathematical structure of the VBS state and the QHE is similar to each other, 
and the pseudo-potential Hamiltonian for QHE can also be obtained by applying the translations between QHE and VBS (Table \ref{correspondenceII}). 
Indeed, the resultant pseudo-potential Hamiltonian for Laughlin-Haldane wave function 
takes the form similar to (\ref{parentHamilgene}):  
\begin{equation}
H=\sum_{ i< j} \sum_{m(N-2) < J }^{m(N-1)} V_J P_J(i,j), 
\end{equation}
where $P_J$ denotes the projection operator to  two-body angular momentum $J$.

\subsubsection{$\mathcal{N}=1$ }\label{trancpseudohamilSVBS}

By replacing the $SU(2)$ operators with $UOSp(1|2)$ ones, we readily construct 
the parent Hamiltonian for the type-I SVBS states which 
are invariant under $UOSp(1|2)$ transformations generated by
the $SU(2)$ bosonic generators $L_i$,  
\begin{equation}
L_{i}=\frac{1}{2}
\Psi^{\dagger}(r)
\begin{pmatrix}
\sigma_i & 0 \\
0 & 0 
\end{pmatrix}\Psi({1}/{r}),  
\end{equation}
and the parameter-dependent fermionic operators $K_{\alpha}$, 
\begin{equation}
K_{\alpha}=\frac{1}{2}
\Psi^{\dagger}(r)
\begin{pmatrix}
0 & \tau_{\alpha} \\
-(i\sigma_2\tau_{\alpha})^t & 0 
\end{pmatrix}\Psi({1}/{r}),   
\end{equation}
where $\Psi(r)$ is defined by (\ref{paraSchwingergrade}).  
Notice that $L_i$ and $K_{\alpha}$ satisfy the $UOSp(1|2)$ algebra (\ref{osp3}). 
With use of the $UOSp(1|2)$ generators, it is straightforward to construct the 
parent Hamiltonian for the SVBS states with
$\it{arbitrary}$ value of the parameter $r$. 
First, we need to derive the projection operators to the subspaces of bond-superspin $J$. 
The decomposition rule for the two superspins $\mathcal{S}$ is given by  
\begin{equation}
J=\mathcal{S}\otimes \mathcal{S}=0\oplus 1/2 \oplus 1 \oplus 3/2 \oplus 2 \oplus \cdots \oplus 2\mathcal{S}.  
\label{uosp12decomporu}
\end{equation}
Note that the $UOSp(1|2)$ decomposition rule is similar to that of the $SU(2)$ except for the bond-superspin decreasing by $1/2$. 
The SVBS state does not contain any $UOSp(1|2)$ bond superspins larger 
than $J_{\text{max}}=(z-1)M$, and is the exact zero-energy ground state of the parent Hamiltonian  
\begin{equation}
H_{\text{type-I}}=\sum_{\langle ij\rangle}\sum_{J=J_{\text{max}}+{1\over 2}}^{2\mathcal{S}}V_{J}\,{\mathbb P}_J(i,j), \label{SUSYparentHamil}
\end{equation}
where $V_J$ are positive coefficients, and 
$\mathbb{P}_{J}(i,j)$ are the projection operator onto the superspin-$J$ representation 
of $UOSp(1|2)$  written in terms of the Casimir operator as:  
\begin{equation}
\mathbb{P}_J(i,j)=\prod_{J'\neq J}^{2\mathcal{S}}\frac{( K_A(i)+K_A(j) )^2-J'(J'+\frac{1}{2}) }{J(J+\frac{1}{2})-J'(J'+\frac{1}{2})} =\prod_{J'\neq
J}^{2\mathcal{S}}\frac{2K_A(i)K_A(j) +2\mathcal{S}(\mathcal{S}+\frac{1}{2})-J'(J'+\frac{1}{2})
}{J(J+\frac{1}{2})-J'(J'+\frac{1}{2})}, \label{projectiongene}
\end{equation}
which projects to the two-site subspace of the bond superspin $J$.
Here,  $K_A(i)K_A(j)\equiv L_i(i)L_i(j)+\epsilon_{\alpha\beta}K_{\alpha}(i)K_{\beta}(j)$ and $K_A(i)K_A(i)\equiv K_A(j)K_A(j)=\mathcal{S}(\mathcal{S}+{1\over 2})$. Since the
projection operators (\ref{projectiongene}) are $UOSp(1|2)$
invariant, the parent Hamiltonian is 
(\ref{SUSYparentHamil}) as well. Following similar discussions about the uniqueness of the VBS state in  Ref.\cite{Afflecketal1988}, we can prove that the SVBS state is the unique zero-energy ground-state of the parent Hamiltonian (\ref{SUSYparentHamil}).  

For concreteness, we demonstrate the derivation of the parent Hamiltonian for the $\mathcal{S}=1$ SVBS chain ($z=2$, $M=1$ and then $J_{\text{max}}=1$). 
 From (\ref{uosp12decomporu}), we obtain  the parent Hamiltonian (\ref{SUSYparentHamil}) 
 for type-I SVBS chain as  
\begin{equation}
\begin{split}
&H_{\text{chain-I}}=\sum_i \left\{
V_{{3\over 2}} \,\mathbb{P}_{{3\over 2}}(i,i+1)+V_2 \,\mathbb{P}_2(i,i+1) 
\right\}  \\
&=\sum_i \biggl\{   \frac{1}{35}(5V_{2}+63V_{{3\over 2}}) K_A(i)K_A(i+1)   
+\frac{2}{45}(9V_{2}-{7}V_{{3\over 2}}) (K_A(i)K_A(i+1) )^2  \\
&~~~~~~~~+\frac{16}{45}(V_{2}-{5}V_{{3\over 2}}) (K_A(i)K_A(i+1) )^3 +\frac{32}{315}(V_{2}-{7}V_{{3\over 2}})(K_A(i)K_A(i+1) )^4
+ 
V_{{3\over 2}}  \biggr\}. 
\end{split}
\label{explicitSUSYAKLT}
\end{equation}
Here, we add several comments.     Since the Casimir operator
$(K_A(i)+K_A(j))^2$ contains pair-creation terms of fermions, such as $f_i^{\dagger}f_j^{\dagger}(a_ib_j-b_ia_j)$, 
 the Hamiltonian (\ref{explicitSUSYAKLT}) does not preserve the total fermion number
$N_{\rm f}=\sum_if_i^\dagger f_i$ (though the total particle number, $i.e.$ the sum of boson number and fermion number, is conserved).  This fermion number non-conserving term is a particular structure of 
the BCS Hamiltonian, and this is in agreement with the SC property of the SVBS model (see Sec.\ref{subsec:scproperty}).

The fermionic generators $K_{\alpha}$ are non-Hermitian, and then the type-I parent Hamiltonian (\ref{SUSYparentHamil}) is  non-Hermitian as well.     
 As a reasonable Hermitian extension of the type-I Hamiltonian, one may adopt 
\begin{equation}
{H'}_{\text{type-I}}=\sum_{\langle ij\rangle}\sum_{J=J_{\text{max}}+{1\over 2}}^{2\mathcal{S}}V_J\,\mathbb{P}_J^{\dagger}(i,j)\,
\mathbb{P}_J(i,j). 
\label{hermitianHamil}
\end{equation}
 The definition (\ref{hermitianHamil}) is a natural generalization of the original type-I parent Hamiltonian, since,
if the projection operators were Hermitian, from the property
$\mathbb{P}_J^{2}=\mathbb{P}_J$ Eq.(\ref{hermitianHamil}) would be reduced to the
original one (\ref{SUSYparentHamil}). 

\subsubsection{$\mathcal{N}=2$ }\label{trancpseudohamilSVBS2}

 The irreducible representation of the $UOSp(2|2)$ group is specified by $(j,b)$, which signify the indices of the largest bosonic subalgebra,  $SU(2) \oplus U(1)$\footnote{We denote the eigenvalues of the $SU(2)$ Casimir operator $\bolS^2$ as  $j(j+1)$ ($j=0,1/2,1,\cdots$.), and arbitrary complex number eigenvalues of the $U(1)$ operator $B$ as $b$.}.   
 The irreducible representation  is classified into two categories; the typical representation and atypical representation. 
For $b\neq \pm j$, the  irreducible representation is called the typical representation with dimension $8j$,  while for $b=\pm j$  the irreducible representation becomes to the  atypical representation with dimension $4j+1$.   
The eigenvalues of the $UOSp(2|2)$ quadratic  
Casimir operator (\ref{osp22casimir}) for $(j,b)$ representation are generally given by 
\begin{equation}
{C}=j^2-b^2.
\label{casimireigensosp22}
\end{equation}
Then for the atypical representation $(b=j)$, the Casimir eigenvalues identically vanish. Since the Casimir corresponds to the square of the radius of fuzzy manifold, it may not be probable to explore fuzzy geometry based on the atypical representation.  
Meanwhile, the typical representation   consists of four $SU(2)$ representations, $|b,j,j_3\rangle$, $|b+{1}/2,j-\frac{1}{2},j_3\rangle$, $|b-{1}/2,j-\frac{1}{2},j_3\rangle$, $|b,j-1,j_3\rangle$, each of which carries the $SU(2)$ spin index $S$ as   
\begin{equation}
\begin{split}
& \text{(i)} \; S=j  ~~~~~~~~~~~~~~~~~~~~\cdots ~~~(2j+1)\text{-dim}\; , \\
& \text{(ii)} \; S=j-1/2 \;\; ~~~~~~~~~ \cdots~~~ 2j\text{-dim} \; ,\\
&\text{(iii)} \; S=j-1/2 \;\;  ~~~~~~~~\cdots~~~ 2j\text{-dim} \; ,\\
&\text{(iv)} \; S=j-1 \;\;  ~~~~~~~~~~~\cdots~~~ (2j-1)\text{-dim}. 
\end{split}
\end{equation}
For instance,  8-dimensional typical representation, $(j,b)=(1,0)$, is constructed with use of the components of the $UOSp(2|2)$ Schwinger operator (\ref{uosp22schwingerope}): 
\begin{equation}
\begin{split}
& \text{(i)} \;
~|+\rangle=\frac{1}{2}{a_i^{\dagger}}^2|\text{vac}\rangle,
\quad |0\rangle={a_i^{\dagger}b_i^{\dagger}}|\text{vac}\rangle, 
\quad |-\rangle=\frac{1}{2}{b_i^{\dagger}}^2|\text{vac}\rangle, \\
& \text{(ii)} \; ~~~~~~~~~|\!\uparrow\rangle=
 a_i^{\dagger}f_i^{\dagger}|\text{vac}\rangle, 
\quad ~~~~~~~~|\!\downarrow \rangle =b_i^{\dagger}f_i^{\dagger}|\text{vac}\rangle,  \\
& \text{(iii)} \;  ~~~~~~~~|\!\uparrow'\rangle=
 a_i^{\dagger}g_i^{\dagger}|\text{vac}\rangle, 
\quad ~~~~~~~~ |\!\downarrow' \rangle =b_i^{\dagger}g_i^{\dagger}|\text{vac}\rangle,  \\
& \text{(iv)} \;  ~~~~~~~~~~~~~~~ ~~~~~~~~~|0'\rangle=
 g_i^{\dagger}f_i^{\dagger}|\text{vac}\rangle.  
\end{split}
\label{uosp22multijb=10}
\end{equation}
They give rise to $\mathcal{N}=2$ SUSY multiplet. The $UOSp(2|2)$ superspin operators can also be given by 
\begin{equation}
S_i=\Psi(r)^{\dagger}l_i\Psi(r),~~~~K_{\alpha}=\Psi^{\dagger}(r)l_{\alpha}\Psi(r),~~~D_{\alpha}=\Psi^{\dagger}(r)l_{\alpha}'\Psi(r),~~~~B=\Psi^{\dagger}(r)\gamma\Psi(r), 
\end{equation}
with $l_i$, $l_{\alpha}$, $l'_{\alpha}$ and $\gamma$  (\ref{osp22typicalmatrices}). 
Obviously,  $S_i$ and $-iB$ are the generators of the subalgebra $su(2)\oplus u(1)$ 
\footnote{ 
The basis states (\ref{uosp22multijb=10}) carry the $SU(2)\oplus U(1)$ indices as 
\begin{equation}
(\text{i}): |b=0,j=1\rangle, ~~~~ (\text{ii})\pm i(\text{iii}): |b=\pm 1/2,j=1/2\rangle, ~~~~ (\text{iv}): |b=0,j=0\rangle.  
\end{equation}
}. 

For a two-site system, the $UOSp(2|2)$ bond superspin operators are constructed as  
$\bolS^{\text{tot}}=\bolS(i)+\bolS(j)$, $K_{\alpha}^{\text{tot}}=K_{\alpha}(i)+K_{\alpha}(j)$, $D_{\alpha}^{\text{tot}}=D_{\alpha}(i)+D_{\alpha}(j)$, $B^{\text{tot}}=B(i)+B(j)$, and the quadratic Casimir operator is expressed as 
\begin{equation}
\begin{split}
{C}_{i,j}&=\bolS^{\text{tot}}\cdot \bolS^{\text{tot}} +\epsilon_{\alpha\beta}K_{\alpha}^{\text{tot}}K_{\beta}^{\text{tot}}-\epsilon_{\alpha\beta}D_{\alpha}^{\text{tot}}D_{\beta}^{\text{tot}}-B^{\text{tot}}B^{\text{tot}}\\
&={C}(i)+{C}(j)+2\biggl\{\bolS(i)\cdot \bolS(j) +\epsilon_{\alpha\beta}K_{\alpha}(i)K_{\beta}(j)-\epsilon_{\alpha\beta}D_{\alpha}(i)D_{\beta}(j)-B(i)B(j) \biggr\}\\
&={C}(i)+{C}(j)+2{L}(i)\cdot {L}(j)\\
&=2\mathcal{S}^2+2{L}(i)\cdot {L}(j),   
\end{split}\label{quadraticcasosp22}
\end{equation}
where ${C}(i)={C}(j)=\mathcal{S}^2$ $(\mathcal{S}=0,1/2,1,3/2,\cdots)$ \footnote{The  graded fully symmetric representation made of the $UOSp(2|2)$ Schwinger operator carries the $SU(2)\oplus U(1)$ indices, $(j,b)=(\mathcal{S},0)$.  } and ${L}(i)\cdot {L}(j)$ 
 is  defined as 
\begin{equation}
{L}(i)\cdot {L}(j)=\bolS(i)\cdot \bolS(j) +\epsilon_{\alpha\beta}K_{\alpha}(i)K_{\beta}(j)-\epsilon_{\alpha\beta}D_{\alpha}(i)D_{\beta}(j)-B(i)B(j).   
\label{N=2superspinspinint}
\end{equation}
Tensor product of two identical typical representations is decomposed as  
\begin{equation}
\begin{split}
(J,B)&=(\mathcal{S},0)\;\otimes \;(\mathcal{S},0) \\
&=\oplus_{n=0}^{2\mathcal{S}-1}\;(2\mathcal{S}-n,0)
\;\oplus_{n=0}^{2\mathcal{S}-1}\;(2\mathcal{S}-1/2-n,1/2)\;\\
& \qquad \qquad 
\oplus_{n=0}^{2\mathcal{S}-1}\;(2\mathcal{S}-1/2-n,-1/2)\;\oplus_{n=1}^{2\mathcal{S}-1}\;(2\mathcal{S}-n, 0) \; .
\end{split} 
\label{generalosp22times}
\end{equation}
For instance,\footnote{%
On the right-hand sides of (\ref{osp221times1}) and (\ref{generalosp22times}),  $(1/2,1/2)\oplus (1/2,-1/2)$ is replaced by not-completely reducible atypical representation consisting of semi-direct sum of atypical representations, $(0,0)$, $(1/2,-1/2)$, $(1/2,1/2)$, $(0,0)$ 
(for details, see Sect.2.53 in Ref.\cite{BookFrappat}). }  
\begin{equation}
(1,0)\otimes (1,0)=(2,0)\oplus  (3/2,1/2) \oplus (3/2,-1/2) \oplus (1,0) \oplus (1,0) \oplus (1/2,1/2) \oplus (1/2,-1/2).  
\label{osp221times1}
\end{equation}
Similar to the $UOSp(1|2)$ decomposition (\ref{uosp12decomporu}), the $UOSp(2|2)$ bond superspins  decrease by 1/2 [see the right-hand side of (\ref{osp221times1})].   
 As also observed in (\ref{osp221times1}), $B$ is specified by $J$: For integer $J$, $B=0$, while for half-integer $J$, $B=1/2$ or $-1/2$. Hence,  the square of $B$ is uniquely determined as a function of $J$:   
\begin{equation}
B(J)^2=\frac{1}{8}(1-(-1)^{2J}). 
\label{bsquareJ}
\end{equation}

Invoking the usual arguments of constructing the parent Hamiltonian, we derive the type-II parent Hamiltonian as    
\begin{equation}
{H}_{\text{type-II}}=\sum_{ (1-{1}/{z})2\mathcal{S} < J}^{2\mathcal{S}} \sum_{<ij>} V_J P_{J}(C_{i,j}), 
\label{N=2parentgeneralhamil}
\end{equation}
where $P_{J}$ stand for  
the projection operators of the bond superspin $J$: 
\begin{equation}
P_{J}({C}_{i,j})=\prod_{J'\neq J} \frac{{C}_{i,j}-({J'}^2-{B}(J')^2)}{(J^2-B(J)^2)-({J'}^2-{B}(J')^2)}. 
\end{equation}
Here, $B(J)^2$ and $B(J')^2$ are given by (\ref{bsquareJ}). 
For $L=1$ type-II SVBS chain,  
the parent Hamiltonian is given by  
\begin{equation}
{H}_{\text{chain-II}}=\sum_{i}\{V_{3/2}P_{3/2}({C}_{i,i+1})+V_2 P_{2}({C}_{i,i+1})\},    
\label{hamiltoniantypeII}
\end{equation}
where the projection operators are 
\begin{equation}
\begin{split}
&P_{3/2}({C}_{i,i+1})= \prod_{J'=2,1,{1}/{2},0} \frac{{C}_{i,i+1}-({J'}^2-{B(J')}^2)}{2-({J'}^2-{B(J')}^2)}=-\frac{1}{8} {C}_{i,i+1}^2({C}_{i,i+1}-1)({C}_{i,i+1}-4),\\
&P_{2}({C}_{i,i+1})=\prod_{J'={3}/{2},1,{1}/{2},0} \frac{{C}_{i,i+1}-({J'}^2-{B(J')}^2)}{4-({J'}^2-{B(J')}^2)}=\frac{1}{96} {C}_{i,i+1}^2({C}_{i,i+1}-1)({C}_{i,i+1}-2). 
\label{projectionsoperators}
\end{split}
\end{equation}
With use of ${C}_{i,i+1}=2{L}(i)\cdot{L}(i+1)+2$, 
 the type-II parent Hamiltonian (\ref{hamiltoniantypeII}) can be rewritten as 
\begin{align}
{H}_{\text{chain-II}}
=\sum_i&\biggl\{\frac{1}{12}(V_2+36V_{3/2}) ({L}(i)\cdot{L}(i+1))+\frac{1}{3}(V_2+3V_{3/2}) ({L}(i)\cdot{L}(i+1))^2\nonumber\\
&+ \frac{1}{12}(5V_2-36 V_{3/2}) ({L}(i)\cdot{L}(i+1))^3  + \frac{1}{6}(V_2-12 V_{3/2}) ({L}(i)\cdot{L}(i+1))^4  +V_{3/2}\biggr\}.
\label{explicitj1parenttypeII}
\end{align}
It is  noticed that, unlike the  type-I parent Hamiltonians (\ref{SUSYparentHamil}), the type-II parent Hamiltonians  (\ref{N=2parentgeneralhamil}) themselves are  Hermitian, since the $UOSp(2|2)$ Casimir itself (\ref{quadraticcasosp22}) is given by a Hermitian operator.  

\section{Supersymmetric Matrix Product State Formalism}
\label{secsmat}
This section reviews the MPS formalism and its supersymmetric version, SMPS. In the formalism,  
the edge degrees of freedom  are naturally incorporated.  
Practically, the MPS formalism provides a powerful method to calculate physical quantities such as 
excitation gap, string order and entanglement spectrum.  

\subsection{Bosonic Matrix Product State Formalism}
\label{sec:bosonic-MPS}
As we have discussed above, the VBS state is expressed as a product of valence bonds defined on two adjacent sites.  
In 1D, the VBS state \eqref{generalsu2vbs} can be rewritten as a product of matrices defined on local site \cite{Fannes-etal1992, Totsuka-Suzuki1995}:  
\begin{align}
|\text{VBS}\rangle_{\alpha\beta} &= 
({\cal R}_{2}\Phi_1^*)^{\alpha} \prod_{i=1}^{L-1}  
(\Phi_i^{\dagger}\mathcal{R}_{2} \Phi^*_{i+1})  
(\Phi^*)^{\beta}_L |\text{vac}\rangle \nonumber\\
& \equiv ({\cal R}_{2}\Phi_1^*)^{\alpha} \Phi_1^{\dagger}
\cdot \left(\prod_{i=2}^{L-1} \mathcal{R}_{2} \Phi_{i}^*\Phi_i^{\dagger} \right)
 \cdot ({\cal R}_{2}\Phi^*_L {\Phi_L^*}^{\beta}) |\text{vac}\rangle \nonumber\\
&=(\mathcal{A}_1 \mathcal{A}_2  \cdots  \mathcal{A}_{L})_{\alpha\beta} ,
\label{eqn:MPS}
\end{align}
 where the ``state-valued'' matrix $\mathcal{A}_{i}$ is given by 
\begin{equation}
\mathcal{A}_i = \mathcal{R}_{2}\Phi^*_i \cdot\Phi_i^{\dagger} |\text{vac}\rangle_{i} \\
=    
\begin{pmatrix}
a_i^{\dagger}b_i^{\dagger}
& (b_i^{\dagger})^2 \\
-(a_i^{\dagger})^2  & -a_i^{\dagger}b_i^{\dagger} 
\end{pmatrix} 
|\text{vac}\rangle_{i} \\
= 
\begin{pmatrix}
|0\rangle_{i}
& \sqrt{2}|{-1}\rangle_{i}  \\
-\sqrt{2}|1\rangle_{i}  & -|0\rangle_{i} 
\end{pmatrix}. 
\label{eqn:g-matrix}
\end{equation}
It is clear, by the Schwinger-boson construction, 
that the row and the column of the matrix $\mathcal{A}_i$ correspond respectively to the valence bonds 
going from the site $i$ to its adjacent left and right sites.   
Sometimes, it is convenient to write \eqref{eqn:g-matrix} in a slightly different way:
\begin{subequations}
\begin{equation}
\mathcal{A}_{i} = \sum_{m=-1}^{1} A(m)|m\rangle_{i} \quad (\text{representation-(i)})
\label{eqn:MPS-rep-1}
\end{equation}
or 
\begin{equation}
\begin{split}
& \mathcal{A}_{i} = \sum_{a=-x,y,z} A^{\prime}(a)|a\rangle_{i} \quad (\text{representation-(ii)}) \\
& |x \rangle = -\frac{1}{\sqrt{2}}\left(
|{+}1\rangle -|{-}1\rangle
\right)
\; , \;\;
|y \rangle = \frac{i}{\sqrt{2}}\left(
|{+}1\rangle +|{-}1\rangle
\right) 
\; , \;\; 
|z \rangle = |0\rangle 
\; .
\end{split}
\label{eqn:MPS-rep-2}
\end{equation}
\end{subequations}
In the first representation, the $c$-number matrices $A(m)$ are given by
\begin{equation}
A(1)= 
\begin{pmatrix}
0 & 0 \\
-\sqrt{2} & 0 
\end{pmatrix} \; , \;\;
A(0)=
\begin{pmatrix}
1 & 0 \\ 0 & -1
\end{pmatrix} \; , \;\;
A(-1)=
\begin{pmatrix}
0 & \sqrt{2} \\
0 & 0 
\end{pmatrix} \; ,
\end{equation}
while, in the second, $A^{\prime}(a)$ is given by the Pauli matrices $\sigma_{a}$ ($a=x,y,z$).    
Using these representation, we can recast \eqref{eqn:MPS} into the form 
where the $c$-number coefficients and the basis part are separated explicitly:
\begin{equation}
(\mathcal{A}_1 \mathcal{A}_2  \cdots  \mathcal{A}_{L})_{\alpha\beta}
= \sum_{\{m_{j}\}}\left\{
A(m_1)A(m_2)\cdots A(m_{L})
\right\}_{\alpha\beta}
|m_1\rangle_{1}{\otimes}|m_2\rangle_{2}{\otimes}
\cdots {\otimes}|m_{L}\rangle_{L} \; .
\label{eqn:MPS2}
\end{equation}
The state \eqref{eqn:MPS} or \eqref{eqn:MPS2} represents a collection of 
the $D^{2}$ states (with $D$ being the matrix size) specified by the matrix indices 
$(\alpha,\beta)$.   In the above case, $(\alpha,\beta)$ have a clear physical meaning 
that they specify the states of the two emergent edge spins. 

\begin{figure}[h]
\begin{center}
\includegraphics[scale=0.65]{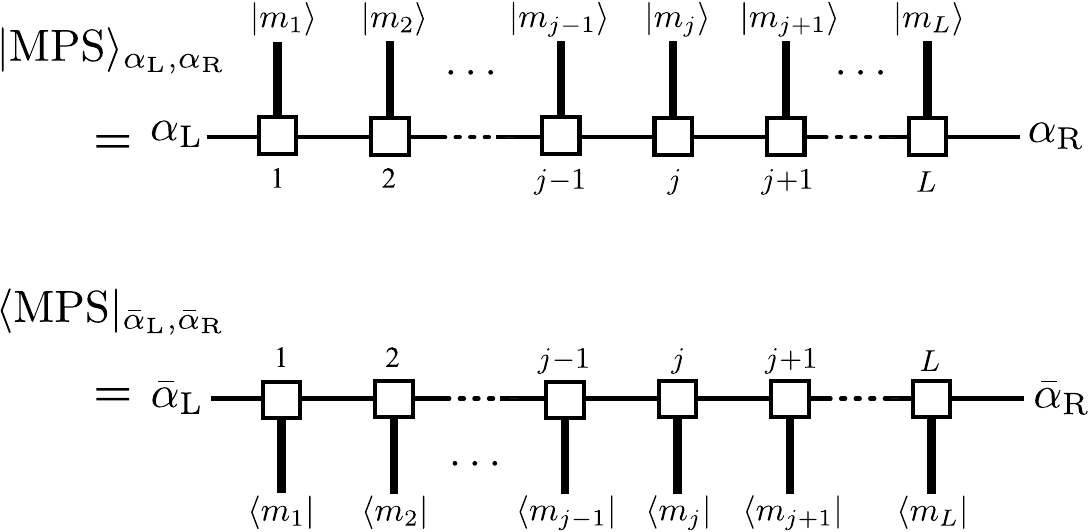}
\end{center}
\caption{Diagrammatic representation of MPS and its dual. 
\label{fig:MPS-string-diag}}
\end{figure}

It is convenient to represent $A(m)$ [and $A^{\ast}(m)$] 
by the following simple {\em tripod} diagrams: 
\begin{equation}
\raisebox{-4.0ex}{\includegraphics[scale=0.65]{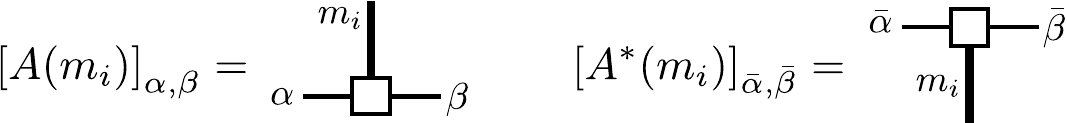}} \; ,
\label{eqn:MPS-tripod-diag}
\end{equation}
where the thick- and the thin lines respectively denote 
the $d$-dimensional {\em physical} Hilbert space (here, $d=3$ spin-1 states labelled by 
$m=-1,0,1$ or $a=x,y,z$) 
and the $D$-dimensional auxiliary space ($D=2$-dimensional space spanned by 
the spinors $a^{\dagger}$ and $b^{\dagger}$, here); the matrix multiplication 
amounts to connecting open thin lines on the adjacent sites. 
Then, the bra and the ket vectors may be depicted by strings of 
these tripods (Fig.~\ref{fig:MPS-string-diag}).  
Quantum states which can be written in the form of \eqref{eqn:MPS} or \eqref{eqn:MPS2} 
are in general called {\em matrix-product states} (MPS).   
As has been mentioned in Sec.~\ref{sec:introduction}, any gapped gapped short-range states 
fall into this category.  We refer the readers to 
recent readable reviews \cite{Verstraete-M-C-08,Schollwoeck-11} 
for more details and the applications. 

We would like to comment on an interesting property of MPS \eqref{eqn:MPS2}. 
When $SU(2)$ rotation acts on the state 
in \eqref{eqn:MPS-rep-2} as $|a\rangle ~\rightarrow~R_{ba}|b\rangle$, 
the local MPS matrix $\mathcal{A}_{i}$ 
transforms like 
\begin{equation}
\mathcal{A}_{i} \xrightarrow{\text{SU(2)}} 
\sum_{a,b=x,y,z}R_{ba}A^{\prime}(a)|b\rangle_{i}
= \sum_{a,b=x,y,z}R_{ba}\sigma_{a}|b\rangle_{i}
= \sum_{b=x,y,z}U^{\dagger}\sigma_{b}U |b\rangle_{i} 
= U^{\dagger} \mathcal{A}_{i}U \; ,
\end{equation}
where $R$ is a three-dimensional rotation matrix and 
$U$ is the corresponding spinor representation.  
Namely, the original $SU(2)$ symmetry for the local spin-1 objects 
``fractionalizes'' into that for the two spin-1/2 objects (spinors).  
From this, it is evident that the spin-1 VBS state on a finite open chain [represented 
by the MPS \eqref{eqn:MPS}] transforms, under $SU(2)$ rotation, as if 
there were two spin-1/2 objects (``quark'' and ``anti-quark'') at the ends of the chain.  
The above is the simplest example of more general symmetry fractionalization property of  
MPS, which will be extensively used in Sec.~\ref{sec:supersymmetryprotected}.  

We can generalize the strategy to construct the parent Hamiltonian for the VBS state 
in Sec.~\ref{sec:parent-Ham} to any MPS.  The idea is to prepare a cluster Hamiltonian 
and tune the parameters so that the Hamiltonian annihilates all the $D^{2}$ states 
({\em i.e.} matrix elements) of the MPS on that cluster. 
In fact, it can be shown that for any given MPS there exists the parent Hamiltonian 
for which the MPS \eqref{eqn:MPS} or \eqref{eqn:MPS2} 
gives the (degenerate) ground states \cite{Perez-Garciaetal2007}.   
By construction, the degree of degeneracy is equal to the number of matrix elements ($D^{2}$).  
For example, the ground state of the parent Hamiltonian of the spin-$S$ VBS state 
(the VBS model) is shown to have $(S{+}1){\times}(S{+}1)$-fold degeneracy, 
when the model is defined on a finite {\em open} chain \cite{Afflecketal1987,Afflecketal1988}.    
This exact degeneracy on a finite chain is peculiar to the VBS model and, 
if we slightly deviate from the solvable VBS point, the emergent edge spins ($S/2$) start 
interacting with each other 
with a coupling constant exponentially small in system size to (partially) resolve the degeneracy.  
For a periodic chain, on the other hand, we have to take the trace over the matrix indices
\begin{equation}
|\text{MPS}\rangle_{\text{PBC}} 
= \text{Tr}\left\{
\bigotimes_{i=1}^{L} \mathcal{A}_{i}  
\right\}
\;, 
\label{eqn:MPS-PBC-1}
\end{equation}
and hence the ground state is not degenerate. 

\subsection{SMPS Formalism and Edge States }\label{section:SMPSEdge}
The Schwinger boson construction described in the previous section 
can be generalized to SUSY cases by 
using the Schwinger operator which contains {\em both} bosons and fermion(s).  
\subsubsection{$\mathcal{N}=1$}
Now let us consider the MPS representation for the type-I VBS chain \cite{Hasebe-Totsuka2011}.  
The SVBS chain (\ref{SVBSstateI}) is written as  
a string of 3$\times$3 matrices $(\alpha,\beta=1,2,3)$:
\begin{equation}
\begin{split}
|\text{SVBS-I}\rangle_{\alpha\beta} &= 
({\cal R}_{1|2}\Psi_1(r)^*)^{\alpha} \prod_{i=1}^{L-1}  
(\Psi_i(r)^{\dagger}\mathcal{R}_{1|2} \Psi_{i+1}(r)^*)  
{\Psi^*(r)}^{\beta}_L |\text{vac}\rangle \\
& \equiv ({\cal R}_{1|2}\Psi_1(r)^*)^{\alpha} \Psi_1(r)^{\dagger}
\cdot \left(\prod_{i=2}^{L-1} \mathcal{R}_{1|2} \Psi_{i}(r)^*\Psi_i(r)^{\dagger} \right)
 \cdot ({\cal R}_{1|2}\Psi_L(r)^* {\Psi_L(r)^{\dagger}})^{\beta} |\text{vac}\rangle \\
&=(\mathcal{A}_1 \mathcal{A}_2  \cdots  \mathcal{A}_{L})_{\alpha\beta} ,
\end{split} 
\label{eqn:SUSY-MPS}
 \end{equation}
 where  
\begin{equation}
\begin{split}
\mathcal{A}_i &= \mathcal{R}_{\text{I}}\Psi_i(r)^* \cdot\Psi_i(r)^{\dagger} |\text{vac}\rangle_{i} \\
&=    
\begin{pmatrix}
a_i^{\dagger}b_i^{\dagger}
& (b_i^{\dagger})^2 &
\sqrt{r}b_i^{\dagger}f_i^{\dagger} \\
-(a_i^{\dagger})^2  & -a_i^{\dagger}b_i^{\dagger} 
& -\sqrt{r} a_i^{\dagger}f_i^{\dagger} \\
-\sqrt{r}f_i^{\dagger}a_i^{\dagger}  & 
-\sqrt{r}f_i^{\dagger}b_i^{\dagger} & 0 
\end{pmatrix} 
|\text{vac}\rangle_{i} \\
&= 
\begin{pmatrix}
|0\rangle_{i}
& \sqrt{2}|{-1}\rangle_{i} &
\sqrt{r}|\!\downarrow\rangle_{i} \\
-\sqrt{2}|1\rangle_{i}  & -|0\rangle_{i}  
& -\sqrt{r} |\!\uparrow\rangle_{i} \\
-\sqrt{r}|\!\uparrow\rangle_{i}  & 
-\sqrt{r}|\!\downarrow\rangle_{i} & 0 
\end{pmatrix}. 
\end{split}
\label{eqn:SUSY-g-matrix}
\end{equation} 
From the expression (\ref{eqn:SUSY-MPS}), it is clear that 
the nine-fold degenerate ground states correspond to  different possible choices 
of the edge states:  
\begin{equation}
\begin{split}
|\text{SVBS-I}\rangle_{\text{open}}
=\bigotimes_{i=1}^{L} \mathcal{A}_{i} 
= 
\begin{pmatrix}
|s_{\text{L}}{=}\downarrow;s_{\text{R}}{=}\uparrow\rangle & 
|s_{\text{L}}{=}\downarrow;s_{\text{R}}{=}\downarrow\rangle & 
|s_{\text{L}}{=}\downarrow;s_{\text{R}}{=}\circ\rangle \\
|s_{\text{L}}{=}\uparrow;s_{\text{R}}{=}\uparrow\rangle & 
|s_{\text{L}}{=}\uparrow;s_{\text{R}}{=}\downarrow\rangle & 
|s_{\text{L}}{=}\uparrow;s_{\text{R}}{=}\circ\rangle \\
|s_{\text{L}}{=}\circ;s_{\text{R}}{=}\uparrow\rangle & 
|s_{\text{L}}{=}\circ;s_{\text{R}}{=}\downarrow\rangle & 
|s_{\text{L}}{=}\circ;s_{\text{R}}{=}\circ\rangle \\
\end{pmatrix} \; .
\end{split}
\label{eqn:MPS-edge}
\end{equation}
The row index specifies the left edge states and the column one the right.  
On the left (right) edge, the matrix indices $\{1,2,3\}$ correspond respectively to 
$\{\downarrow,\uparrow,\text{hole}\}$ ($\{\uparrow,\downarrow,\text{hole}\}$).  

By looking at the form of $\mathcal{A}_{i}$ \eqref{eqn:SUSY-g-matrix}, one sees that 
the matrix has a block structure
\begin{equation}
\begin{pmatrix}
B(2,2) & F(2,1) \\ F(1,2) & B(1,1) 
\end{pmatrix} \; ,
\end{equation}
where $B(m,n)$ and $F(m,n)$ respectively denote ``bosonic'' and ``fermionic'' 
({\em i.e.} anti-commuting) matrices of the dimension $m{\times}n$.  
Therefore, it is convenient to regard $\mathcal{A}_{i}$ as a {\em supermatrix}. 
Thus, the SVBS chain can be expressed in the form of {\em supermatrix-product state} (SMPS), 
and the matrix size of the SMPS is directly related to the number of edge degrees of freedom.  
The (S)VBS states with different edge states have 
{\em finite} overlaps with each other,  
which are exponentially decreasing as the system size $L$. 
That is, two (S)VBS states with different edge states are orthogonal 
to each other {\em only} in the infinite-size limit.  

In constructing the SVBS state on a {\em periodic} chain, one  
has to treat the fermion sign carefully and one sees that 
the trace operation used in the standard MPS representation 
(\ref{eqn:MPS-PBC-1}) should be 
replaced with the {\em supertrace}: 
\begin{subequations}
\begin{equation}
|\text{SVBS-I}\rangle_{\text{periodic}}
= \text{STr}\left\{
\bigotimes_{i=1}^{L} \mathcal{A}_{i}
\right\} \; ,
\label{eqn:MPS-PBC-2}
\end{equation}
where the supertrace is defined as 
\begin{equation}
\text{STr}({\cal M}) \equiv {\cal M}_{11}+{\cal M}_{22}-{\cal M}_{33} \; .
\label{eqn:STr-1}
\end{equation}
\end{subequations}
From these $\mathcal{A}$-matrices, we obtain the following 9$\times$9 $T$-matrices 
({\em transfer matrix}):  
\begin{equation}
\begin{split}
T_{\bar{\alpha},\alpha;\bar{\beta},\beta} \equiv \mathcal{A}^{\ast}(\bar{\alpha},\bar{\beta})
\mathcal{A}(\alpha,\beta) 
&= 
\left(
\begin{array}{lllllllll}
 1 & 0 & 0 & 0 & 2 & 0 & 0 & 0 & r \\
 0 & -1 & 0 & 0 & 0 & 0 & 0 & 0 & 0 \\
 0 & 0 & 0 & 0 & 0 & 0 & 0 & -r & 0 \\
 0 & 0 & 0 & -1 & 0 & 0 & 0 & 0 & 0 \\
 2 & 0 & 0 & 0 & 1 & 0 & 0 & 0 & r \\
 0 & 0 & 0 & 0 & 0 & 0 & r & 0 & 0 \\
 0 & 0 & 0 & 0 & 0 & -r & 0 & 0 & 0 \\
 0 & 0 & r & 0 & 0 & 0 & 0 & 0 & 0 \\
 r & 0 & 0 & 0 & r & 0 & 0 & 0 & 0
\end{array}
\right) \\
& \quad  (\bar{\alpha},\alpha,\bar{\beta},\beta=1,2,3)  \; . 
\end{split}
\label{eqn:transfortypeI}
\end{equation}  
Here, $\mathcal{A}^{\ast}$ is obtained from $\mathcal{A}$ by 
$|\cdot\rangle \mapsto \langle \cdot|$ and complex conjugation. 

\begin{figure}[h]
\begin{center}
\includegraphics[scale=0.6]{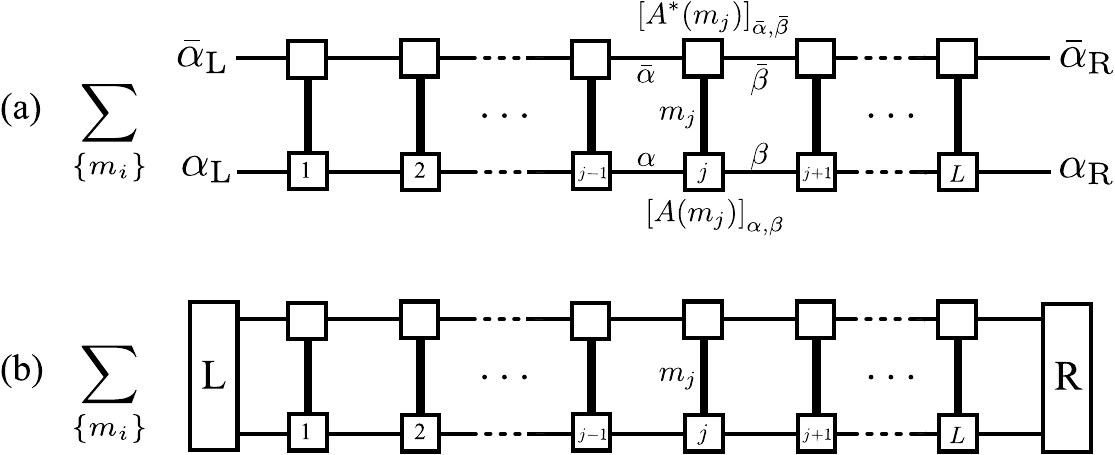}
\end{center}
\caption{(a) Diagrammatic representation of overlap of two (S)MPSs. 
(b) More general boundary conditions which are linear combinations of different 
$(\bar{\alpha}_{\text{L/R}},\alpha_{\text{L/R}})$s may be used. In those cases, 
edge states are expressed by $D^{2}$-dimensional vectors `L' and `R' 
(for instance, in the simplest case (a), the edge-state vector has the components 
$\delta_{\bar{\alpha},\bar{\alpha}_{\text{L}}}\delta_{\alpha,\alpha_{\text{L}}}$). 
\label{fig:MPS-norm}}
\end{figure}

Using the matrices $A(m)$ and the diagrammatic representation introduced in Sec.~\ref{sec:bosonic-MPS}, 
the transfer matrix may be expressed as
\begin{equation}
T_{\bar{\alpha},\alpha;\bar{\beta},\beta} \equiv \sum_{m=1}^{d}
\left[A^{\ast}(m)\right]_{\bar{\alpha},\bar{\beta}}
\left[A(m)\right]_{\alpha,\beta}=
\raisebox{-4.0ex}{\includegraphics[scale=0.6]{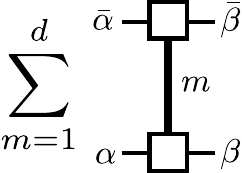}}
\end{equation}
The transfer matrix naturally appears in the calculations of the MPS formalism. 
For instance, by using the diagrammatic representation Fig.~\ref{fig:MPS-string-diag} 
and the orthogonality of 
the local basis states $\langle m|n\rangle_{i}=\delta_{mn}$, 
one can show that the overlap of the two (S)MPSs with edge states 
$\{\bar{\alpha}_{\text{L}},\bar{\alpha}_{\text{R}}\}$ and 
$\{\alpha_{\text{L}},\alpha_{\text{R}}\}$ can be written as\footnote{%
For the bosonic MPS, this is straightforward.  For the SMPS, one has to treat the fermion sign 
carefully but, in the end of the day, we can check that the final result is the same.}
(see Fig.~\ref{fig:MPS-norm}(a))
\begin{equation}
{}_{(\bar{\alpha}_{\text{L}},\bar{\alpha}_{\text{R}})}\VEV{\text{MPS}|
\text{MPS}}_{(\alpha_{\text{L}},\alpha_{\text{R}})} 
= \left[ T^{L} \right]_{\bar{\alpha}_{\text{L}},\alpha_{\text{L}};\bar{\alpha}_{\text{R}},
\alpha_{\text{R}}} \; ,
\label{eqn:MPS-overlap}
\end{equation}
where the matrix multiplication is taken over the tensor index $(\bar{\alpha},\alpha)$.  

\begin{figure}[h]
\begin{center}
\includegraphics[scale=0.7]{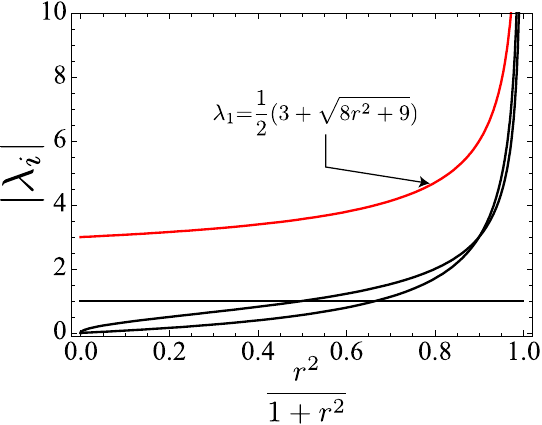}
\caption{(Color online) Plot of absolute values of the five 
different eigenvalues of $T$. The largest eigenvalue is 
always unique and non-degenerate (except for $r\rightarrow \infty$). 
(Figure and Caption are taken from Ref.\cite{Hasebe-Totsuka2011}.) 
\label{fig:eigenV-SUSY-AKLT1}}
\end{center}
\end{figure}

In the periodic case, the above expression is modified:
\begin{equation}
\VEV{\text{SVBS-I}|\text{SVBS-I}}_{\text{PBC}} 
= \sum_{\alpha,\beta}\text{sgn}(\alpha)\text{sgn}(\beta)
\left\{T^{L}\right\}_{(\alpha,\beta;\alpha,\beta)} \; ,
\label{eqn:norm-PBC-2}
\end{equation}
where 
\begin{equation} 
\text{sgn}(\alpha) =
\begin{cases}
1 & \text{for } \alpha=1,2 \\
-1 & \text{for } \alpha=3 \; .
\end{cases}
\end{equation}
Notice that the transfer matrix directly appears in the right-hand side of (\ref{eqn:norm-PBC-2}), and hence  
 the calculation is boiled down to that of the power of the transfer matrix.  
The eigenvalues of the transfer matrix \eqref{eqn:transfortypeI} are computed as  
\begin{equation}
\biggl\{-1(\times 3),~-i r(\times 2),~i r(\times 2),~ 
\frac{1}{2} \left(3-\sqrt{8 r^2+9}\right),~\frac{1}{2}
   \left(3+\sqrt{8 r^2+9}\right) \biggr\} \; , 
\end{equation}
and plotted in Fig. \ref{fig:eigenV-SUSY-AKLT1}. The largest eigenvalue of the transfer matrix,  $\frac{1}{2}\left(3+\sqrt{8 r^2+9} \right)$,  will be relevant in the thermodynamic limit. 

\subsubsection{$\mathcal{N}=2$ }

Similar to the type-I SVBS chain, we can express the type-II SVBS chain $(M=1)$ in the form of SMPS:  
\begin{equation}
|\text{SVBS-II}\rangle_{\alpha\beta}=(\mathcal{A}_1\mathcal{A}_2\cdots \mathcal{A}_L)_{\alpha\beta}
\end{equation}
where 
\begin{equation}
\begin{split}
\mathcal{A}_{i} =\mathcal{R}_{2|2}\Psi_i^*(r)\Psi_i(r)^{\dagger}\vac_i 
= 
\begin{pmatrix}
a^{\dagger}_{i}b^{\dagger}_{i} & (b^{\dagger}_{i})^{2} 
& \sqrt{r}b^{\dagger}_{i}f^{\dagger}_{i}
& \sqrt{r}b^{\dagger}_{i}g^{\dagger}_{i} \\
-(a^{\dagger}_{i})^{2} & -a^{\dagger}_{i}b^{\dagger}_{i} 
& -\sqrt{r}a^{\dagger}_{i}f^{\dagger}_{i}
& -\sqrt{r}a^{\dagger}_{i}g^{\dagger}_{i} \\
-\sqrt{r}f^{\dagger}_{i}a^{\dagger}_{i}
& -\sqrt{r}f^{\dagger}_{i}b^{\dagger}_{i} 
& 0 & -r f^{\dagger}_{i}g^{\dagger}_{i}\\
-\sqrt{r}g^{\dagger}_{i}a^{\dagger}_{i}
& -\sqrt{r}g^{\dagger}_{i}b^{\dagger}_{i} 
& r f^{\dagger}_{i}g^{\dagger}_{i} & 0
\end{pmatrix}
| \text{vac} \rangle_{i}.  
\end{split}
\end{equation}
As in the type-I SVBS state, the supertrace is necessary for the periodic 
system:
\begin{equation}
|\text{SVBS-II}\rangle = \text{STr}\left\{
\bigotimes_{i=1}^{L} \mathcal{A}_{i}
\right\} \; ,
\label{eqn:STr-2}
\end{equation}
where $\text{STr}({M}) \equiv {M}_{11}+{M}_{22}
-{M}_{33}-{M}_{44}$.   
The corresponding transfer matrix is a 16$\times$16 matrix 
and has seven different eigenvalues: 
\begin{equation} 
\bigl\{-1(\times 3),~-i r(\times 4),~+i r(\times 4),~
-r^2(\times 2),~r^2,~ \frac{1}{2} \left(r^2+3-f(r)\right),~
\frac{1}{2} \left(r^2+3+f(r)\right)\bigr\} \; ,
\end{equation}
where $f(r)\equiv \sqrt{r^4+10r^2+9}$.  
Regardless of the value of $r$, the largest eigenvalue  is 
$\frac{1}{2} \left(r^2+3+f(r)\right).$  
\subsection{Excitations}
In this section, we delve into dynamical properties, $i.e.$, low-lying excitations on the SVBS chain.  
In Sec.~\ref{sec:parent-Ham}, 
we have already obtained the parent Hamiltonian {\em from} the VBS state. 
Given the form of the Hamiltonian, we can, in principle, investigate dynamical properties of the system.  
Unfortunately, however, even though we have the exact ground state in hand, only limited 
(exact or rigorous) information about the excitations is available \cite{Knabe-88,Arovas-89}.  
Nevertheless, when the explicit form of the (whether exact or approximate) ground-state 
wave function is known, the {\em single-mode approximation} (SMA) gives reasonably 
good results \cite{Girvin-M-P-86,Arovasetal1988}: 
\begin{equation}
~~~~~~~~~~\text{VBS~ state}~~~ \xrightarrow{\text{Exact} }~~~\text{Parent ~Hamiltonian}~~~ 
\xrightarrow{\text{SMA}}~~~ \text{Excitation}. 
\end{equation}
The SMA not only provides us with a simple transparent way of calculating (approximate) 
excitation spectrum but also sets a rigorous upper bound for the true spectrum. 
\subsubsection{Fixing parent Hamiltonian}
Since the type-I Hamiltonian (\ref{explicitSUSYAKLT}) contains one extra parameter up to the overall factor, $i.e.$ the ratio of $V_{3/2}$ to $V_2$,  we begin with fixing the form of 
the parent Hamiltonian.   
One way to fix the remaining coupling is to require that the SUSY 
parent Hamiltonian (\ref{explicitSUSYAKLT}) should reduce 
to the original $SU(2)$ VBS Hamiltonian (\ref{basicparentharewritten}) in the limit $r\rightarrow 0$. 
This naturally fixes the two coupling 
constants in the type-I parent Hamiltonian (\ref{explicitSUSYAKLT}) as 
\begin{equation}
V_{3/2}=\tanh r \; , \;\; V_{2}=\sqrt{2} \; , 
\label{eqn:VBS-coupling}
\end{equation}
and we have 
\begin{equation}
\mathcal{H}=\sum_i \biggl\{ \tanh(r)\mathbb{P}^{\dagger}_{\frac{3}{2}}(i,i+1)\mathbb{P}_{\frac{3}{2}}(i,i+1)+\sqrt{2}\mathbb{P}^{\dagger}_2(i,i+1)\mathbb{P}_2(i,i+1)\biggr\}.
\label{fixedhamil}
\end{equation}
Some of the matrix elements 
in the fermionic sector have a factor $1/r$, and in the limit 
$r\rightarrow 0$ they are divergent. 
However, they are harmless in the limit, since the ground states contain no fermion 
in the $r\rightarrow 0$ limit. Consequently,   
 the type-I parent Hamiltonian  {\em projected onto 
the bosonic sector} coincides with the spin-1 VBS 
Hamiltonian (\ref{basicparentharewritten}). 

\subsubsection{Crackion Excitation }\label{subsec:crackion}
\begin{figure}[h]
\begin{center}
\includegraphics[scale=0.75]{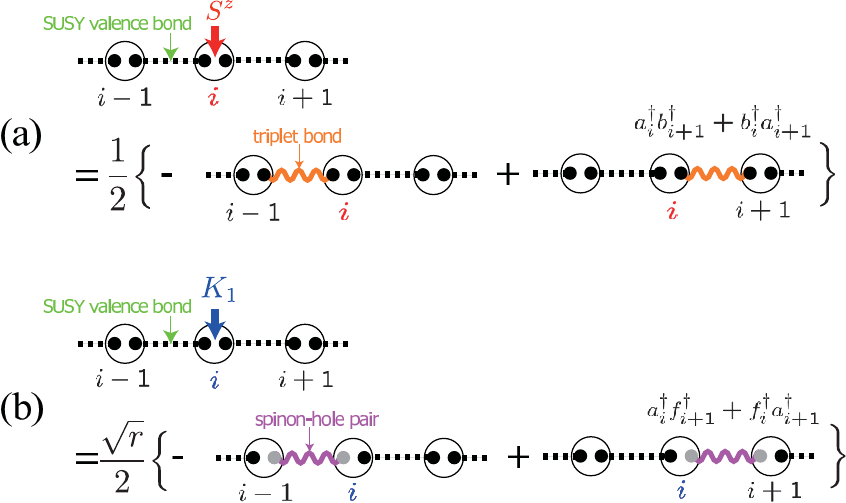}
\caption{(Color online) Action of bosonic (spin) operator $S^{z}$ (a) and 
fermionic generator $K_{1}$ (b) onto 
the SVBS state. The local operators $S^{a}(i)$ ($a=x,y,z$) 
and $K_{1,2}(i)$ respectively create 
a triplet bond and a spinon-hole pair  
on either of the two adjacent bonds $(i-1,i)$ and  
$(i,i+1)$. [Figure and Caption are taken from Ref.\cite{Hasebe-Totsuka2011}.] %
\label{fig:crackion}}
\end{center}
\end{figure}

Now we are ready to derive the excitation spectra by using SMA.  
The paradigmatic picture of the low-lying excitations in 
the half-odd-integer-spin chains is provided by the so-called 
Lieb-Schultz-Mattis construction \cite{LiebSchhultzMattis1961}, 
where we apply a slow twist along one of the symmetry axis (say, $z$-axis) of 
the spin Hamiltonian.  Physically, this boosts the quasi-particles in the system and thereby 
creates low-lying excitations [with energies of the order of $\sim (\text{chain length})^{-1}$] at 
a special momentum determined solely by the total magnetization.  
Unfortunately, this construction does not work 
in the usual VBS state \cite{Totsuka-Suzuki1995}. 
Instead, as we will see, an excited triplet bond 
({\em crackion} \cite{FathSolyom1993}, {\em i.e.}, a ``crack'' created in a ``solid'' of valence bonds;  
See Fig.~\ref{fig:crackion}) 
in the VBS gives, to good approximation, 
a physical low-lying excitation.   Due to the simple structure of the VBS states, 
the excitations considered in the SMA essentially coincide with the crackions 
\cite{FathSolyom1993,Totsuka-Suzuki1995}.  
In the Schwinger-boson construction of the VBS states \eqref{generalsu2vbs}, 
the crackion excitation is obtained by replacing one of the singlet valence bonds 
$(a_{i}^{\dagger}b_{j}^{\dagger}-b_{i}^{\dagger}a_{j}^{\dagger})$ by a triplet one 
[either $a^{\dagger}_{i}a^{\dagger}_{j}$ or 
$(a_{i}^{\dagger}b_{j}^{\dagger} + b_{i}^{\dagger}a_{j}^{\dagger})$ or 
$b^{\dagger}_{i}b^{\dagger}_{j}$].  Since a single Schwinger boson $a^{\dagger}$ or 
$b^{\dagger}$ describes a spin-1/2 spinon, we may thought of the triplet crackion 
as the {\em confined} triplet pair of two spinons.  
The notion of crackion excitations can be generalized in other VBS-type models (states) 
with higher symmetries [{\em e.g.}, $SU(N)$], where an intriguing picture on the relation 
between spinon confinement and the existence of `Haldane gaps' has been proposed 
\cite{Greiteretal2007II,Rachel-spinon-2009}

Now let us consider the crackions in the SVBS chains \cite{Hasebe-Totsuka2011}.    
Since we are dealing with a SUSY system, 
we consider two different types of excitations (see Fig.~\ref{fig:crackion}) 
that may be regarded as super-partners of each other: 
\begin{itemize}
\item Spin excitation: \\~~~~~~Spin triplet excitation  ($S=1$) created by $UOSp(1|2)$ bosonic operators
\item Spinon-hole excitation: \\~~~~~~Spin doublet excitation ($S=1/2$) 
paired with a hole created by $UOSp(1|2)$ fermionic operators 
\end{itemize}
They are schematically represented in Fig.~\ref{fig:crackion}. 
In the SMA, the (unnormalized) excited-state wave function is assumed to be given 
(for the spin excitation) by 
\begin{equation}
| k,a \rangle= S^a(k)|\text{SVBS}\rangle \; , 
\label{eqn:Bijl-Feynman}
\end{equation}
where $S^a(k)$ denotes the Fourier transform of the local spin operator $S^a_j$.  
Then, the excitation spectrum (or, the Bijl-Feynman frequency \cite{Bijl1940,Feynman1954}) 
is obtained by calculating the following quantity: 
\begin{equation}
\omega^{a}_{\text{SMA}}(k) =\frac{\langle k,a | \mathcal{H}| k,a\rangle}{\langle k,a | k,a\rangle} -E_{0}
=\frac{\langle k,a | (\mathcal{H}-E_{0})| k,a\rangle}{\langle k,a | k,a\rangle} \; , 
\label{crackionenergyformula} 
\end{equation}
where $\mathcal{H}$ is given by (\ref{fixedhamil}) and $E_{0}$ is the ground-state energy.  We can consider other types of excitations 
by changing $S^a(k)$ to other operators.  
\subsubsection{Spin Excitation}
\begin{figure}[h]
\begin{center}
\includegraphics[scale=0.6]{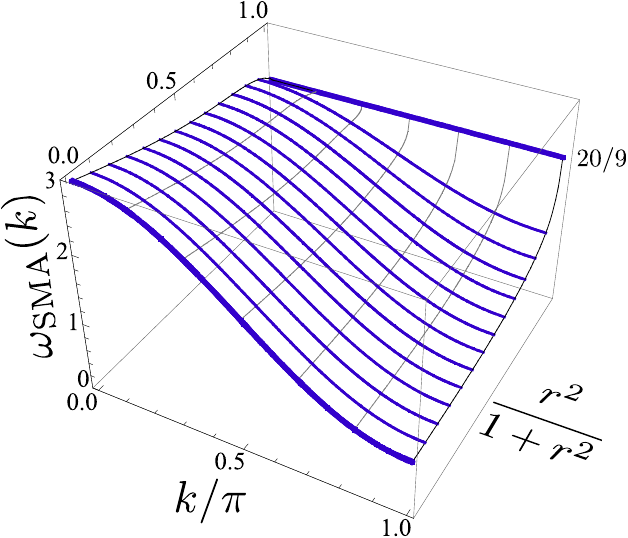}
\caption{(Color online) The spin excitation (triplon) 
spectrum $\omega^{\text{s}}_{\text{SMA}}(k)$ obtained by SMA.  
At $r=0$, it reduces to the well-known 
dispersion $\omega_{\text{SMA}}(k)= 10(5+3 \cos k)/27$ of the spin-1 VBS model. 
When $r\nearrow \infty$ (Majumdar-Ghosh limit), 
dispersion becomes flat. [Figure and Caption are taken from Ref.\cite{Hasebe-Totsuka2011}.] 
\label{fig:SMA-spectrum}}
\end{center}
\end{figure}

Let us start by investigating the action of local spin operators 
\begin{equation}
S^{+}(i)=a^{\dagger}_{i}b_{i} \, , \; 
S^{-}(i)=b^{\dagger}_{i}a_{i} \, , \; 
S^{z}(i)=\frac{1}{2}(a^{\dagger}_{i}a_{i}-b^{\dagger}_{i}b_{i}) 
\end{equation}
on the SVBS state.  A little algebra shows that these spin operators 
create triplet bonds around the site $i$ [see Fig.~\ref{fig:crackion}]:
\begin{subequations}
\begin{align}
& S^{+}_{i}|\text{SVBS-I}\rangle 
= |\psi^{(1)}_{i-1}\rangle -|\psi^{(1)}_{i}\rangle,  
\label{eqn:transverse-crackion}  \\
& S^{z}_{i}|\text{SVBS-I}\rangle 
= \frac{1}{2}\left\{
- |\psi^{(0)}_{i-1}\rangle + |\psi^{(0)}_{i}\rangle 
\right\} \; ,
\label{eqn:z-crackion}
\end{align}
\end{subequations}
where $|\psi^{(1)}_{i}\rangle$ and $|\psi^{(0)}_{i}\rangle$ are 
obtained by replacing the SUSY valence bond 
$(a^{\dagger}_{i}a^{\dagger}_{i+1}-b^{\dagger}_{i}b^{\dagger}_{i+1}
-r\, f^{\dagger}_{i}f^{\dagger}_{i+1})$ by triplet bonds 
$a^{\dagger}_{i}a^{\dagger}_{i+1}$ and 
$(a^{\dagger}_{i}b^{\dagger}_{i+1}+b^{\dagger}_{i}a^{\dagger}_{i+1})$,  
respectively.  
By taking the Fourier transform of eqs.\eqref{eqn:transverse-crackion} and 
\eqref{eqn:z-crackion}, one immediately sees that 
the triplon-crackion equivalence (except for the momentum-dependent form factor) 
found in the ordinary VBS states 
\cite{FathSolyom1993,Totsuka-Suzuki1995} holds in the SVBS case as well.   
By a simple algebra, it is easy to show the following bound  
for the true spin-excitation spectrum $\omega_{\text{true}}^{\text{s},a}(k)$:
\begin{equation}
\begin{split}
\omega_{\text{SMA}}^{\text{s},a}(k) &=  \frac{\langle\text{SVBS-I}| S^{a}(k)({\cal H}-E_{0})
S^{a}(-k)|\text{SVBS-I}\rangle}{\langle\text{SVBS-I}|S^{a}(k)
S^{a}(-k)|\text{SVBS-I}\rangle} \\
& = \frac{1}{2}
\frac{\langle\text{SVBS-I}|\, [S^{a}(-k), [{\cal H}, S^{a}(k)]]
\, |\text{SVBS-I}\rangle}{\langle\text{SVBS-I}|S^{a}(k)
S^{a}(-k)|\text{SVBS-I}\rangle} \\
& \geq \omega_{\text{true}}^{\text{s},a}(k) \; .
\end{split}
\label{eqn:SMA-formula-1}
\end{equation}
The last inequality is proven by noting that the left-hand side can be rewritten as 
the following average
\begin{equation}
\frac{%
\int_{0}^{\infty}\!d \omega \, \omega S^{aa}(k,\omega) }{%
\int_{0}^{\infty}\!d \omega S^{aa}(k,\omega)}
\end{equation}
and using the spectral decomposition of the dynamical structure factor $S^{aa}(k,\omega)$. 
The spin-excitation spectrum obtained \cite{Hasebe-Totsuka2011} in this way is shown 
in Fig.~\ref{fig:SMA-spectrum}.  
At $r=0$, the dispersion reduces to the well-known 
result of the original VBS chain \cite{Arovasetal1988}: 
\begin{equation}
\omega_{\text{SMA}}^{\text{s},a}(k) = 
\frac{10}{27} (5 + 3 \cos k)  \;\; (a=x,y,z) \; .
\label{SMAspnexcite}
\end{equation}
In the limit $r\nearrow \infty$, on the other hand, the spin excitation loses 
its dispersion.  This is easily understood by noticing that the ground-state reduces 
to the  Majumdar-Ghosh dimer states on which excitations cannot move.  

\subsubsection{Spinon-Hole Excitations}
The dynamics of doped holes in the spin-gapped background is in its own right 
interesting \cite{Xu-Y2BaNiO5-00}.  In the context of the VBS models, 
some (both exact and approximate) results have been 
obtained.  For instance, in Ref.~\cite{ZhangArovas1989}, 
the motion of spin-0 holes in the spin-1 VBS background is considered and the exact one-hole 
spectrum is obtained.  Motivated by the experiments carried out for hole-doped spin-1 
compound, 
Ref.~\cite{PencShibaPRB1995} introduced a realistic model and investigated the motion of 
a single spin-1/2 hole immersed in the gapped spin-1 VBS background.     

\begin{figure}[h]
\begin{center}
\includegraphics[scale=0.6]{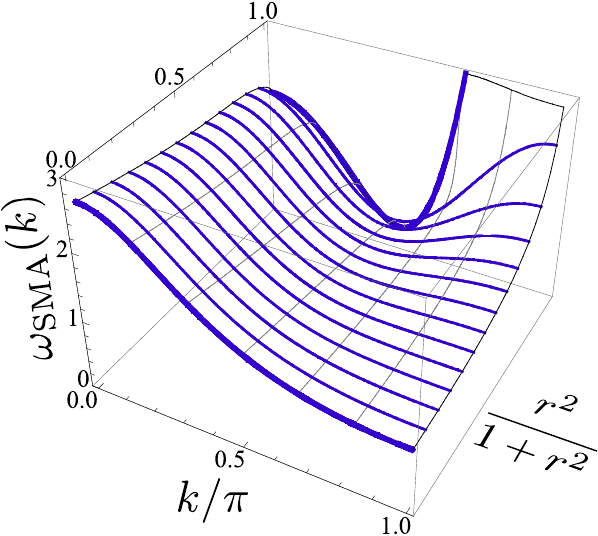}
\caption{(Color online) The excitation spectrum $\omega^{\text{h}}_{\text{SMA}}(k)$ of 
a spinon-hole pair obtained by SMA.   
This spinon-hole pair state is created by fermionic generator 
$K_{1}$ except at $r=0$, where the transition matrix elements of $K_{1}$ 
from the ground state vanish. [Figure and Caption are taken from Ref.\cite{Hasebe-Totsuka2011}.] 
\label{fig:SMA-spectrum-hole}}
\end{center}
\end{figure}

A similar strategy can be used to obtain the spectrum of the charged (hole, $f^{\dagger}$) excitation  
which is always paired with the $S=1/2$ spinon ($a^{\dagger}$ or $b^{\dagger}$). 
These excitations are created by applying the two fermionic generators 
of $UOSp(1|2)$
\begin{equation}
\begin{split}
& K_{1}(i)= \frac{1}{2}(\frac{1}{\sqrt{r}}f_{i}a_{i}^{\dagger}+\sqrt{r} f_{i}^{\dagger}b_{i}) \; , \\
& K_{2}(i)= \frac{1}{2}(\frac{1}{\sqrt{r}}f_{i}b_{i}^{\dagger}-\sqrt{r}
 f_{i}^{\dagger}a_{i}), \;\;
\end{split}
\end{equation}
to the VBS ground state. 
By using the explicit form of the ground-state wave function, 
it is easy to show 
\begin{equation}
K_{1}(i)|\text{SVBS-I}\rangle 
= \frac{\sqrt{r}}{2}\left\{
|\psi_{i-1}^{(1/2)}\rangle - |\psi_{i}^{(1/2)}\rangle 
\right\} \; ,
\end{equation}
where the crackion state 
$|\psi_{i}^{(1/2)}\rangle$ is obtained by replacing 
the SUSY valence bond 
$(a_{i}^{\dagger}b_{i+1}^{\dagger}-b_{i}^{\dagger}a_{i+1}^{\dagger}
-r f_{i}^{\dagger}f_{i+1}^{\dagger})$  with 
a spinon-hole pair 
$(a_{i}^{\dagger}f_{i+1}^{\dagger}+f_{i}^{\dagger}a_{i+1}^{\dagger})$ 
[see Fig.~\ref{fig:crackion}(b)]. 
The excited state $K_{2}|\text{SVBS}\rangle$ is defined similarly 
with $a^{\dagger}$ in the above expression replaced with 
$b^{\dagger}$. 
For $r=0$, the spectrum is given by 
\begin{equation}
\omega_{\text{SMA}}^{\text{h}}(k)
= \frac{8}{3(2-\cos k)} \; .
\end{equation}
The behavior of the spectrum as a function of $r$ \cite{Hasebe-Totsuka2011} is plotted in 
Fig.~\ref{fig:SMA-spectrum-hole}. 

We add some comments about distinctions between   
$\omega^{\text{s}}_{\text{SMA}}(k)$ and $\omega^{\text{h}}_{\text{SMA}}(k)$.  
Since SUSY relates the bosonic generators $\bold{S}$ and the two fermionic generators $K_{\alpha}$, 
one might naively expect the same spectra for their corresponding excitations.  
However, this expectation relies 
on the existence of a `unitary' transformation which linearly transforms 
the set of the SUSY generators onto themselves. 
Since no such transformation exists in the present SUSY, the spectra 
for the spin and charge sectors indeed exhibit different behaviors.

\section{Topological Order}
\label{Sectopological}
As has been discussed in Sec.~\ref{sec:introduction}, no true topological order 
is possible in 1D systems \cite{Chenetal2011}.  
However, if we impose a certain kind of symmetries, 
there can be topologically non-trivial phases protected by the symmetries dubbed 
{\em symmetry-protected topological phases}.  
One of the typical examples would be a non-trivial topological phase with the Majorana edge mode  
in 1D interacting fermions \cite{Kitaev-01}.  We already know that  
there exists an analogous ``topological'' phase with (almost) free edge spins ($S=1/2$) at 
the edges in 1D spin systems as well.   
 In this section, we give detailed discussions about topological properties of the SVBS states 
 \cite{Hasebe-Totsuka2011, Hasebe-Totsuka2012}.  
 In particular, we investigate the string order and the entanglement spectrum 
 of type-I and type-II VBS states. 
Then we generalize the MPS argument of the symmetry-protected topological order 
\cite{Pollmannetal2010,Pollmannetal2012} to SUSY cases 
to understand the degeneracy structure in the entanglement spectrum.    

\subsection{Hidden Antiferromagnetic Order and String Order Parameter}
\label{sec:string-ordinary-VBS}
Before proceeding to the SUSY cases, we briefly recapitulate the hidden non-local order 
in 1D spin systems.  
The concept of hidden order is an {\em isotropic} generalization of the N\'{e}el order. 
As we have seen in Sec.~\ref{sec:VBS-introduction}, the N\'{e}el order 
for $S=1$ antiferromagnetic spin chains looks like (if we assume that the ordering occurs 
in the $z$-axis) 
\begin{equation}
\cdots ~ + ~ - ~ + ~ - ~ + ~ - ~ + ~ - ~ + ~  \cdots .
\end{equation}
Here, $+$ stands for $S^z=+1$, and $-$ for $S^z=-1$.  
Clearly, the spin-spin correlation
\begin{equation}
\langle S^{z}_{i}S^{z}_{j} \rangle
\end{equation}
depends on the parity of the number $n$ of sites between $i$ and $j$.  
A simple way of turning this position-dependent (or, alternating) correlation 
into the smooth ferromagnetic one would be to insert a phase factor $(-1)^{n}$ 
between the two spins
\begin{equation}
\langle S^{z}_{j}S^{z}_{j+n} \rangle \mapsto 
\langle S^{z}_{j}(-1)^{n}S^{z}_{j+n} \rangle 
\label{eqn:Neel-string}
\end{equation}
in order to cancel the sign factor coming from the alternating $+1$ and $-1$ between 
the sites $j$ and $j+n$.  Namely, the N\'{e}el antiferromagnetic correlation translates to 
the ferromagnetic correlation in $\langle S^{z}_{j}(-1)^{n}S^{z}_{j+n} \rangle$.  

On the other hand, as is seen in the expansion of the VBS state [see (\ref{expVBSeq})], 
a typical  $S^z$ sequence appearing in the state reads as    
\begin{equation}
\cdots  ~ + ~ - ~ + ~ 0 ~ - ~ + ~ - ~  0 ~ 0 ~ + ~ - ~ 0 ~ + ~\cdots .
\end{equation}
As has been pointed out already in Sec.~\ref{sec:VBS-introduction}, 
by removing zeros in the sequence, we can reproduce the usual 
N\'{e}el order.  In this sense, there still exists a certain kind of N\'{e}el order, 
though ``disordered'' by randomly inserted 0s, called 
the hidden string order\footnote{%
However, there is a striking difference from the usual N\'{e}el order.  
As has been mentioned in Sec.~\ref{sec:VBS-introduction}, 
the string order exists regardless of the choice of the quantization axis, while 
the N\'{e}el AF order is observed only in a particular direction.} \cite{NijsRommelse-1989,Tasaki1991}.  
However, the trick used above does not work since, due to the intervening 0s, 
the positions of $+1$ and $-1$ are random (though they still appear in an alternating way) and 
the phase $(-1)^{n}$ can not cancel the sign factor.  
Nevertheless, a little thought tells that the following choice will do the job:
\begin{equation}
\exp\left( i \pi S^{z}_{\text{tot}}(j,j+n) \right) 
= (-1)^{\text{\# of }\pm 1\text{ between $j$ and $j+n$}} \; ,
\end{equation}
where 
\begin{equation}
S^{z}_{\text{tot}}(i,j) \equiv 
\sum_{k=i}^{j}S^{z}_k  
\label{sumspintoi}
\end{equation}
stands for the partial sum of $S^{z}_k$ between an arbitrary pair of sites $i$ and $j$.  
Therefore, it is suggested that we should use, instead of eq.\eqref{eqn:Neel-string}, 
the following pair of {\em non-local} order parameters 
\begin{equation}
\begin{split}
{\cal O}_{\text{string}}^{x,\infty}
\equiv     
\lim_{n\nearrow \infty}
\Biggl\langle S^{x}_{j} \, 
\exp\{
i\pi S^{x}_{\text{tot}}(j+1,j+n)\} S^{x}_{j+n} \Biggr\rangle  \\
{\cal O}_{\text{string}}^{z,\infty}
\equiv     \lim_{n\nearrow \infty}
\Biggl\langle S^{z}_{j} \, 
\exp\{
i\pi S^{z}_{\text{tot}}(j,j+n-1)\} S^{z}_{j+n} \Biggr\rangle
\end{split}
\label{stringorderquan}
\end{equation}
known as the string order parameters \cite{NijsRommelse-1989,Tasaki1991}
in order to characterize the non-trivial spin order in the ($S=1$) VBS state.   
In the first line, $S^{x}_{\text{tot}}(j+1,j+n)$ is defined similarly to $S^{z}_{\text{tot}}(j,j+n-1)$ 
[see eq.\eqref{sumspintoi}].  

The above expressions have been guided by a simple physical intuition. 
However, as has been pointed out in Refs.~\cite{KennedyTasaki1992I,KennedyTasaki1992II}, 
the string order parameters have in fact a deeper meaning than we expect from the above 
simple argument.  
To see this, first we note that the string correlation functions can be recasted in a suggestive 
way:
\begin{equation}
\begin{split}
{\cal O}_{\text{string}}^{x}(j,j+n)
& \equiv
\Biggl\langle S^{x}_{j} \, 
\exp\{
i\pi S^{x}_{\text{tot}}(j+1,j+n)\} S^{x}_{j+n} \Biggr\rangle \\
&= \left\langle \widetilde{S}^{x}_{j} \widetilde{S}^{x}_{j+n} \right\rangle \\
{\cal O}_{\text{string}}^{z}(j,j+n)
& \equiv     
\Biggl\langle S^{z}_{j} \, 
\exp\{i\pi S^{z}_{\text{tot}}(j,j+n-1)\} 
S^{z}_{j+n} \Biggr\rangle  \\
&= \left\langle \widetilde{S}^{z}_{j} \widetilde{S}^{z}_{j+n} \right\rangle  \; ,
\end{split}
\end{equation}
where $\widetilde{S}^{x}$ and $\widetilde{S}^{z}$ are defined as
\begin{equation}
\widetilde{S}^{x}_{j} \equiv S^{x}_{j} \exp\{i\pi S^{x}_{\text{tot}}(j+1,L)\}  \; , \quad 
\widetilde{S}^{z}_{j} \equiv \exp\{i\pi S^{z}_{\text{tot}}(1,j-1)\} S^{z}_{j}  \; .
\end{equation}
The physical meaning of these operators may be best understood by considering 
the `height plot' of the VBS spin configurations \cite{NijsRommelse-1989,Totsuka-Suzuki1995},   
where the value $S^{z}_{\text{tot}}(1,j)$ is represented by the ``height'' between the sites 
$j$ and $j+1$ (hence $S^{z}_{j}$ itself is a ``step'' at the site $j$).  
Fig.~\ref{fig:step-VBS-1}(a) is a plot of a typical $S^{z}$-configuration of the spin-1 VBS state 
when the left edge state is $\uparrow$.  One can clearly see that the meandering steps 
are always confined between the heights 0 and $+1$.   A similar analysis in the case with 
the left edge state $\downarrow$ shows that the heights are either 0 or $-1$.  
From these observations, it is concluded that the string 
$\exp\{i\pi S^{z}_{\text{tot}}(1,j-1)\}$ attached to the left of the operator 
$S^{z}_{j}$ somehow suppresses the strong fluctuations in $S^{z}$ and that 
$\widetilde{S}^{z}$ takes either 0 or $+1$ (0 or $-1$) when the left edge state is $\uparrow$ 
($\downarrow$).   Similarly, one can show that $\widetilde{S}^{x}$ becomes weakly ferromagnetic 
depending on the {\em right} edge states.  
In short, non-zero string order parameters translate to the existence of a certain kind of 
weakly ferromagnetic order in the $x$ and the $z$ directions 
(i.e. $\langle \widetilde{S}^{a}\rangle \neq 0$ for $a=x,z$).  

Remarkably, the following unitary transformation \cite{Oshikawa1992} 
\begin{equation}
U_{\text{KT}} = \exp\left\{
i\pi \sum_{k<j}S^{z}_{k}S^{x}_{j} \right\} 
=\prod_{k<j} \exp\{ i\pi S_{k}^{z}S^{x}_{j} \} \; ,
\end{equation}
relating the two operators $S^{a}$ and $\widetilde{S}^{a}$ as
\begin{equation}
\widetilde{S}_{j}^{x} = U_{\text{KT}} \, S^{x}_{j} \, U_{\text{KT}}^{-1} \; , \;\;
\widetilde{S}_{j}^{z} = U_{\text{KT}} \, S^{z}_{j} \, U_{\text{KT}}^{-1} \; ,
\end{equation}
transforms the original [SU(2)-invariant] Hamiltonian into the one 
$U_{\text{KT}} \, \mathcal{H} \, U_{\text{KT}}^{-1}$ which is invariant only under 
the dihedral group $D_{2}$ (or $\mathbb{Z}_{2}{\times}\mathbb{Z}_{2}$) 
consisting of two $\pi$ rotations with respect to the $x$ and the $z$ axes 
\cite{KennedyTasaki1992I,KennedyTasaki1992II}.  
Therefore, we can interpret the existence of the string order (both in $x$ and $z$) 
in the original system 
as a consequence of the spontaneous breakdown of the $\mathbb{Z}_{2}{\times}\mathbb{Z}_{2}$-symmetry 
in the {\em transformed} system $U_{\text{KT}} \, \mathcal{H} \, U_{\text{KT}}^{-1}$ and 
the resulting (weak) ferromagnetic order in $\widetilde{S}^{x,z}$.  

The idea of non-local hidden order and edge states has been to some extent 
generalized \cite{Hatsugai1992,Oshikawa1992,Totsuka-Suzuki1995,Schollwock}  
to other values of integer-spin-$S$ although the hidden $\mathbb{Z}_2{\times} \mathbb{Z}_2$-symmetry is never broken \cite{Oshikawa1992} in the case of even-$S$ 
(see Fig.~\ref{fig:String-Order-S=even,odd})\footnote{%
This does not mean that $\mathbb{Z}_2{\times} \mathbb{Z}_2$-symmetry never breaks down 
in {\em any} even-$S$ chains.  In fact, even when $S=$even, 
one can construct the model ground states which have non-vanishing string order parameters.}. 
In the course of these studies, it has been recognized that there are some differences 
\cite{Oshikawa1992,Totsuka-Suzuki1995} in the ground-state properties according to the 
parity of $S$.  
Nevertheless, by analogy with the quantum-Hall systems \cite{Girvin-Arovas1989}, 
the ground state of generic integer-spin antiferromagnetic chains, 
including the original VBS state and its higher-spin 
generalizations \cite{Arovasetal1988}, characterized by certain kinds of 
non-local correlations and emergent edge states have been called `topological' 
in a rough sense.  

To illustrate distinct behaviors depending on the parity of spin $S$, 
we may introduce the following generalized (angle-dependent) string order 
parameter \cite{Oshikawa1992}:  
\begin{equation}
{\cal O}_{\text{string}}^{z,\infty}(\theta)
\equiv \lim_{n\nearrow \infty} 
\Biggl\langle S^{z}_{j} \, 
\exp\{
i\theta S^{z}_{\text{tot}}(j,j+n-1)\} S^{z}_{j+n} \Biggr\rangle \; , 
\label{stringtopologicalorder}
\end{equation}
where the parameter $\theta$ has been introduced for convenience\footnote{%
The introduction of the $\theta$ parameter is mainly motivated by the idea that the intermediate 
string might somehow cancel the fluctuations between the two spins $S^{z}(i)$ and $S^{z}(i+n)$. 
For $\theta=\pi$, this works {\em perfectly} in the $S=1$ VBS state.   
Except for $\theta=0$ (ordinary spin-spin correlation) and $\theta=\pi$ (string correlation), 
no symmetry-related reason has been found so far.}.  
When $\theta=0$, it reduces to the usual spin-spin correlation and, when $\theta=\pi$, 
it coincides with the string order parameter discussed above.  
The behaviors of the generalized string order parameter \eqref{stringtopologicalorder} 
are shown for several values of bulk spin $S$ in Fig.~\ref{fig:String-Order-S=even,odd} 
\cite{Totsuka-Suzuki1995}.  
The string order parameters are symmetric with respect to $\theta=\pi$ and generally 
have $S$ peaks for the spin-$S$ VBS state.  
As is expected from that the ground state is magnetically disordered, 
the infinite-distance limit of the usual spin-spin 
correlation function ($\theta=0$) vanishes regardless of the value of $S$.  
On the other hand, Fig. \ref{fig:String-Order-S=even,odd} demonstrates  
that  ${\cal O}_{\text{string}}^{z,\infty}$ takes finite values for the odd-$S$ VBS states 
while it vanishes for even-$S$.    
Therefore, in the sense of the $\mathbb{Z}_{2}{\times}\mathbb{Z}_{2}$-symmetry 
argument \cite{KennedyTasaki1992I,KennedyTasaki1992II} mentioned above, 
this hidden symmetry is never broken \cite{Oshikawa1992} in the even-$S$ VBS models. 
\begin{figure}[h]
\begin{center}
\includegraphics[scale=0.9]{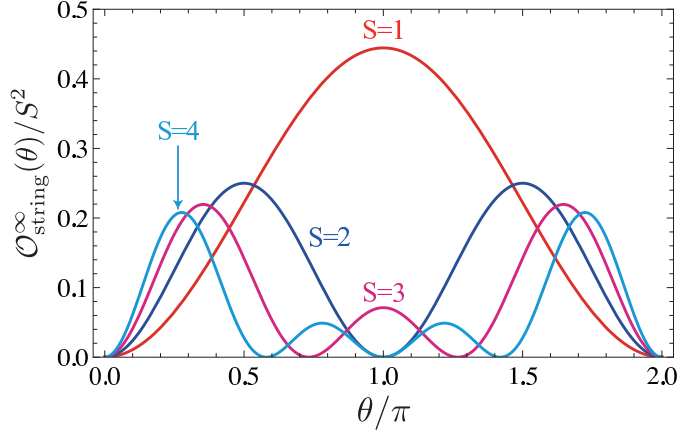}
\caption{(Color online) The behaviors of string order parameters in correspondence 
with magnitude of bulk spins.  In particular, at $\theta=\pi$, the string order parameters 
of even spin VBS $(S=2,4)$ vanish, while those of odd spin VBS $(S=1,3)$ have  finite values. 
\label{fig:String-Order-S=even,odd}}
\end{center}
\end{figure}

\subsection{Generalized Hidden String Order in SVBS Chain}
\label{sec:stringOP-in-SVBS}

\subsubsection{$\mathcal{N}=1$ }\label{subsc:hiddenorder1}
\begin{figure}[h]
\begin{center}
\includegraphics[scale=0.6]{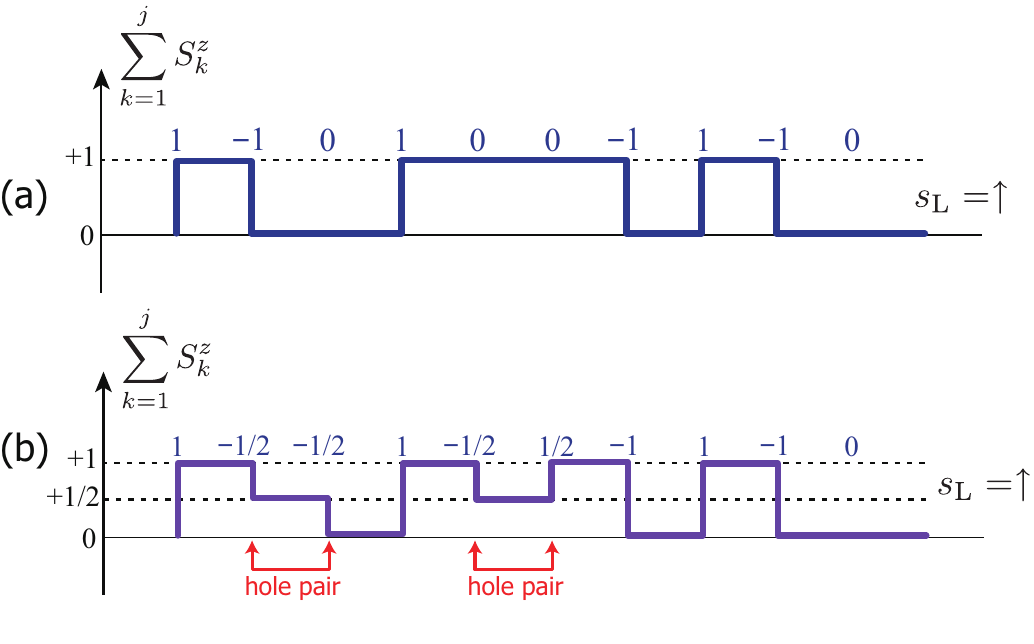}
\caption{(Color online) Height plot of typical spin configurations 
in $S=1$ VBS chain (a) and $\mathcal{S}=1$ SVBS chain (b).  
Note that heights are confined within a region of width 1.  
Although a simple ``diluted'' N\'{e}el picture does not hold 
because of the presence of hole pairs, still we can find 
{\em string order} when hole pairs are grouped together in (b). 
(Figure and Caption are taken from Ref.\cite{Hasebe-Totsuka2011}.) 
\label{fig:step-VBS-1}}
\end{center}
\end{figure}

\begin{figure}[H]
\begin{center}
\includegraphics[scale=0.6]{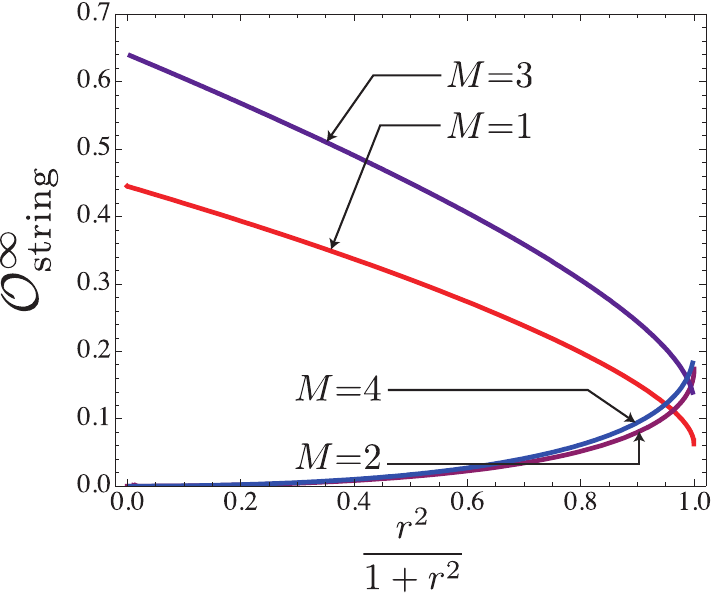}
\caption{(Color online) The string correlation function 
${\cal O}^{\infty}_{\text{string}}$ (\ref{stringorderquan})  of the SVBS infinite-chain 
for several values of the superspin $\mathcal{S}=M$ is plotted as a function of $r$.  
Notice that, in the limit $r\rightarrow 0$, the string order parameter ${\cal O}^{\infty}_{\text{string}}$ 
for the $\mathcal{S}=M$  SVBS chain reproduces that of 
the $S=M$ VBS chain. 
(Figure and Caption are taken from Ref.\cite{Hasebe-Totsuka2011}.)
\label{fig:Ostring-model1}}
\end{center}
\end{figure}

Unlike the original VBS chains,  $S^z=1/2$ 
and $-1/2$ generally appear in the $S^z$ sequence of the SVBS chain, and a     
 typical $S^z$ sequence of the SVBS chains is given by\footnote{%
It may be worthwhile to give some comments on the relation 
 to  the ferrimagnetic spin chains that also consist of {\em alternating} spin 1 
 and spin 1/2 \cite{Kolezhuketal1997}.  
Though both the ferrimagnetic chains and the present SVBS chains contain 
spin-1 and spin-1/2 degrees of freedom, 
in the SVBS chains the spin 1 and spin 1/2 are not necessarily alternating 
[see (\ref{sequenceofszsusyori}) and Fig.~\ref{expSUSY.fig}].  
More importantly, while ferrimagnetic spin chain can exhibit a long-range magnetic order as 
the order parameter commutes with the Hamiltonian, 
the ground state of the SVBS chain itself is spin-singlet and the $SU(2)$-symmetry 
is never broken (spontaneously).}  
\begin{equation}
\cdots  ~ 0 ~ \underbrace{\uparrow  ~ \uparrow} ~ 0 ~ 0 ~ 
\underbrace{\downarrow ~ \downarrow} ~ 
+ ~ - ~  0 ~ 0 ~ \underbrace{\uparrow ~ \downarrow} ~ 
+ ~ \underbrace{\downarrow ~  \uparrow} ~ 
\underbrace{\downarrow ~ \downarrow} ~ 0 ~\cdots  .
\label{sequenceofszsusyori}
\end{equation}
From the sequence, one can ``derive'' the ordinary hidden oder.  
First, we search the spin-half sites from the left, and whenever we encounter a pair of 
spin-half sites we sum the two $S^{z}$ values to have the effective $S^{z}(=+,0,-)$ (e.g. $\downarrow \; \downarrow \, \mapsto - $):
\begin{equation}
\cdots  ~ 0 ~ + ~ 0 ~ 0 ~ - ~ 
+ ~ - ~  0 ~ 0 ~ 0 ~ 
+ ~ 0 ~ - ~ 0 ~\cdots .  
\label{sequenceofszsusy}
\end{equation}
Next we remove the zeros in the sequence to recover the standard N\'{e}el pattern:     
\begin{equation}
\cdots ~ + ~ - ~ + ~ -  ~ + ~ -  ~\cdots .
\end{equation}
This observation leads to the existence of (generalized) hidden order 
of the SVBS chain. 
Fig.~\ref{fig:step-VBS-1} shows a typical 
height configuration corresponding to the usual VBS chain 
and its SUSY counterpart.  
Reflecting the existence of  hidden order, the height configuration 
is always meandering between the height 0 and the height 1 \footnote{ 
The same reasoning applies to the general spin-$S$ VBS cases and  the height configurations are 
confined within a region of width $S$ \cite{Totsuka-Suzuki1995}.}.    
It should also be noted that the height-configuration  (\ref{sumspintoi}) is directly reflected in the components of the SMPS (\ref{eqn:MPS-edge}): 
\begin{equation}
\begin{pmatrix}
|{\cal S}^{z}_{\text{tot}}(i){=}0\rangle 
& |{\cal S}^{z}_{\text{tot}}(i){=}-1\rangle &
|{\cal S}^{z}_{\text{tot}}(i){=}-1/2\rangle \\
|{\cal S}^{z}_{\text{tot}}(i){=} 1\rangle 
& |{\cal S}^{z}_{\text{tot}}(i){=}0\rangle &
|{\cal S}^{z}_{\text{tot}}(i){=} 1/2\rangle \\
|{\cal S}^{z}_{\text{tot}}(i){=} 1/2\rangle 
& |{\cal S}^{z}_{\text{tot}}(i){=}-1/2\rangle &
|{\cal S}^{z}_{\text{tot}}(i){=}0\rangle 
\end{pmatrix}.
\label{eqn:left-subsys}
\end{equation}
To substantiate the existence of the hidden order,  
we explicitly calculate the string parameter for the SVBS chains.  The behaviors of the string order are depicted in Fig.~\ref{fig:Ostring-model1} with respect to the hole doping parameter.  
From Fig.~\ref{fig:Ostring-model1}, one can find distinct behaviors of the string parameter in terms of the parity of bulk superspin. 
The string parameters of odd superspin SVBS chains generally decrease with increase of the hole doping, while those of the even superspin SVBS chains increase.   Since the hole-doping simply reduces the spin degrees of freedom on the spin chain, the decrease of the string order of the odd superspin SVBS chains may be naturally understood.  
On the other hand, the string order behavior of the even superspin SVBS chains is quite interesting, since  the string order revives with the hole doping.     
 Intuitive explanation may go as follows. For instance, consider $\mathcal{S}=2$ SVBS chain.  At $r=0$, the $\mathcal{S}=2$ SVBS chain is exactly identical to the $S=2$  VBS chain which is essentially constituted of two $S=1$ VBS chains, upper and lower chains. By doping the holes to the $S=2$ VBS chain, the spin degrees of freedom, say, on the upper $S=1$ VBS chain decrease to have  deficits in the chain. Below the deficits on the upper chain,  the spin degrees of freedom on the lower $S=1$ VBS chain emerge and come into effect. In this way, the spin degrees of freedom of the lower $S=1$ VBS chain contribute to generate the finite string order with increase of hole doping to the $S=2$ VBS chain.    This intuitive explanation can be applicable to  the revival of the string orders of general even superspin SVBS chains.  
Consequently, the SVBS states bear  finite string order with a finite amount of hole-doping $\it{regardless}$ of the parity of bulk superspin. This is the salient SUSY effect to the topological stability of quantum spin chains. We revisit this effect in the context of the symmetry protected topological order in Sec.\ref{sec:enetanglement} and \ref{sec:supersymmetryprotected}. 

\subsubsection{$\mathcal{N}=2$ }\label{subsc:hiddenorder2}

\begin{figure}[H]
\begin{center}
\includegraphics[scale=0.6]{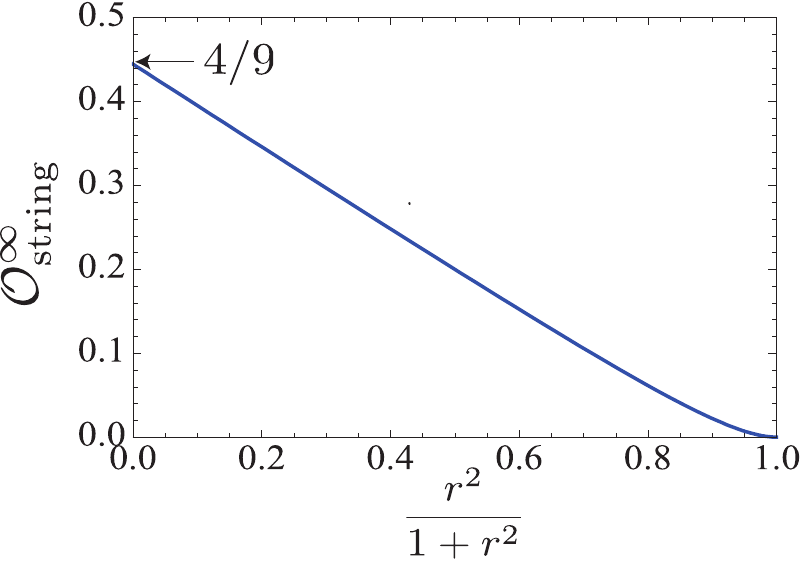}
\caption{The infinite-distance limit of 
the string correlation function (\ref{stringorderquan}) as a function of $r$.  
The value of string correlation smoothly decreases from the original VBS value 
$4/9$ to 0 (no spins left). (Figure and Caption are taken from Ref.\cite{Hasebe-Totsuka2011}.)
\label{fig:string-AKLT-2c}}
\end{center}
\end{figure}

For the $\mathcal{S}=1$ type-II SVBS  infinite chain, the string correlation is computed as 
\begin{equation}
{\cal O}^{\infty}_{\text{string}}
=\frac{4}{\left(r^2+1\right) \left(r^2+9\right)}, 
\label{eqn:string-AKLT-2a}
\end{equation}
which is plotted in Fig.~\ref{fig:string-AKLT-2c}.  
The string order (\ref{eqn:string-AKLT-2a}) takes the value of the $S=1$ VBS chain, $4/9$,  
at  $r=0$ like the $\mathcal{S}=1$ type-I chain, while approaches to zero  in the  $r\nearrow \infty$ limit unlike the  $\mathcal{S}=1$ type-I chain.  
Since spin degrees of freedom completely disappear from the type-II chain  
in the  $r\nearrow \infty$ limit [Fig.~\ref{expNewSUSY.fig}],  the string order of the type-I and II chains shows qualitatively different behaviors.

\subsection{Entanglement Spectrum and Edge States }\label{sec:enetanglement}
As discussed above, the string order of type-I SVBS chains with even bulk-superspin 
$\mathcal{S}$ revives upon hole doping.  
This suggests that, in contrast to their bosonic counterpart, 
even the SVBS states with even superspin can host the stable topological phase. 
Though the string order parameter is appealing in its similarity to the order parameter in FQHE 
\cite{Girvin-Arovas1989} and its relation to the hidden $\mathbb{Z}_{2}{\times}\mathbb{Z}_{2}$-symmetry, 
its fragility under perturbations has also been discussed recently 
\cite{AnfusoRosch2007,GuWen2009} and the alternative ``order parameter'' has been sought for. 

Recent development in quantum-information-theoretic approaches to quantum many-body 
problems enables us to extract information on the bulk topological order from 
the entanglement properties of the {\em ground-state} 
wave function \cite{Levin-W-06,Kitaev-P-06,LiHaldane2008}.  
The topological states in one-dimensional  (1D) spin systems have been 
reconsidered \cite{GuWen2009,Pollmannetal2010,Pollmannetal2012} 
from the modern point of view and the precise meaning of 
the topological Haldane phase has been clarified.    
In these studies, the string order parameters and the edge states, which in general 
are not robust against small perturbations, are replaced by more robust objects 
({\em i.e.} the structure of the entanglement spectrum or the structure of tensor-network).  
In particular, it has been shown in Ref.~\cite{Pollmannetal2010,Pollmannetal2012} 
that the existence of (at least one of) the discrete symmetries 
(time-reversal, link-inversion and $\mathbb{Z}_2 \times \mathbb{Z}_2 $ symmetry) divides  
all states of matter in 1D into two categories: topologically-non-trivial ones and the rest. 
Generic odd-integer-$S$ spin chains belong to the former while even-$S$ chains to 
the latter. The hallmark of the topological phase protected by the above discrete symmetries 
is that all entanglement levels are even-fold degenerate.   
In this formulation, the difference between the odd-$S$ VBS states and the even-$S$ ones 
is naturally understood 
in terms of the entanglement structure; the degenerate structure exists only for odd-$S$ 
cases\footnote{%
This does not mean that {\em all} the odd-$S$ spin chains have the degenerate entanglement spectrum. 
We can construct an odd-$S$ spin state {\em without} the even-fold degeneracy.}.   
It should also be mentioned that the topological phases of one-dimensional gapped spin systems have been 
classified by group cohomology, \cite{Chenetal2011,Schuchetal2011} 
and the detailed analyses based on the Lie group symmetries are reported 
in Ref.~\cite{DuivenvoordenandQuella2012}.

\begin{figure}[H]
\begin{center}
\includegraphics[scale=0.8]{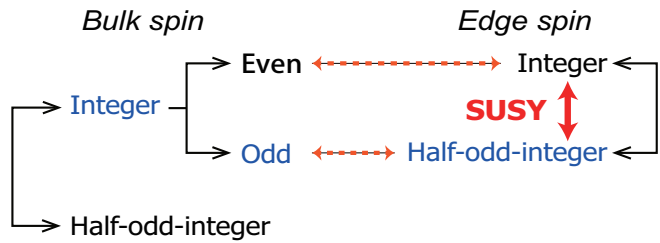}
\caption{(Color online) SUSY relates the edge spins with different parity.  
\label{fig:topoedge}}
\end{center}
\end{figure}

Following the proposal of Li and Haldane\footnote{%
Li and Haldane proposed to take the degeneracy of entanglement spectrum as the hallmark of topological phases, which can be applicable to general topological phases beyond QAFM.} \cite{LiHaldane2008}, 
we use the structure of the entanglement spectrum ({\em e.g.} degeneracy of the entanglement levels) 
as the fingerprint of topological phases. 
Then, the problem of the topological stability of the SVBS chains translates 
to the stability of the degenerate structure of the entanglement spectrum. 
Before proceeding to the details, we briefly introduce characteristic features 
of the entanglement entropy of the original $SU(2)$-invariant VBS states.  It has been reported 
that the entanglement entropy of the $SU(2)$ spin-$S$ VBS state on an infinite chain 
is given by a constant\footnote{%
On the other hand, for gapless spin chains, 
the entanglement entropy diverges as $\log (L)$ with $L$ the length of a subsystem  
for which entanglement entropy is defined \cite{Vidaletal2003}.} 
determined essentially by the degrees of freedom 
of the edge spins $S/2$ \cite{Katsura2007,Katsuraetal2008}:    
\begin{equation}
S_{\text{E.E.}}~~\sim~~\log(S+1)   ~~~~~(L~\rightarrow~ \infty) \; .
\label{behaviorSee}
\end{equation}
Corresponding to the parity of the bulk spin of the VBS chains, there appear either integer or half-integer spin at the edge. Meanwhile, in the presence of SUSY there necessarily appear both integer and half-integer spins at the edge, since SUSY relates integer and half-integer spin degrees of freedom [Fig.~\ref{fig:topoedge}].   Such particular feature of the edge spins is crucial in understanding of the salient structures of the entanglement spectrum of the SVBS chains.      
 \subsubsection{Schmidt Decomposition and Canonical Form of MPS}
 \label{sec:Schmidt-MPS}
\begin{figure}[h]
\centering 
\includegraphics[width=12cm]{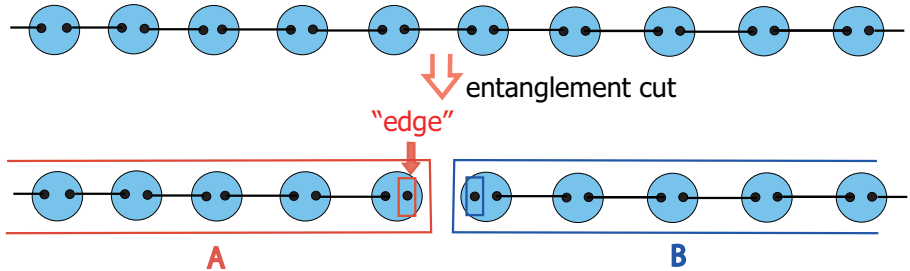}
\caption{The VBS state is divided into two parts and the edge degrees of freedom emerge at the cut.   
\label{examplesAKLT-cut.fig} }
\end{figure}

Before going into the detailed discussion, 
it would also be worthwhile here to give the derivation of the entanglement spectrum using MPS.   
Suppose we divide a system into the two parts A and B (see Fig.~\ref{examplesAKLT-cut.fig}).  
Then, we can express any wave function (state) $|\Psi\rangle$ as \cite{Nielsen-Chuang}
\begin{equation}
|\Psi\rangle =\sum_{\alpha=1}^{\chi} \lambda_{\alpha}|\alpha\rangle_\mathrm{A} 
\otimes |\alpha\rangle_\mathrm{B} 
\label{defSchmidteq}
\end{equation}
with non-zero coefficients $\lambda_{\alpha}(\geq 0)$. 
Here, $\{|\alpha\rangle_\mathrm{A}\}$ ($1\leq \alpha \leq \text{dim}\mathcal{H}_{\text{A}}$; 
$\mathcal{H}_{\text{A}}$: Hilbert space of A)
and $\{|\beta\rangle_\mathrm{B}\}$ ($1\leq \beta \leq \text{dim}\mathcal{H}_{\text{B}}$) 
are orthonormal basis states in the subspaces A and B, respectively: 
\begin{equation}
\langle \alpha|\alpha^{\prime}\rangle_\mathrm{A}=\delta_{\alpha,\alpha^{\prime}}\; , \;\;
\langle \beta |\beta^{\prime} \rangle_\mathrm{B}
=\delta_{\beta,\beta^{\prime}}.
\label{eqn:orthonormality-cond}
\end{equation}
The expansion \eqref{defSchmidteq} called the {\em Schmidt decomposition} 
defines the Schmidt coefficients $\lambda_{\alpha}$ and describes the entanglement 
between the two subsystems.  
The {\em Schmidt number} $\chi$ is the number of non-zero Schmidt coefficients 
and never exceeds the minimum of the dimensions of the Hilbert subspaces.  
From the normalization of $|\Psi\rangle$, the Schmidt coefficients satisfy 
$\sum_{\alpha=1}^{\chi}\lambda_{\alpha}^2=1$. 
The ``spectrum'' of the {\em entanglement energy} $\epsilon_{\alpha}$ defined by
\begin{equation}
{\lambda_{\alpha}}^{2} = \be^{- \epsilon_{\alpha}} \;\;
(\epsilon_{\alpha} \geq 0)
\end{equation}
is called the {\em entanglement spectrum} \cite{LiHaldane2008}.  
In terms of the Schmidt coefficients, 
the (von Neumann) entanglement entropy is given by \cite{Nielsen-Chuang}
\begin{equation}
S_{\text{E.E.}}= - \sum_{\alpha}\lambda_{\alpha}^2 \log \lambda_{\alpha}^2 
= \sum_{\alpha}\epsilon_{\alpha} \be^{- \epsilon_{\alpha}} \; .
\end{equation}
It is interesting to observe that $S_{\text{E.E.}}$ may be viewed as the ordinary thermodynamic 
entropy if we introduce a fictitious ``temperature'' $T$ and ``partition function'' as
\begin{equation}
Z(T) \equiv \sum_{\alpha} \be^{- \frac{1}{T} \epsilon_{\alpha}} \; , \;\;
Z(T=1)=1
\end{equation}
and define
\begin{equation}
S_{\text{E.E.}}=-\frac{\partial}{\partial T}\{-T \log Z(T)\}\Bigl|_{T\rightarrow 1} \; .
\end{equation}

 In the MPS formulation, the derivation of the Schmidt coefficients are rather 
 straightforward.     
 Since the MPS  is represented as the product of matrices $\mathcal{A}_{j}$ defined on each site, 
 one can easily write down an analogue of the Schmidt decomposition \eqref{defSchmidteq}  
 (see Sec.~\ref{section:SMPSEdge} for the physical meaning of the matrix indices):  
\begin{equation}
(\mathcal{A}_1 \mathcal{A}_2\cdots  \mathcal{A}_N)_{\alpha_L,\alpha_R}
=\sum_{\alpha=1}^D 
(\underbrace{\mathcal{A}_1 \mathcal{A}_2 \cdots \mathcal{A}_i}_{\mathrm{A}})_{\alpha_L,\alpha}\cdot 
(\underbrace{\mathcal{A}_{i+1} \cdots \mathcal{A}_N}_{\mathrm{B}})_{\alpha,\alpha_R} , 
\label{pre-Schmidtdecomp}
\end{equation}
where $D$ is the size of the matrix $\mathcal{A}_{j}$ and each state-valued matrix $\mathcal{A}$ is 
expanded explicitly in terms of the $c$-numbered matrices $A(m)$ and 
the orthogonal local basis states $|m\rangle_j$ as
\begin{equation}
\mathcal{A}_{j} = \sum_{m=1}^{d} A_{j}(m) |m\rangle_{j}  
\end{equation}  
($d$: dimension of local physical Hilbert space).  
For simplicity of argument, we assume that the system is uniform [$A_{j}(m)=A(m)$] and 
defined on a open chain.  

One may think that eq.\eqref{pre-Schmidtdecomp} already completes the Schmidt decomposition \eqref{defSchmidteq} 
with the identification 
\begin{equation}
\begin{split}
& |\alpha\rangle_\mathrm{A} = (\mathcal{A}_1 \mathcal{A}_2 \cdots \mathcal{A}_i)_{\alpha_L,\alpha}  \\
& |\alpha\rangle_\mathrm{B} = (\mathcal{A}_{i+1} \cdots \mathcal{A}_N)_{\alpha,\alpha_R} \; .
\end{split}
\end{equation}
However, the orthonormality condition \eqref{eqn:orthonormality-cond} are not always satisfied 
for the above choice.  
Normally, we use the ``gauge ambiguity'' \cite{Perez-Garciaetal2007,Verstraete-M-C-08} 
to find an appropriate set of edge states in such a way that 
the following overlap matrices may equal to identity:
\begin{equation}
\begin{split}
& \mathcal{L}_{\alpha,\beta} = 
(\mathcal{A}^{\ast}_1 \mathcal{A}^{\ast}_2 \cdots \mathcal{A}^{\ast}_i)_{\alpha_\text{L},\alpha} 
(\mathcal{A}_1 \mathcal{A}_2 \cdots \mathcal{A}_i)_{\alpha_\text{L},\beta} \\
& \mathcal{R}_{\alpha,\beta} = 
(\mathcal{A}^{\ast}_{i+1} \cdots \mathcal{A}^{\ast}_N)_{\alpha,\alpha_\text{R}}
(\mathcal{A}_{i+1} \cdots \mathcal{A}_N)_{\beta,\alpha_\text{R}}  \; .
\end{split}
\label{eqn:overlap-L-R}
\end{equation}
This procedure can be carried out both for finite systems and for infinite-size systems 
by using the singular-value decomposition (SVD) \cite{Vidal2003PRL,Vidal2004PRL,OrusVidal2008PRB}.  
As a result, we obtain, instead of a single matrix $A(m)$,  
two different MPS matrices $\Lambda \Gamma(m)$ for the left subsystem A and 
$\Gamma(m) \Lambda$ for the right subsystem B.  The (diagonal) matrix elements 
of the $D{\times}D$ diagonal matrix $\Lambda$ coincide with the Schmidt coefficients:
\begin{equation}
[\Lambda]_{\alpha\alpha} = \lambda_{\alpha}  \; .
\end{equation}
The MPS characterized by the set of matrices $\{\Lambda,\Gamma(m)\}$ ($m=1,\ldots,d$) is called 
{\em canonical} \cite{Perez-Garciaetal2007} 
and automatically completes the Schmidt decomposition \eqref{defSchmidteq} as 
\cite{Vidal2003PRL,Vidal2004PRL,OrusVidal2008PRB}
\begin{equation}
\begin{split}
|\Psi\rangle & = \sum_{\alpha=1}^{D} \lambda_{\alpha}|\alpha\rangle_\mathrm{A} 
\otimes |\alpha\rangle_\mathrm{B} \\
&= \sum_{\alpha=1}^{D}\sum_{\{m_{i}\}} \lambda_{\alpha}
\left[ \cdots \Lambda\Gamma(m_{i-1})\Lambda\Gamma(m_{i})\right]_{\alpha_\text{L},\alpha} 
\left[\Gamma(m_{i+1})\Lambda\Gamma(m_{i+2})\Lambda \cdots \right]_{\alpha,\alpha_\text{R}}  \\
&\phantom{\left[ \cdots \Lambda\Gamma(m_{i-1})\Lambda\Gamma(m_{i})\right]_{\alpha_\text{L},\alpha} } 
\times |m_1\rangle{\otimes}|m_2\rangle{\otimes} \cdots {\otimes}|m_{L}\rangle 
\end{split}
\label{Schmidt-MPS}
\end{equation}
(for a rather comprehensive account of the use of SVD in MPS, see, for instance, \cite{Schollwoeck-11}).  

\begin{figure}[h]
\begin{center}
\includegraphics[scale=0.5]{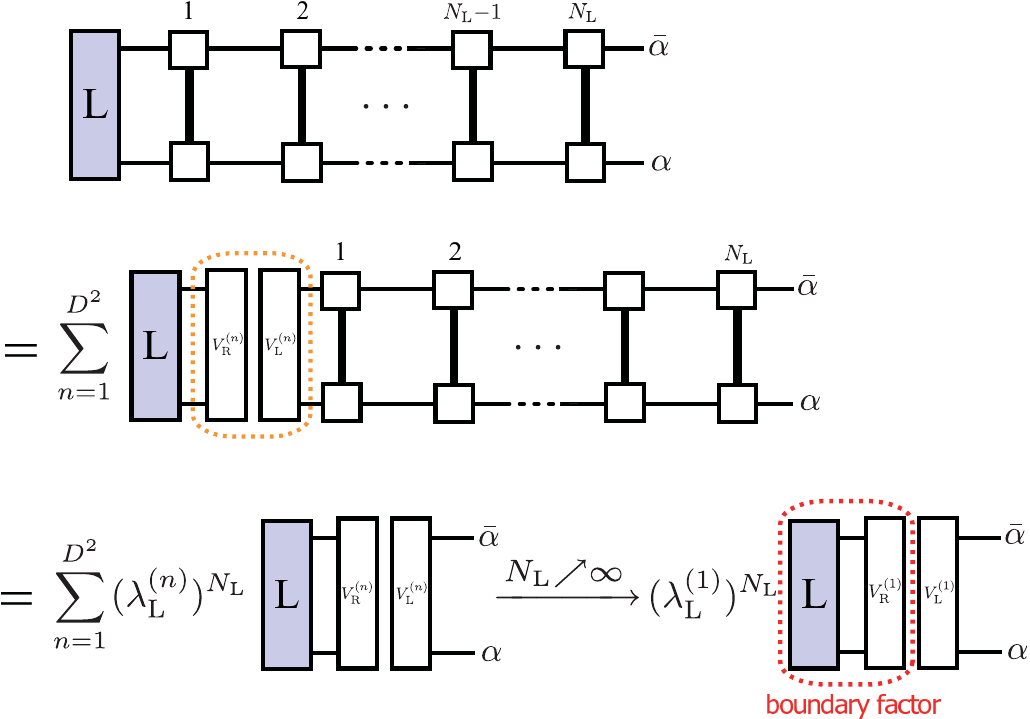}
\caption{(Color online) Norm of MPS on a semi-infinite system. 
If the MPS is pure,  the norm is essentially determined only by the dominant 
eigenvector $\mathbf{V}_{\mathrm{L}}^{(1)}$ (except for the boundary factor which depends on 
the boundary condition imposed on the edge states). 
Similar relation holds for the right semi-infinite system. 
\label{fig:MPS-norm-semi-infinite}}
\end{center}
\end{figure}

In infinite-size systems, the overlap calculation \eqref{eqn:overlap-L-R} simplifies a lot 
since $\mathcal{L}$ ($\mathcal{R}$) reduces essentially to the left (right) eigenvector 
$\mathbf{V}_{\mathrm{L}}^{(1)}$ ($\mathbf{V}_{\mathrm{R}}^{(1)}$) 
with the largest eigenvalue ({\em dominant eigenvector}; 
see Fig.~\ref{fig:MPS-norm-semi-infinite} for a diagrammatic representation)\footnote{%
When the MPS is not of the canonical form, we can take 
$\Gamma(m)=A(m)$ and $\Lambda=\mathbf{1}$ (and hence $T_{\text{L}}=T_{\text{R}}$).} 
of the transfer matrix $T_{\text{L}}$ ($T_{\text{R}}$) defined by
\begin{subequations}
\begin{align}
& (T_{\mathrm{L}})_{\bar{\alpha},\alpha;\bar{\beta},\beta} 
\equiv \sum_{m}(\Lambda\Gamma^{\ast}(m))_{\bar{\alpha}\bar{\beta}}
(\Lambda\Gamma(m))_{\alpha\beta} \\
& (T_{\mathrm{R}})_{\bar{\alpha},\alpha;\bar{\beta},\beta} 
\equiv \sum_{m}(\Gamma^{\ast}(m)\Lambda)_{\bar{\alpha}\bar{\beta}}
(\Gamma(m)\Lambda)_{\alpha\beta} \quad 
(\bar{\alpha},\bar{\beta},\alpha,\beta=1,\ldots,D)  \; .
\end{align}
\end{subequations}
Specifically, except for the unimportant factors determined solely by the edge states, 
the overlap matrices (\ref{eqn:overlap-L-R}) coincide with the ($D^{2}$-dimensional) eigenvectors 
$\mathbf{V}_{\mathrm{L}}^{(1)}$ and $\mathbf{V}_{\mathrm{R}}^{(1)}$ of 
$T_{\text{L}}$ and $T_{\text{R}}$, respectively (see Fig.~\ref{fig:MPS-norm-semi-infinite}):
\begin{equation}
\mathcal{L}_{\bar{\alpha}\beta} \propto [\mathbf{V}_{\mathrm{L}}^{(1)}]_{\bar{\alpha},\beta} \; , \;\;
\mathcal{R}_{\bar{\alpha}\beta} \propto [\mathbf{V}_{\mathrm{R}}^{(1)}]_{\bar{\alpha},\beta}  \; .
\end{equation}

\begin{figure}[H]
\begin{center}
\includegraphics[scale=0.6]{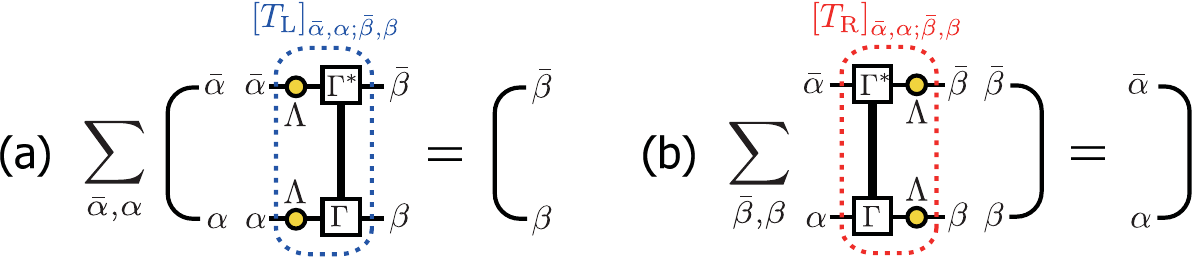}
\caption{(Color online) Graphical representation of the two conditions, (\ref{eqn:canonical-iMPS-TL}) and (\ref{eqn:canonical-iMPS-TR}), for 
canonical iMPS.  In Ref.~\cite{Garcia-W-S-V-C-08}, the action of the type (a) ((b)) is 
denoted by $\mathcal{E}^{\ast}(\Lambda^{2})=\Lambda^{2}$ 
($\mathcal{E}(\mathbf{1})=\mathbf{1}$). 
\label{fig:canonical-cond}}
\end{center}
\end{figure}

If our infinite-size MPS (iMPS) assumes the canonical form, 
$[\mathcal{L}]_{\bar{\alpha}\beta}\propto \delta_{\bar{\alpha}\beta}=(\mathbf{1}_{D})_{\bar{\alpha}\beta}$ 
and 
$[\mathcal{R}]_{\bar{\alpha}\beta}\propto \delta_{\bar{\alpha}\beta}=(\mathbf{1}_{D})_{\bar{\alpha}\beta}$ 
by definition (with $\mathbf{1}_{D}$ being the $D$-dimensional identity matrix), 
and therefore the two transfer matrices satisfy\footnote{%
By the assumption of pure MPS, the largest eigenvalue is unique 
\cite{Fannes-etal1992,Garcia-W-S-V-C-08}. (MPS is called a pure MPS when transfer matrix has non-degenerate maximal eigenvalue. In the infinite-size limit, the pure MPS is reduced to a pure state, and hence the name pure MPS.) 
When the value of the largest eigenvalue is not 1, we can rescale the matrices $A(m)$ 
so that it may be 1.} \cite{Fannes-etal1992,Garcia-W-S-V-C-08}
\begin{subequations}
\begin{equation}
\begin{split}
& \sum_{\bar{\alpha},\alpha} (\mathbf{1}_{D})_{\bar{\alpha},\alpha}
\left[T_{\text{L}}\right]_{\bar{\alpha},\alpha;\bar{\beta},\beta} 
= \sum_{\alpha=1}^{D}\sum_{m=1}^{d}
\left[\Lambda \Gamma^{\ast}(m)\right]_{\alpha,\bar{\beta}}
\left[\Lambda \Gamma(m)\right]_{\alpha,\beta} \\
& \phantom{\sum_{\bar{\alpha},\alpha} (\mathbf{1}_{D})_{\bar{\alpha},\alpha}
\left[T_{\text{L}}\right]_{\bar{\alpha},\alpha;\bar{\beta},\beta}}
= \sum_{m=1}^{d}\left[\Gamma^{\dagger}(m)\Lambda^{2}\Gamma(m)\right]_{\bar{\beta},\beta}
= (\mathbf{1}_{D})_{\bar{\beta},\beta} \quad (\text{left action})
\end{split}  
\label{eqn:canonical-iMPS-TL} 
\end{equation}
and
\begin{equation}
\begin{split}
& \sum_{\bar{\beta},\beta} 
\left[T_{\text{R}}\right]_{\bar{\alpha},\alpha;\bar{\beta},\beta} (\mathbf{1}_{D})_{\bar{\beta},\beta}
= \sum_{\beta=1}^{D}\sum_{m=1}^{d}
\left[\Gamma^{\ast}(m)\Lambda \right]_{\bar{\alpha},\beta}
\left[\Gamma(m)\Lambda \right]_{\alpha,\beta} \\
& \phantom{\sum_{\bar{\alpha},\alpha} (\mathbf{1}_{D})_{\bar{\alpha},\alpha}
\left[T_{\text{R}}\right]_{\bar{\alpha},\alpha;\bar{\beta},\beta}}
= \sum_{m=1}^{d}\left[
\left\{\Gamma(m)\Lambda^{2}\Gamma^{\dagger}(m)\right\}^{\text{t}}\right]_{\bar{\alpha},\alpha}
= (\mathbf{1}_{D})_{\bar{\alpha},\alpha} \quad (\text{right action}) 
\end{split}
\label{eqn:canonical-iMPS-TR}
\end{equation}\label{eqn:canonical-iMPS}
\end{subequations}
(see Fig.~\ref{fig:canonical-cond}). Notice that these equations can also be regarded 
as the eigenvalue equations for the transfer matrices:  $\mathbf{1}_D$ is 
the left (right) eigenstate of the transfer matrix $T_{\text{L}}$  ($T_{\text{R}}$) 
with the eigenvalue $1$.    
Eq.(\ref{eqn:canonical-iMPS}) may be thought of as the conditions in order that the iMPS is of the canonical form. 

On the other hand, when the MPS in question satisfies\footnote{%
This is the case for all the (S)MPSs treated in this paper.} 
\begin{equation}
\mathcal{L}_{\alpha\beta} \propto \delta_{\alpha,\beta} \; , \;\;
\mathcal{R}_{\alpha\beta} \propto \delta_{\alpha,\beta} 
\end{equation}
in the infinite-size limit, the Schmidt decomposition for the infinite chain 
is obtained very easily just by rescaling the MPSs for the subsystems.  
The normalized state $|\Psi\rangle$ is constructed as 
\begin{equation}
|\Psi(\alpha,\beta)\rangle
=\frac{1}{\sqrt{A(\alpha,\beta)}} (\mathcal{A}_1\mathcal{A}_2 \cdots   \mathcal{A}_N)_{\alpha,\beta}, 
\end{equation}
where $A(\alpha,\beta)$ is the magnitude of $|\Psi(\alpha,\beta)\rangle$: 
\begin{equation}
A(\alpha,\beta)= |(\mathcal{A}_1\mathcal{A}_2 \cdots   \mathcal{A}_N)_{\alpha,\beta}|^2=(\mathcal{A}_N^{\dagger}\cdots\mathcal{A}_2^{\dagger}    \mathcal{A}_1^{\dagger})_{\beta,\alpha}(\mathcal{A}_1\mathcal{A}_2 \cdots   \mathcal{A}_N)_{\alpha,\beta}.
\label{magnitudeA}
\end{equation}
With these normalization constants, the normalized MPS is written as 
\begin{equation}
|\Psi (\alpha_L,\alpha_R)\rangle=\sum_{\alpha=1}^D 
\sqrt{\frac{{A(\alpha_L,\alpha)A(\alpha,\alpha_R)}}{{A(\alpha_L,\alpha_R)}}}|\Psi (\alpha_L,\alpha)\rangle \cdot |\Psi (\alpha,\alpha_R)\rangle.
\end{equation}
Comparing this expansion with (\ref{defSchmidteq}), one may read off the Schmidt coefficients as  
\begin{equation}
\lambda_{\alpha}=
\sqrt{\frac{{A(\alpha_L,\alpha)A(\alpha,\alpha_R)}}{{A(\alpha_L,\alpha_R)}}}.
\label{formulaschimidt}
\end{equation}
In the infinite limit, $\lambda_{\alpha}$ is not relevant to the polarization of the edge spins, $\alpha_L$ and $\alpha_R$. 
Therefore for the infinite (S)VBS chain, we only need to evaluate the magnitude of the matrix product to obtain the Schmidt coefficients, and thus the derivation of  Schmidt coefficients is boiled down to the computation of the normalization constants, $A(\alpha,\beta)$ (\ref{magnitudeA}).   

\subsubsection{$\mathcal{N}=1$ }
To substantiate the topological stability of the SVBS chain, we investigate two type-I chains with distinct bulk superspins, $\mathcal{S}=1$ and $\mathcal{S}=2$.  

\begin{figure}[!t]
\centering
\includegraphics[width=7cm]{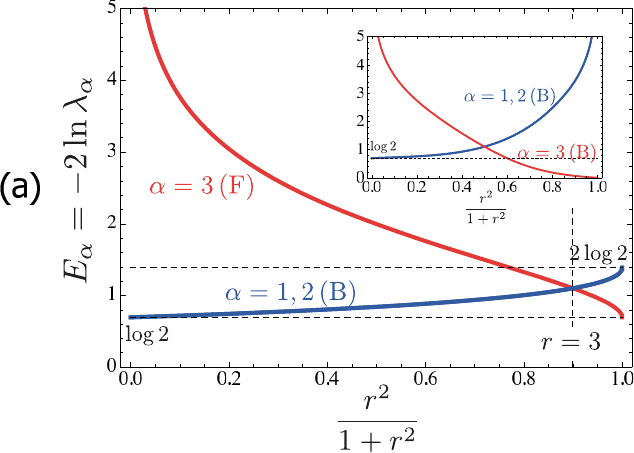}
~~~\includegraphics[width=7cm]{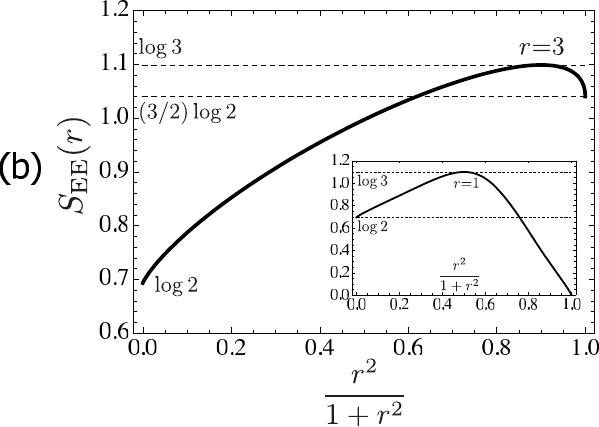}
\caption{The change of the entanglement spectra as a function of $r$ (left panel) and 
entanglement entropy (right panel) of the $\mathcal{S}=1$ type-I chain.  
The insets are the results for the bosonic-pair VBS chain.  
(Figure and Caption are taken from Ref.\cite{Hasebe-Totsuka2012}.) \label{entropySL11.fig} }
\end{figure}

\noindent%
{\bf (i) Superspin $\mathcal{S}=1$}\\
From the formula (\ref{formulaschimidt}), the entanglement spectrum of the $\mathcal{S}=1$ type-I 
state (on an infinite chain) is derived as 
\begin{subequations}
\begin{align}
&{\lambda_{\mathrm{B}}}^2\equiv {\lambda_1}^2={\lambda_2}^2=\frac{1}{4}+\frac{3}{4\sqrt{9+8r^2}},\label{bosontype1schmit}\\
&{\lambda_{\mathrm{F}}}^2\equiv {\lambda_3}^2=\frac{1}{2}-\frac{3}{2\sqrt{9+8r^2}}.\label{fermiontype1schmit} 
\end{align}
\end{subequations}\label{s=1osp(12)Schmidt}
They are shown in Fig.~\ref{entropySL11.fig} (left-figure) with the  entanglement entropy (right-figure).  At $r=0$, the SVBS state reduces to the $S=1$ VBS state and reproduces  both entanglement spectra and entanglement entropy of the $S=1$ VBS chain, $i.e.$, ${\lambda_{\mathrm{B}}}^2 \rightarrow 1/2$ and $S_{EE}\rightarrow \text{ln}2$, which should be compared with  (\ref{behaviorSee}) for $S=1$. Similarly, in the limit $r\rightarrow \infty$, the SVBS chain  reduces to the MG chains, and  the entanglement entropy of SVBS chain also reproduces the finite entanglement entropy of the MG chain (the right-figure of Fig.~\ref{entropySL11.fig}).

\begin{figure}[H]
\begin{center}
\includegraphics[scale=1.2]{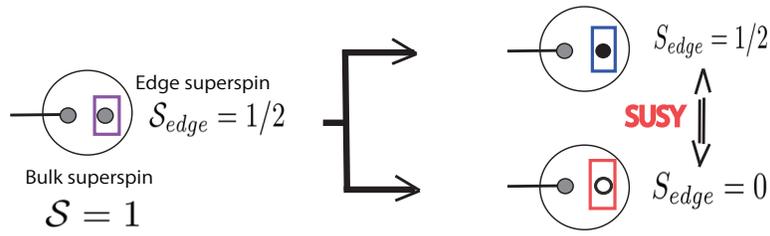}
\caption{(Color online) There always exit half-integer and interger edge spin states as the super-partner of the SUSY. 
The half-integer edge spin states play a crucial role in the stability of topological phases. 
\label{fig:Edgespins}}
\end{center}
\end{figure}

 As in the entanglement spectra of the left of Fig.~\ref{entropySL11.fig}, we have two distinct entanglement spectra, one of which is doubly degenerate spectrum (blue curve) for the ``bosonic'' Schmidt coefficients corresponding to those of the original $S=1$ VBS chain (\ref{bosontype1schmit}) and the other is the non-degenerate spectrum (red curve) for the ``fermionic'' Schmidt coefficient (\ref{fermiontype1schmit}). 
The existence of  such two types of entanglement spectra is a salient feature of the SUSY state and can be readily understood based on the following edge state picture. 
For $\mathcal{S}=1$ SVBS chain, its edge superspin states are given by the $UOSp(1|2)$  multiplet with superspin  $\mathcal{S}_{\text{edge}}=1/2$ that consists of the ordinary $SU(2)$ states with spin, $1/2\oplus 0$ (Fig.~\ref{fig:Edgespins}) :  
\begin{equation}
(\mathcal{S}_{\text{edge}}=1/2) = (S_{\text{edge}}=1/2)\oplus (S_{\text{edge}}=0)
\end{equation}
or 
\begin{equation}
\bold{3}=\bold{2}\oplus \bold{1}. 
\end{equation}
The  $S=1/2$ $SU(2)$ edge spin states generate the double degeneracy in the entanglement spectra, while the 
 $S=0$ $SU(2)$ edge spin state gives the non-degenerate one.  
Due to the existence of $1/2$ spin edge degrees of freedom, the double degeneracy  is guaranteed (we will give a detail discussion in Sec.\ref{sec:supersymmetryprotected}), and hence we find that the type-I SVBS chain is in the topological phase. 
To highlight the effect of SUSY, we consider the non-SUSY ({\em i.e.} purely bosonic) cases and 
replace the fermion operator $f_i$ with the boson operator $c_i$ ($[c_i,c_j^{\dagger}]=\delta_{ij}$) to define the following boson-pair VBS chain, 
\begin{equation}
|\Phi_{\text{b.p.}}\rangle =\prod_i (a_i^{\dagger}b_{i+1}^{\dagger}-b_i^{\dagger}a_{i+1}^{\dagger}-rc_{i}^{\dagger}c_{i+1}^{\dagger})|\text{vac}\rangle. 
\end{equation}
As the insets of Fig.~\ref{entropySL11.fig}, we depicted the entanglement spectra and entanglement entropy.  
The crucial difference to the SUSY case will be apparent in the limit $r\rightarrow \infty$, 
$|\Phi_{\text{b.p.}}\rangle$, is reduced to a simple product state,  
\begin{equation}
|\Phi_{\text{b.p.}}\rangle~\rightarrow ~\prod_{i}c_i^{\dagger}|\text{vac}\rangle, 
\end{equation}
and the entanglement entropy vanishes (the inset of the right figure of Fig.~\ref{entropySL11.fig}). 

\begin{figure}[!t]
\centering
\includegraphics[width=7.6cm]{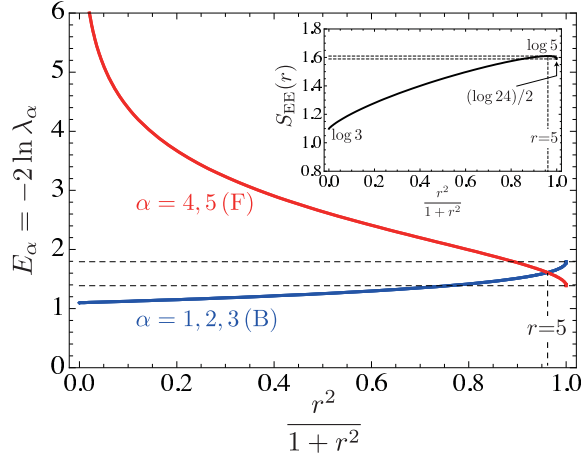} 
\caption{The behaviors of the Schmidt coefficients (\ref{s=2osp(12)Schmidtdoubtrip}) and the entanglement entropy (inset) of the ${S}=2$ type-I SVBS chain. 
(Figure and Caption are taken 
from Ref.\cite{Hasebe-Totsuka2012}.) \label{entropySL12.fig} }
\end{figure}

\begin{figure}[H]
\begin{center}
\includegraphics[scale=1.2]{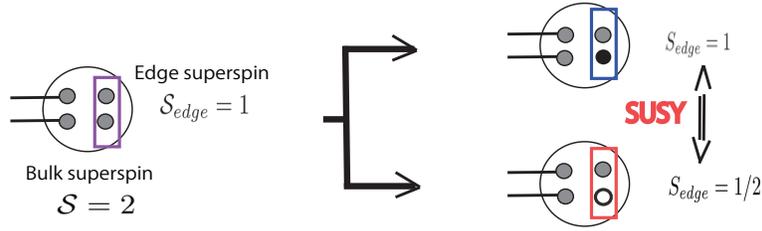}
\caption{(Color online) ${S}_{\text{edge}}=1/2$  generates the stability of topological phase of the $\mathcal{S}=2$ type II SVBS chain. \label{fig:Edgespins2}}
\end{center}
\end{figure}

\noindent%
{\bf (ii) Superspin $\mathcal{S}=2$}\\
Next, we examine the entanglement spectrum of the $\mathcal{S}=2$ type-I SVBS chain.  
Though the topological phase of its bosonic counterpart, $S=2$ VBS chain, is fragile under perturbation   \cite{AnfusoRosch2007}, the $\mathcal{S}=2$ type-I SVBS chain itself is topologically stable with finite amount of hole doping.   As we shall see below, SUSY plays a crucial role for the stability of the topological phase.     
At $r=0$,  $\mathcal{S}=2$ SVBS chain is reduced to the $S=2$ VBS chain, while in the 
limit $r\rightarrow \infty$ the SVBS chain is reduced to the partially dimerized chain. 
 We have the following five Schmidt coefficients for the $\mathcal{S}=2$ SVBS chain: 
\begin{subequations}
\begin{align}
&{\lambda_\mathrm{B}}^2\equiv {\lambda_1}^2={\lambda_2}^2
={\lambda_3}^2=\frac{1}{6}+ \frac{5(4+\sqrt{25+24r^2})}{6(25+24r^2+4\sqrt{25+24r^2})},\label{s=2osp(12)Schmidttirple}\\
&{\lambda_\mathrm{F}}^2\equiv {\lambda_4}^2={\lambda_5}^2= \frac{1}{4}- \frac{5(4+\sqrt{25+24r^2})}{4(25+24r^2+4\sqrt{25+24r^2})}.  \label{s=2osp(12)Schmidtdoub}
\end{align} \label{s=2osp(12)Schmidtdoubtrip}
\end{subequations}
 Thus, the five Schmidt coefficients are split into the triply degenerate (\ref{s=2osp(12)Schmidttirple}) and doubly degenerate (\ref{s=2osp(12)Schmidtdoub}) spectra showing distinct behaviors in Fig.~\ref{entropySL12.fig}.  
Again, such splitting of the Schmidt coefficients are readily understood by the edge state picture for the SUSY chain. For the $\mathcal{S}=2$ type-I SVBS chain, the edge superspin is given by $\mathcal{S}_{\text{edge}}=1$ that consists of the $SU(2)$ edge spin $S_{\text{edge}}=1$ and $S_{\text{edge}}=1/2$ [Fig.~\ref{fig:Edgespins2}]:  
\begin{equation}
(\mathcal{S}_{\text{edge}}=1) = (S_{\text{edge}}=1)\oplus (S_{\text{edge}}=1/2)
\end{equation}
or 
\begin{equation}
\bold{5}=\bold{3}\oplus \bold{2}. 
\end{equation}
The $S_{\text{edge}}=1$ degrees of freedom generate the triple degeneracy in the entanglement spectra, while the $S_{\text{edge}}=1/2$ degrees of freedom give the double degeneracy.  
Due to the existence of SUSY, the $\mathcal{S}=2$ SVBS chain necessarily contains the $S_{\text{edge}}=1/2$ edge spin degrees of freedom that do not originally exist in the  $S=2$ VBS chain [Fig.~\ref{fig:Edgespins2}], 
 and they guarantee the double degeneracy in the entanglement spectra, $i.e.$ the stability of the topological phase \cite{Pollmannetal2010,Pollmannetal2012}.   

From the above demonstrations for $\mathcal{S}=1$ and $\mathcal{S}=2$ type I SVBS chains, one may see that   regardless of the parity of bulk superspin, the SUSY introduces the  half-integer edge-spin states that necessitate at least double degeneracy in the entanglement spectrum.    
%

\subsubsection{$\mathcal{N}=2$}

\begin{figure}[!t]
\centering
\includegraphics[width=7.9cm]{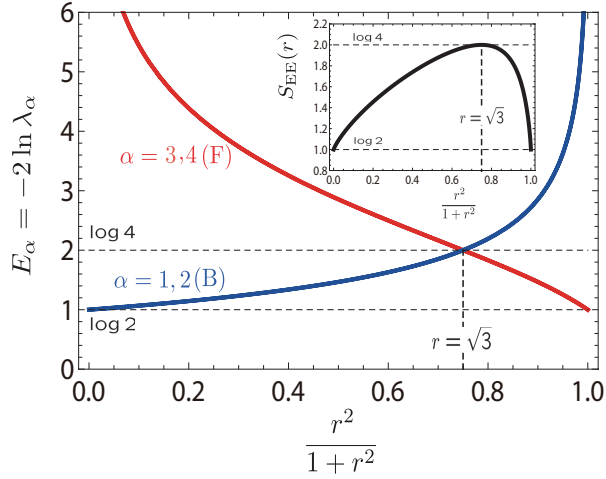} 
\caption{The behaviors of the Schmidt coefficients  and the entanglement entropy (inset) of the $\mathcal{S}=1$ type-II SVBS chain. \label{entropySII.fig4} }
\end{figure}

It is also straightforward to  calculate the Schmidt coefficients for the type-II SVBS chain: 
\begin{align}
&\!\!\!{\lambda_1}^2={\lambda_2}^2=\frac{1}{4}-\frac{3-r^2}{4\sqrt{9+10r^2+r^4}},\nonumber\\
&\!\!\!{\lambda_3}^2={\lambda_4}^2=\frac{1}{4}+\frac{3-r^2}{4\sqrt{9+10r^2+r^4}}.
\end{align}
The four Schmidt coefficients are split into two  groups showing distinct behaviors [Fig.~\ref{entropySII.fig4}] according to the $SU(2)$ decomposition of the $UOSp(1|2)$ edge superspin state: 
\begin{equation}
(\mathcal{S}_{\text{edge}}=1/2) = (S_{\text{edge}}=1/2)\oplus (S_{\text{edge}}=0)\oplus (S_{\text{edge}}=0), 
\end{equation}
or 
\begin{equation}
\bold{4}=\bold{2}\oplus \overbrace{\bold{1}\oplus \bold{1}}^{\bold{2}}. 
\label{N=2dimedgedecom}
\end{equation}
The first $\bold{2}$ on the right-hand side of (\ref{N=2dimedgedecom}) corresponds to the doubly degenerate blue curve in Fig.~\ref{entropySII.fig4}, while the remaining $\bold{1}\oplus \bold{1}(=\bold{2})$ represents the doubly degenerate red curve. 

\subsection{Supersymmetry-Protected Topological Order}
\label{sec:supersymmetryprotected}
In this section, we clarify the relation between the structure of the entanglement spectra discussed in 
the previous sections and the symmetry-protected topological order \cite{GuWen2009}.   
To this end, we use the MPS representation established in Sec.~\ref{secsmat}.  
Because of its simplicity in the entanglement structure and the wide applicability to gapped 
states in 1D, the MPS approach provides us with a powerful tool in investigating the topological phases 
in 1D 
\cite{Pollmannetal2010,Pollmannetal2012,Turner-P-B-11,Fidkowski-K-11,Schuchetal2011,Chenetal2011}. 
\subsubsection{Symmetry Operation and MPS}
\label{sec:symmetry-in-MPS}
For later convenience, we give a quick summary of some tools used in the MPS approach. 
As has been discussed in Sec.~\ref{secsmat} and \ref{sec:Schmidt-MPS}, any MPS 
on an open chain may be written as
\begin{equation}
|\text{MPS}\rangle 
= \bigotimes_{i=-\infty}^{\infty} \mathcal{A}_{i} 
= \cdots\otimes \mathcal{A}_{-1}\otimes\mathcal{A}_{0}\otimes\mathcal{A}_{1}\otimes\mathcal{A}_{2}
\otimes \cdots \; ,
\end{equation}
or in terms of the $c$-number matrices $A(m)$ [$\mathcal{A}_{j}=\sum_{m=1}^{d}A_{j}(m)|m\rangle_{j}$, 
$d$ being the dimension of the local physical Hilbert space], as 
\begin{equation}
|\text{MPS}\rangle 
= \sum_{\{m_{i}\}} \left\{
\cdots A(m_{-1})A(m_{0})A(m_{1})A(m_{2})\cdots \right\}
  |m_1\rangle{\otimes}|m_2\rangle{\otimes} \cdots {\otimes}|m_{L}\rangle 
\end{equation}
As has been mentioned in Sec.~\ref{sec:Schmidt-MPS}, we can ``gauge-transform'' 
this into the canonical form which conforms with the Schmidt decomposition; 
on the left of the entanglement cut, we use the MPS matrix 
$A_{\mathrm{L}}(m)=\Lambda \Gamma(m)$, 
and on the right $A_{\mathrm{R}}(m)=\Gamma(m)\Lambda$. 
In the following sections, our arguments will be based on the canonical form characterized 
by the MPS data $\{\Lambda,\Gamma(m)\}$ ($m=1,\ldots,d$).  

Now let us consider unitary operations on MPS. 
A given MPS ($|\Psi_{\text{MPS}}\rangle$) is said to be invariant under a unitary operation 
if the product of local unitary operators $\otimes \hat{u}$ leaves the MPS invariant 
(up to an overall phase) \cite{Garcia-W-S-V-C-08,Sanz-W-G-C-09}:
\begin{equation}
\hat{u}\otimes \hat{u} \otimes \cdots \otimes \hat{u}|\Psi_{\text{MPS}}\rangle 
= \be^{i \alpha_{\mathrm{g.s.}}} |\Psi_{\text{MPS}}\rangle  \; .
\label{eqn:def-unitary-invariance}
\end{equation}
The local unitary operation $\hat{u}$ acts on $\mathcal{A}$ in a site-wise manner: 
\begin{equation}
\mathcal{A}_{i} \mapsto 
\mathcal{A}^{\prime}_i=\sum_{m}\Lambda \Gamma'(m)|m\rangle_i
= \sum_{m}\Lambda \left\{ \sum_{n} \langle m | \hat{u}|n\rangle_i\Gamma(n) \right\}|m\rangle_i 
\; , 
\label{eqn:MPS-unitary-op} 
\end{equation}
where we have used the completeness relation on each site, $\sum_m |m\rangle_i \langle m|_i=1$. 
If the symmetry operation is anti-unitary (like time-reversal), the complex-conjugation  
$\Gamma(m) \mapsto \Gamma^{\ast}(m)$ should be taken in eq.\eqref{eqn:MPS-unitary-op}.   

Then, it can be shown that 
the above unitary invariance is equivalent to the existence of the following unitary operator $U$ 
acting on $\Gamma(m)$ \cite{Garcia-W-S-V-C-08}:
\begin{equation}
\hat{u}\otimes \hat{u} \otimes \cdots \otimes \hat{u}|\Psi_{\text{\rm MPS}}\rangle 
= \be^{i \alpha_{\mathrm{g.s.}}}  |\Psi_{\text{\rm MPS}}\rangle 
\quad \Longleftrightarrow \quad 
\sum_{n=1}^{d}\langle m |\hat{u}|n\rangle \Gamma(n) 
= \be^{i\theta_{u}} \, U^{\dagger} \Gamma(m) U \; ,
\label{eqn:U-Gamma-U}
\end{equation}
where $\theta_{u}$ denotes a $\hat{u}$-dependent phase.  
Once the unitary $\hat{u}$ is given, $U$ is uniquely determined \cite{Garcia-W-S-V-C-08} 
up to an overall phase. 
To be more precise, if we define a generalized transfer matrix 
\begin{equation} 
[T^{(u)}]_{\bar{\alpha},\alpha;\bar{\beta},\beta} \equiv \sum_{m,n=1}^{d}
\left[ \Lambda\Gamma^{\ast}(m)\right]_{\bar{\alpha},\bar{\beta}}
\left[ \Lambda\Gamma(n)\right]_{\alpha,\beta} 
\langle m|\hat{u}|n\rangle \; ,
\end{equation}
the $D$-dimensional unitary matrix $U$ is given essentially 
by its left eigenvector with the largest eigenvalue\footnote{%
We have assumed that $\Gamma(m)$ is rescaled in such a way that the largest (or, dominant) 
right-eigenvalue of $T_{\text{L,R}}$ is unity.} $\be^{i\theta_{u}}$ (see Eq.(\ref{lefteigentrans}) and the explanations around it). 
Since $U$ leaves the MPS $|\Psi_{\text{\rm MPS}}\rangle$ invariant, it is natural 
to assume that $U$ does not change the physical entanglement spectrum $\Lambda$, 
{\em i.e.} 
\begin{equation}
[ U \, , \, \Lambda ] =0 \; .
\end{equation}
Physically, the above relation \eqref{eqn:U-Gamma-U} implies that the original symmetry operation 
(acting on the {\em physical} Hilbert space at each site) 
``fractionalizes'' into the ones ($U$ and $U^{\dagger}$) which act on the edge 
states on both ends of the system.  
The equation \eqref{eqn:U-Gamma-U} plays a crucial role in the following discussions. 
\subsubsection{Case of SMPS}
\label{sec:determining-U-SMPS}
Now we extend the arguments developed by Pollmann et al. \cite{Pollmannetal2012,Pollmannetal2010} 
for the bosonic MPS to the SUSY case. 
First remember that degeneracy in energy spectra of quantum mechanical Hamiltonian can be attributed to some symmetry of the Hamiltonian. 
Meanwhile, since the entanglement spectrum is solely determined by a ground-state wave function, the degeneracy of entanglement spectrum is expected to stem from some symmetry of the ground-state wave function.   
Indeed, several discrete symmetries are identified to guarantee degeneracy in the entanglement spectrum. 
 Since the degeneracy of the entanglement spectrum is the hallmark of the topological order, it is said that the topological order is protected by the symmetry of the ground-state wave function, 
 and hence the name symmetry protected topological order.    
In the following, we will show 
that the SUSY guarantees the existence of at least two-fold degeneracy of entanglement spectrum $\it{regardless}$ $\it{of}$ the parity of superspin, provided that at least 
one of the three symmetries, inversion, time-reversal and $\mathbb{Z}_2\times \mathbb{Z}_2$, is present.    
We will also find that the (S)MPS formalism plays a crucial role in the discussions.  

As has been discussed in Sec.\ref{section:SMPSEdge}, the SMPS is generally represented as 
\cite{Hasebe-Totsuka2011}   
\begin{equation}
|\Psi \rangle =\mathcal{A}_1\mathcal{A}_2\cdots\mathcal{A}_L,   
\end{equation}
where $\mathcal{A}_i$ $(i=1,2,\cdots,L)$ are supermatrices defined on the sites $i$.  
Then, we follow the same steps as in Sec.~\ref{sec:symmetry-in-MPS} to transform 
the SMPS into the canonical form.  When a given unitary operation leaves the MPS 
invariant [in the sense of eq.\eqref{eqn:def-unitary-invariance}], 
the unitary operation $\hat{u}$ fractionalizes into $U$ and $U^{\dagger}$, and acts like
\begin{equation}
\sum_{n=1}^{d}\langle m |\hat{u}|n\rangle \Gamma(n) 
= \be^{i\theta_{U}} \, U^{\dagger} \Gamma(m) U \; .
\label{eqn:U-Gamma-U-SUSY}
\end{equation}
In fact, the local symmetry operation need not be unitary (as will be in the following sections). 
In this case, the left-hand side may be replaced with the general form $\Gamma^{\prime}(m)$. 
Therefore, the most general form reads as
\begin{equation}
\Gamma^{\prime}(m) 
= \be^{i\theta_{U}} \, U^{\dagger} \Gamma(m) U \; ,
\label{eqn:U-Gamma-U-SUSY-2}
\end{equation}
where the phase $\theta_{U}$ depends on the symmetry operation considered.  
In the above equations, $m$ labels both bosonic and fermionic states, i.e. $m=i,\alpha$,  
and $\Gamma(m)$ are given, for $UOSp(1|2)$, by 
\begin{equation}
\begin{split}
&\Gamma(i)=\begin{pmatrix}
M_1(i) & 0 \\
0 & M_2(i)
\end{pmatrix} \;\; (i=x,y,z)\\
&\Gamma(\alpha)=\begin{pmatrix}
0 & N_1(\alpha)\\
N_2(\alpha) & 0
\end{pmatrix} \;\; (\text{$\alpha=\theta_1,\theta_2$}) \; ,  
\end{split}
\end{equation}
where $M_1,M_2,N_1$ and $N_2$ are {\em c-number} matrices. 

Here, a remark is in order about the form of $U$. 
The ($c$-number) unitary matrix $U$ in \eqref{eqn:U-Gamma-U-SUSY} may be postulated as: 
\begin{equation}
U =
\begin{pmatrix}
U_1 & 0 \\
0 & U_2
\end{pmatrix}, 
\label{eqn:block-form-U}
\end{equation}
where $U_1$ and $U_2$ are unitary matrices that act on 
the two bosonic subspaces having different fermion numbers. 
The reason for choosing the above form may be seen as follows. 
First we note that eq.\eqref{eqn:U-Gamma-U-SUSY} implies that the MPS 
on a periodic chain transforms like
\begin{equation}
|\Psi\rangle \xrightarrow{\otimes \hat{u}}  
\str ( U^{\dagger}\mathcal{A}_1 \mathcal{A}_2\cdots \mathcal{A}_{2n+1}U) \; , 
\label{eqn:Str-UAAAU}
\end{equation}
where the supertrace $\str$ is defined for general supermatrices as 
[see also eq.\eqref{eqn:STr-1}]
\begin{equation}
\str 
\begin{pmatrix}
A_{\text{B}}^{(1)} & A_{\text{F}}^{(1)} \\
A_{\text{F}}^{(2)} & A_{\text{B}}^{(2)} 
\end{pmatrix}
= \tr A_{\text{B}}^{(1)} - \tr A_{\text{B}}^{(2)} \; .
\end{equation}
We expect that for a periodic chain, which does not have edges, is {\em fully} invariant 
under the unitary operation, i.e., the expression \eqref{eqn:Str-UAAAU} coincides with 
the original MPS (up to an overall phase).   
While in the case of bosonic MPS, this, combined with $\tr(AB)=\tr(BA)$, 
immediately implies $\otimes \hat{u} |\Psi\rangle \propto |\Psi\rangle$, 
the relation $\str(AB)=\str(BA)$ 
holds only when $A$ and $B$ are super-matrices 
(that contain the Grassmann-odd blocks in their off-diagonal parts):
\begin{equation}
|\Psi\rangle \xrightarrow{\otimes \hat{u}}  
\str ( U^{\dagger}\mathcal{A}_1 \mathcal{A}_2\cdots \mathcal{A}_{2n+1}U) 
\overset{\text{?}}{=} 
\str ( \mathcal{A}_1 \mathcal{A}_2\cdots \mathcal{A}_{2n+1}U U^{\dagger})
\; .
\label{eqn:Str-UAAAU-2}
\end{equation} 
In fact, if an arbitrary pair of two super matrices $A$ and $B$ were merely $c$-number matrices, 
$A$ and $B$, in general, would not commute inside $\str({\cdot})$: $\str(AB)\neq \str(BA)$.  
In order to satisfy $\str(AB)=\str(BA)$ only with $c$-number matrices, 
either $A$ or $B$ is forbidden to have $c$-number components 
in the off-diagonal blocks.  

For later convenience, we derive a useful property of pure canonical MPSs 
\cite{Perez-Garciaetal2007,OrusVidal2008PRB}.  In the following sections, we assume that 
the MPS in question is defined on an infinite-size system.  
The equation \eqref{eqn:U-Gamma-U-SUSY-2} is the most general statement 
about how a given symmetry of MPS is realized by a projective representation.  
However, sometimes it happens that $\Gamma^{\prime}(m) = \Gamma(m)$ for a certain 
symmetry operation and in these cases we can draw an interesting conclusion 
about the properties of $U$ \cite{Pollmannetal2010,Pollmannetal2012,Hasebe-Totsuka2012}.    
Suppose that we have a pure MPS whose canonical form is characterized by the MPS 
data \cite{Perez-Garciaetal2007,OrusVidal2008PRB} 
$\{\Lambda,\Gamma\}$ and that it satisfies the following relation for some unitary 
matrix $U$:
\begin{equation}
\Gamma(m) = \be^{i\theta_{U}} \, U^{\dagger} \Gamma(m) U \; .  
\label{eqn:VeqVGammaV}
\end{equation}
Since the MPS is canonical, the following holds [see eq.\eqref{eqn:canonical-iMPS-TL}]:
\begin{equation}
\sum_{m}\Gamma^{\dagger}(m)\Lambda^{2}\Gamma(m) = \mathbf{1}_{D} \; .
\label{eqn:canonical-condition-2}
\end{equation}
Physically, it states that the $D^{2}$-dimensional vector $\mathbf{V}_{\mathrm{L}}^{(1)}$ 
\begin{equation}
(\mathbf{V}_{\mathrm{L}}^{(1)})_{a,b} \equiv \delta_{ab} \quad 
(1 \leq a,b \leq D) 
\end{equation}
is the dominant left-eigenvector of the left transfer matrix (see Fig.~\ref{fig:canonical-cond})
\begin{equation}
(T_{\mathrm{L}})_{\bar{\alpha},\alpha;\bar{\beta},\beta} 
\equiv \sum_{m}(\Lambda\Gamma^{\ast}(m))_{\bar{\alpha}\bar{\beta}}
(\Lambda\Gamma(m))_{\alpha\beta} \; .
\end{equation}
Plugging   
$\Gamma^{\dagger}(m) = \be^{-i\theta_{U}} \, U^{\dagger} \Gamma^{\dagger}(m) U$ into 
(\ref{eqn:canonical-condition-2}), we obtain: 
\begin{equation}
\be^{-i\theta_{U}} \,\sum_{m} U^{\dagger} \Gamma^{\dagger}(m) U
\Lambda^{2}\Gamma(m)
=  \mathbf{1}_{D} \; , 
\end{equation}
or equivalently
\begin{equation}
\sum_{m} \Gamma^{\dagger}(m) \Lambda U
\Lambda\Gamma(m)
=\be^{i\theta_{U}} \,  U \; . 
\end{equation}
This implies that the $D{\times}D$ unitary matrix 
\begin{subequations}
\begin{equation}
(U)_{\bar{b}b}=\sum_{a}
\left\{ \mathbf{1}{\otimes} U \right\}_{aa;\bar{b}b}
\equiv \sum_{a}\delta_{a\bar{b}}(U)_{ab} \; ,
\end{equation}
when viewed as a $D^{2}$-dimensional 
vector, is the left-eigenvector of $T_{\mathrm{L}}$ with the eigenvalue $\be^{i\theta_{U}}$: 
\begin{equation}
UT_{\mathrm{L}} 
= \be^{i\theta_{U}}U  \; .
\end{equation}\label{lefteigentrans}
\end{subequations}
Since, by assumption of canonical MPS, $\mathbf{1}_{D}$ is the only left-eigenvector 
having the eigenvalue $|\lambda|=1$, we conclude 
\begin{equation}
\be^{i\theta_{U}} =1 \; , \;\; U = \be^{i \Phi_{U}} \mathbf{1}_{D} \; . 
\label{eqn:V-as-dominanteigenv}
\end{equation}
Since, in deriving the above, we have only assumed that the (infinite-system) MPS in question  
is pure and takes the canonical form, (\ref{eqn:V-as-dominanteigenv}) holds 
for any MPS (including SMPS) satisfying the assumption. 

\subsubsection{Inversion Symmetry}
\label{sec:inversion}
Now let us look at what inversion symmetry $\mathcal{I}$ with respect to a given link 
implies about the structure of the entanglement spectrum.  
It is convenient to consider an inversion transformation  on a circle:  
\begin{equation}
\Psi =\text{STr}(\mathcal{A}_1\mathcal{A}_2\cdots \mathcal{A}_{2n+1}). 
\end{equation}
The inversion on a given link transforms the SMPS chain as 
\begin{equation}
\mathcal{I} \Psi= \text{STr}(\mathcal{A}_{2n+1}\mathcal{A}_{2n}\cdots \mathcal{A}_{1}). 
\label{inversionpsi}
\end{equation}
By the relation  
$\text{STr}(M_1M_2)= \text{STr}((M_1M_2)^{\text{st}})= \text{STr}(M_2^{\text{st}}M_1^{\text{st}})$, 
(\ref{inversionpsi}) can be rewritten as 
\begin{equation}
\mathcal{I} \Psi= \text{STr}(\mathcal{A}_1^{\st}\mathcal{A}_2^{\st}\cdots \mathcal{A}_{2n+1}^{\st}), 
\end{equation}
where  `$\st$' stands for {\em supertranposition}  defined by  
\begin{equation}
\begin{pmatrix}
M_1 & N_1 \\
N_2 & M_2
\end{pmatrix}^{\st}\equiv 
\begin{pmatrix}
M_1^\text{t} & N_2^\text{t} \\
-N_1^\text{t} & M_2^\text{t}
\end{pmatrix}. \label{definitionsupertranspose}
\end{equation}
It is easy to verify that applying the supertrace twice does {\em not} return 
a supermatrix to the original one:
\begin{equation}
\left\{
\begin{pmatrix}
M_1 & N_1 \\
N_2 & M_2
\end{pmatrix}^{\st}\right\}^{\st}
= 
\begin{pmatrix}
M_1 & -N_1 \\
-N_2 & M_2 
\end{pmatrix} = P \begin{pmatrix}
M_1 & N_1 \\
N_2 & M_2 
\end{pmatrix} P
\; ,
\label{supertranspose-twice}
\end{equation}
where the matrix
\begin{equation}
P \equiv 
\begin{pmatrix}
\mathbf{1}_{1} & 0 \\
0 & -\mathbf{1}_{2}
\end{pmatrix}  
\label{eqn:def-P-matrix}
\end{equation}
has been defined in such a way that its adjoint action $P(\cdot)P$ multiples 
the fermionic blocks by a factor $(-1)$.  [$\mathbf{1}_{1}$ and $\mathbf{1}_{2}$ correspond to the unit matrices of two bosonic subspaces and  should not be confused with the $D \times D$ unit matrix $\mathbf{1}_{D}$.] 
 
Therefore, at the level of the local MPS matrix, the inversion $\mathcal{I}$ is realized as
\begin{equation}
\mathcal{A}_{j} \xrightarrow{\mathcal{I}} \mathcal{A}_{j}^{\text{st}} \; .
\end{equation}
Finally, from eq.\eqref{eqn:U-Gamma-U-SUSY-2}, one sees that 
when  $\Psi$ has the inversion symmetry, the $\mathcal{A}$-matrix should satisfy the relation 
\cite{Hasebe-Totsuka2012}
\begin{equation}
 \Gamma(m)^{\st}=e^{i\theta_{I}} U_{I}^{\dagger}\Gamma(m)U_{I}.  
\label{sinversionsymm2}
\end{equation}
Note that at this stage, we can not apply the argument in Sec.~\ref{sec:determining-U-SMPS} 
since eq.\eqref{sinversionsymm2} does not assume exactly the same form as 
eq.\eqref{eqn:VeqVGammaV} [the LHS is not $\Gamma(m)$].  

In order to obtain the relation of the form \eqref{eqn:VeqVGammaV}, 
we combine eq.\eqref{sinversionsymm2} with 
the fact that the link-inversion squared to unity $\mathcal{I}^{2}=1$.  
Applying supertransposition $(\cdot)^{\st}$ to  
(\ref{sinversionsymm2}) once again and using $(A^{\st})^{\st}=PAP$ 
[eq.\eqref{supertranspose-twice}], 
we obtain\footnote{%
In the bosonic case, the matrix $P$ is not necessary \cite{Pollmannetal2010,Pollmannetal2012}.}  
\begin{equation}
\Gamma(m) = \be^{2i\theta_{I}}\, (U_{I}P U_{I}^{\ast})^{\dagger}
\Gamma(m)\, (U_{I} P U_{I}^{\ast})  \; .
\end{equation}
This is of the form of eq.\eqref{eqn:VeqVGammaV} and 
then \eqref{eqn:V-as-dominanteigenv} immediately implies that 
\begin{equation}
(U_{I} P U_{I}^{\ast}) = \be^{i\Phi_{I}}\mathbf{1}_{D} \; , \;\;
\be^{2i\theta_{I}}=1 \Leftrightarrow 
\be^{i\theta_{I}} = \pm 1 \; .
\label{eqn:UI-P-UI}
\end{equation}    
After multiplying $U^{\text{t}}_{I}$ from the right and making transposition, we deduce 
\begin{equation}
U_{I} =  \be^{-i\Phi_{I}}P {U_{I}}^{\text{t}}
= \be^{-2i\Phi_{I}}P^{2} U_{I}
= \be^{-2i\Phi_{I}}U_{I}  \;  \Leftrightarrow \; \be^{-i\Phi_{I}} = \pm 1
\end{equation}
[note $PU_{I}=U_{I}P$ by eqs.\eqref{eqn:block-form-U} and \eqref{eqn:def-P-matrix}].  
Therefore, we obtain \cite{Hasebe-Totsuka2012}
\begin{equation}
U^{\text{t}}_{I} = \pm P U_{I}  \; .
\label{eqn:UI-transpose} 
\end{equation}
Eq.~\eqref{eqn:UI-transpose} states that when $U_{1}$ is symmetric (anti-symmetric), 
$U_{2}$ is anti-symmetric (symmetric).  
It should be noted that unlike in the bosonic case \cite{Pollmannetal2010,Pollmannetal2012}, 
the symmetry constraint is  imposed 
on each of the  spin-$S$ ``bosonic'' sector $U_1$ 
and $S-1/2$ ``fermionic'' sector $U_2$ in the case of SUSY.   

When $U_{1}$ is anti-symmetric, for instance, the sector with $(-1)^{F}=+1$ 
must have a special structure in its entanglement spectrum. 
In fact, by computing the determinant of $U_{1}$
\begin{equation}
\det U_{1} = \det U_{1}^{\text{t}}=\det(-U_{1}) = (-1)^{d_{1}}\det U_{1} \;\; 
(d_{1}: \text{ dimension of }U_{1}) \; ,
\end{equation}
one can immediately see that $d_{1}$ should be even (the same argument applies to 
$U_{2}$ as well). 
From this, one can conclude that 
either fermionic (when the sign + occurs) or bosonic ($-$) sector has even-fold degeneracy 
in each entanglement level, which we can use as the fingerprint 
\cite{GuWen2009,Pollmannetal2010,Pollmannetal2012} of 
the SUSY-protected topological order \cite{Hasebe-Totsuka2012}.  

\subsubsection{Time-Reversal Symmetry}
\label{sec:time-reversal}
We can draw a similar conclusion for the time-reversal symmetry $\mathcal{T}$. 
Let us recall the time-reversal operation for the superspin, $S_i$ and $S_{\alpha}$  
\begin{equation}
\begin{split}
&S_i \xrightarrow{\mathcal{T}} (e^{i\pi S_y} K)S_i (K e^{-i\pi S_y}) =-S_i,    \\
&S_{\alpha} \xrightarrow{\mathcal{T}} 
(e^{i\pi S_y} K)S_{\alpha} (K e^{-i\pi S_y})=\epsilon_{\alpha\beta}S_{\beta}, 
\end{split}
\label{timereversalsymss}
\end{equation}
with complex conjugation operator $K$. 
The set of equations (\ref{timereversalsymss}) implies that, 
for integer superspins $\mathcal{S}$, 
$\mathcal{T}=K e^{-i\pi S_y}$ satisfies\footnote{%
Recall that, in the $SU(2)$ case, ${\mathcal{T}}^{2}=+\mathbf{1}$ for integer $S$.  
When superspin $\mathcal{S}$ is half-odd-integer, the time reversal-operator satisfies  
$\mathcal{T}^{2} =-P$. (This is indeed a generalization of  
${\mathcal{T}}^{2}=-\mathbf{1}$ for the  $SU(2)$ half-integer spin case.) }
\begin{equation}
\mathcal{T}^{2} = \mathcal{P} = (-1)^{F} \; .
\end{equation}
where the $d$-dimensional matrix $\mathcal{P}$, which acts on the {\em physical} Hilbert 
space and multiples a minus sign when the state is fermionic ({\em i.e.} when 
the fermion number $F=\text{odd}$), 
is analogous to the $D$-dimensional matrix $P$ [see eq.\eqref{eqn:def-P-matrix}] 
acting on the {\em auxiliary} space.   

From (\ref{timereversalsymss}), 
one sees that, in terms of the canonical matrix, the time reversal transformation 
is represented as 
\begin{equation}
\Gamma(m) \xrightarrow{\mathcal{T}} 
\Gamma(m)'=\sum_{n}[R^{y}(\pi)]_{mn} \Gamma(n)^*,  
\end{equation}
where $m,n=i,\alpha$ and $R^{y}(\pi)=e^{i\pi S_y}$. 
For instance, $R^{y}(\pi)$ is given, for superspin $\mathcal{S}=1$, by   
\begin{equation}
R^y(\pi)=\begin{pmatrix}
-1 & 0 & 0 & 0 & 0 \\
0 & 1 & 0 & 0 & 0 \\
0 & 0 & -1 & 0 & 0 \\
0 & 0 & 0 & 0 & 1 \\
0 & 0 & 0 & -1 & 0 
\end{pmatrix}.
\end{equation}
Again by eq.\eqref{eqn:U-Gamma-U-SUSY-2}, 
the time-reversal invariance implies that there exists 
a unitary matrix $U_{{T}}$ that satisfies 
\begin{equation}
\Gamma(m) \xrightarrow{\mathcal{T}} 
\sum_n R_{mn}^y(\pi)\Gamma^*(n)=e^{i\theta_{T}}U_{{T}}^{\dagger}\Gamma(m)U_{{T}} \; .
\label{susytimereversal}
\end{equation}
Applying this twice and using $\mathcal{T}^{2}=\mathcal{P}$, we obtain
\begin{equation}
\begin{split}
& (\mathcal{P})_{ll} \,  \Gamma(l) =  
\sum_{m=1}^{d}[R^{y}(\pi)]_{lm} \left\{
\sum_{n=1}^{d}[R^{y}(\pi)]_{mn}\Gamma^{\ast}(n) \right\}^{\ast}  \\
& \phantom{(\mathcal{P})_{ll} \,  \Gamma(l)}
= \sum_{m=1}^{d}[R^{y}(\pi)]_{lm} \left\{
\be^{-i\theta_{T}} U^{\text{t}}_{T} \Gamma^{\ast}(m) U^{\ast}_{T}
\right\}   \\
& \phantom{(\mathcal{P})_{ll} \,  \Gamma(l) }
= \be^{-i\theta_{T}} U^{\text{t}}_{T} \left\{
\sum_{m=1}^{d}[R^{y}(\pi)]_{lm} \Gamma^{\ast}(m) 
\right\}
U^{\ast}_{T}  \\
& \phantom{(\mathcal{P})_{ll} \,  \Gamma(l) }
= \left\{ U_{T}U^{\ast}_{T} \right\}^{\dagger} 
\Gamma(l) 
\left\{U_{T}U^{\ast}_{T}\right\}  \\
& \Leftrightarrow  \quad 
\Gamma(l) = 
\left\{ U_{T}U^{\ast}_{T} \right\}^{\dagger} 
(\mathcal{P})_{ll} \, \Gamma(l) 
\left\{U_{T}U^{\ast}_{T}\right\}\; .
\end{split}
\label{eqn:TR-gamma-adjoint-SUSY}
\end{equation}
Using the relation $(\mathcal{P})_{ll} \Gamma(l) = P \, \Gamma(l) P$, 
we may rewrite \eqref{eqn:TR-gamma-adjoint-SUSY} into the following form
\begin{equation}
\Gamma(l) = 
\left\{ U_{T}PU^{\ast}_{T} \right\}^{\dagger} 
 \Gamma(l) 
\left\{U_{T}PU^{\ast}_{T}\right\} \; ,
\end{equation}
to which we can apply the argument presented in Sec.~\ref{sec:determining-U-SMPS} 
[see eq.\eqref{eqn:VeqVGammaV}].  
Thus we obtain
\begin{equation}
U_{T} P U^{\ast}_{T}  = \be^{i\Phi_{T}}\mathbf{1} \; .
\label{eqn:condition-U-TR-SUSY}
\end{equation}
(the value of $\theta_{T}$ can not be determined by this method).  
This is of exactly the same form as \eqref{eqn:UI-P-UI} and we deduce the same 
conclusion:
\begin{equation}
U_{{T}}^{t}=\pm PU_{{T}} \; , \;\;
\be^{i\Phi_{T}} = \pm 1 \; .
\label{eqn:UT-transpose}
\end{equation}
Therefore, as in the previous case [see eq.\eqref{eqn:UI-transpose}], 
we see that time-reversal symmetry guarantees 
the existence of at least double degeneracy in the entanglement spectrum of 
(either of the bosonic or the fermionic sector).  

\subsubsection{$\mathbb{Z}_2 \times \mathbb{Z}_2$ Symmetry}
\label{sec:Z2timesZ2}
Lastly, we consider the two independent $\pi$ rotations around the $x$ [$\hat{u}_{x}(\pi)$] 
and the $z$ axes [$\hat{u}_{z}(\pi)$].\footnote{%
Note that $\hat{u}_{y}(\pi)$ is redundant since 
$\hat{u}_{x}(\pi)\hat{u}_{z}(\pi)=\hat{u}_{z}(\pi)\hat{u}_{x}(\pi)=\hat{u}_{y}(\pi)$.}  
As will be seen below, $\hat{u}_{x}(\pi)$ or $\hat{u}_{z}(\pi)$ alone does not 
lead to any significant conclusion.  
However, the combination of the two \cite{Pollmannetal2010,Pollmannetal2012} 
leads to a similar conclusion about the entanglement spectrum. 
Under the $\pi$ rotation around the $x$ ($z$) axis $\hat{u}_{x}(\pi)$ 
($\hat{u}_{z}(\pi)$),  SMPS transforms as   
\begin{equation}
\Gamma(m) \xrightarrow{\hat{u}_{a}(\pi)}
\Gamma(m)'= \sum_n R_{mn}^{a}(\pi)\Gamma(n) \;\; (a=x,z) \; . 
\end{equation}
When the SMPS respects such a symmetry, we have 
\begin{equation}
\sum_n [R^{a}(\pi)]_{mn}\Gamma(n)
=\be^{i\theta_{a}}U_{a}^{\dagger}\Gamma(m)U_{a} 
\; ,
\label{eqn:U-Gamma-U-rotation}
\end{equation}
where $(R^{a}(\pi))^{2}=\mathcal{P}$ for integer superspin $\mathcal{S}$.   
Again we follow the same steps as before to show  
\begin{equation}
\begin{split}
&e^{2i\theta_a}=1 \Rightarrow e^{i\theta_a}=\pm 1, \\
&U_a P U_a =e^{i\Phi_a}\mathbf{1} \quad (a=x,z) \; . 
\end{split}
\end{equation}
Note that the phase factor $e^{i\Phi_{a}}$ can be absorbed in the definition of $U_{{a}}$ 
and hereafter we assume $U^{\dagger}_{a} = PU_{a}$ ($a=x,z$). 
Unlike in the previous cases, this relation alone does not give any useful information about $U_{a}$.   

On the other hand, 
the combination of the ``commutation relation'' 
$\hat{u}_{x}(\pi)\hat{u}_{z}(\pi)=\mathcal{P}\hat{u}_{z}(\pi)\hat{u}_{x}(\pi)$ 
and eq.\eqref{eqn:U-Gamma-U-rotation} implies, in terms of $\Gamma(m)$, 
\begin{equation}
\Gamma(m) = (U_{x}U_{z}U_{x}^{\dagger}PU_{z}^{\dagger})\Gamma(m) 
(U_{z}PU_{x}U_{z}^{\dagger}U_{x}^{\dagger})
\label{eqn:U-Gamma-U-for-UxUz}
\end{equation}
and hence gives 
\begin{equation}
(U_z P U_x)(U^{\dagger}_{z}U^{\dagger}_{x})=e^{i\Phi_{xz}}\mathbf{1}  \; .
\label{secondrelationforz2}
\end{equation}
In obtaining eq.\eqref{eqn:U-Gamma-U-for-UxUz}, the phase factor 
$\be^{i\theta_{xz}}=\be^{i\theta_{x}}\be^{i\theta_{z}}$ appears.  
However, it cancels out in the final expression \eqref{eqn:U-Gamma-U-for-UxUz}.  

Since the phases of $U_x$ and $U_z$ have been already fixed, 
the phase of $U_xU_z$ cannot be arbitrary and has a definite physical meaning. 
By multiplying $U_{z}P$ and $U_{x}$ from the left and the right of \eqref{secondrelationforz2}, 
respectively, and 
using $U_{a}^{\dagger} = PU_{a}=U_{a}P$ repeatedly, 
one can show
\begin{equation}
U_{x}U_{z} = \be^{i\Phi_{xz}} U_{z}PU_{x} \;, 
\end{equation}
which is combined with \eqref{secondrelationforz2} to give $\be^{i\Phi_{xz}}=\pm 1$.  
To summarize, the two unitary matrices $U_{x}$ and $U_{z}$ obey 
the following commutation relation:
\begin{equation}
U_x U_z  =\pm P U_{z}U_{x} \; . 
\label{eqn:UxUz-exchange-SUSY}
\end{equation}
In  terms of the block components $U_{a,1}$ and $U_{a,2}$,   
eq.(\ref{eqn:UxUz-exchange-SUSY}) reads as 
\begin{equation}
U_{x,1}U_{z,1} = \pm U_{z,1}U_{x,1} \; , \;\;
U_{x,2}U_{z,2} = \mp U_{z,2}U_{x,2} \; . 
\label{uxuzrelations}
\end{equation}
As in the previous cases, 
(\ref{uxuzrelations}) immediately implies at least double degeneracy of each entanglement 
levels in the sector (1 or 2) taking the minus sign in the above equation.    
\subsubsection{String Order Parameters and Entanglement Spectrum}
In the previous sections, we have shown 
that, as in the purely bosonic cases \cite{GuWen2009,Pollmannetal2010,Pollmannetal2012}, 
such elementary symmetries as inversion and time-reversal protect the Haldane phase 
from collapsing into trivial gapped phases 
on the basis of the assumption that the even-fold degeneracy in 
the entanglement spectrum either in the fermionic or in the bosonic sector is 
the entanglement fingerprint of the topological ``Haldane phase'' in SUSY systems.    
On the other hand, the string order parameters [see Sec.~\ref{sec:string-ordinary-VBS}] 
have been used traditionally to characterize the Haldane phase 
\cite{NijsRommelse-1989,Tasaki1991}.  
Then, a natural question arises: is there any connection between the description by 
the string order parameters and the modern characterization in terms of the entanglement spectrum? 
In this section, we give the answer to this question.  

Let us first consider the structure of the string order parameters 
${\cal O}_{\text{string}}^{x,\infty}$ and ${\cal O}_{\text{string}}^{z,\infty}$
[eq.\eqref{stringorderquan}] from the MPS viewpoint 
\cite{Totsuka-Suzuki1995,Garcia-W-S-V-C-08}.  
When we evaluate them using MPSs, we encounter the following matrices 
\cite{Totsuka-Suzuki1995}: 
\begin{equation}
\begin{split}
& [T^{a}]_{\bar{\alpha},\alpha;\bar{\beta},\beta} \equiv \sum_{m,n=1}^{d}
\left[A^{\ast}(m)\right]_{\bar{\alpha},\bar{\beta}}
\left[A(n)\right]_{\alpha,\beta} 
\langle m|S^{a}|n\rangle \quad (a=x,z) \\
& [T_{\text{string}}]_{\bar{\alpha},\alpha;\bar{\beta},\beta} \equiv \sum_{m,n=1}^{d}
\left[A^{\ast}(m)\right]_{\bar{\alpha},\bar{\beta}}
\left[A(n)\right]_{\alpha,\beta} 
\langle m|\exp(i \pi S^{a})|n\rangle   \\
& [T_{\text{string}}^{a}]_{\bar{\alpha},\alpha;\bar{\beta},\beta} \equiv \sum_{m,n=1}^{d}
\left[A^{\ast}(m)\right]_{\bar{\alpha},\bar{\beta}}
\left[A(n)\right]_{\alpha,\beta} 
\langle m|S^{a}\exp(i \pi S^{a})|n\rangle \quad (a=x,z)
\end{split}
\end{equation}
as well as the usual transfer matrix $T$.  
By using these matrices, the MPS expression of the string order parameter 
$\mathcal{O}^{z}_{\text{string}}$ is given as
\begin{equation}
T^{N_{\text{L}}}T_{\text{string}}^{z} (T_{\text{string}})^{|i-j|}T^{z}\,T^{N_{\text{R}}}  \; ,
\end{equation}
where we have omitted the denominator necessary to normalize the MPS.  
The parts $T^{N_{\text{L}}}$ and $T^{N_{\text{R}}}$ are easy; 
for the canonical MPS, they reduces, in the infinite-size limit, to 
(see Sec.~\ref{sec:Schmidt-MPS}):
\begin{equation}
[T^{N_{\text{L}}}]_{\bar{\alpha}_{\text{L}},\alpha_{\text{L}};\bar{\beta},\beta}
 \xrightarrow{N_{\text{L}}\nearrow \infty} 
 \delta_{\bar{\alpha}_{\text{L}},\alpha_{\text{L}}}
\delta_{\bar{\beta},\beta} 
\quad , \quad 
[T^{N_{\text{R}}}]_{\bar{\alpha},\alpha;\bar{\beta}_{\text{R}},\beta_{\text{R}}}
 \xrightarrow{N_{\text{R}}\nearrow \infty} 
 \delta_{\bar{\alpha},\alpha} \delta_{\bar{\beta}_{\text{R}},\beta_{\text{R}}} \; .
\end{equation}
The boundary dependent factors $\delta_{\bar{\alpha}_{\text{L}},\alpha_{\text{L}}}$ and 
$\delta_{\bar{\beta}_{\text{R}},\beta_{\text{R}}}$ are cancelled by those coming from 
the denominator.  Therefore, the quantity which we have to compute is
\begin{equation}
\sum_{\alpha,\beta}\left[ T_{\text{string}}^{z} 
(T_{\text{string}})^{|i-j|}T^{z} \right]_{\alpha,\alpha;\beta,\beta}  \; .
\label{eqn:Ostring-by-Ts}
\end{equation}

Since we are interested in the long-distance limit $|i-j|\nearrow \infty$, 
we need know the asymptotic behavior of the string $(T_{\text{string}})^{|i-j|}$.  
First we note that $T_{\text{string}}$ may be thought of as the overlap
\begin{equation}
\langle \text{MPS}|\hat{u}\otimes \hat{u} \otimes \cdots \otimes \hat{u}|\Psi_{\text{MPS}}\rangle
\end{equation}
with $\hat{u}= \exp(i \pi S^{a})$ ($a=x,z$) [see eq.\eqref{eqn:def-unitary-invariance}]. 
Then, it can be shown \cite{Garcia-W-S-V-C-08} that 
in order for the string $(T_{\text{string}})^{|i-j|}$ not to vanish in the long-distance limit, 
the MPS should be invariant under both 
of the $\pi$-rotations $\hat{u}_{x}(\pi)$ and $\hat{u}_{z}(\pi)$.  
  
Then, from the discussion in Sec.\ref{sec:Z2timesZ2} (we just replace $P \mapsto 1$ to obtain 
the results for the purely bosonic case), we see that there exists a pair of unitary matrices 
satisfying \eqref{eqn:U-Gamma-U-rotation} and 
\begin{equation}
U_x U_z  =\pm U_{z}U_{x}  
\label{eqn:UxUz-exchange}
\end{equation}
[see eq.\eqref{eqn:UxUz-exchange-SUSY}].  
Unfortunately, we cannot tell whether the even-fold degeneracy in the entanglement spectrum, 
which is the fingerprint of the topological Haldane phase, happens 
or not since we do not know which of the $\pm$ signs is chosen.  

\begin{figure}[h]
\begin{center}
\includegraphics[scale=0.5]{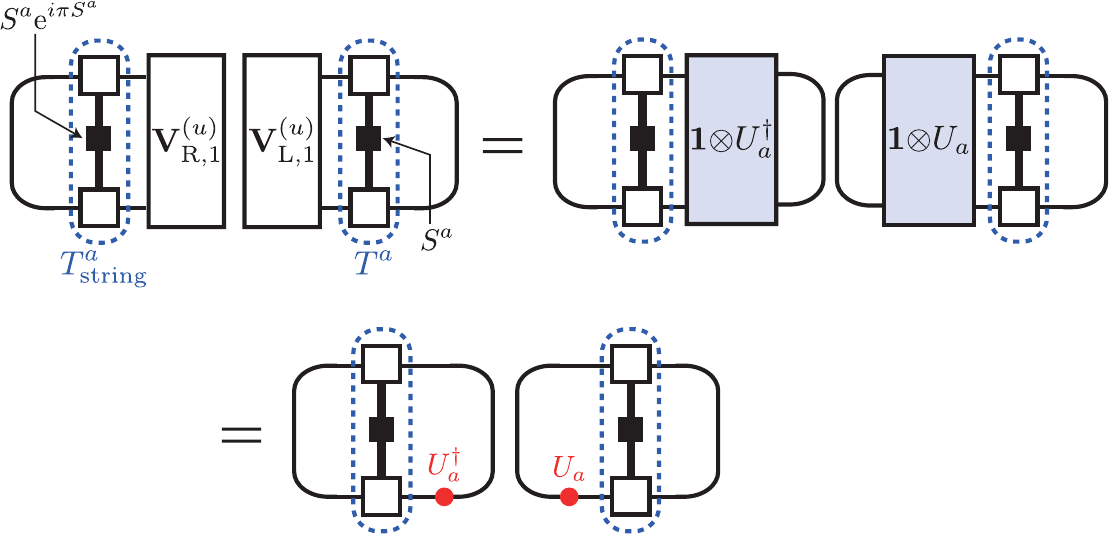}
\end{center}
\caption{(Color online) Diagrammatic representation of the main part of 
string correlation function $\left\{
(T_{\text{string}}^{z} \mathbf{V}^{(u)}_{\text{R},1})
(\mathbf{V}^{(u)}_{\text{L},1}T^{z})
\right\}$. 
$\mathbf{V}^{(u)}_{\text{L,R},1}$ denotes the dominant eigenvector of 
$T_{\text{string}}$. (Figure has been adapted from Ref.~\cite{Hasebe-Totsuka2012}.) 
\label{fig:string-in-MPS}}
\end{figure}

Now we show that when the string order parameters 
are non-vanishing $\mathcal{O}_{\text{string}}^{z,x}\neq 0$, 
the minus sign in fact realizes ({\em i.e.} $U_{x}$ and $U_{z}$ anti-commute) 
in eq.\eqref{eqn:UxUz-exchange} and the entanglement spectrum exhibits  
the degenerate structure. 
To this end, we investigate eq.(\ref{eqn:Ostring-by-Ts}). 
Since the largest (right)eigenvalue of $T_{\text{string}}$ is $\be^{i\theta_{a}}$ 
[see Sec.~\ref{sec:symmetry-in-MPS}], $(T_{\text{string}})^{|i-j|}$ reduces essentially to 
a phase $(\be^{i\theta_{a}})^{|i-j|}=(\pm 1)^{|i-j|}$.  
The price to pay is the following boundary factors appearing at the two end points 
of the string correlators (see Fig.\ref{fig:string-in-MPS}):
\begin{equation}
\begin{split}
& \sum_{\alpha,\beta}\left\{T_{\text{string}}^{z} 
\left(
\sum_{n=1}^{D^{2}}\mathbf{V}^{(u)}_{\text{R},n}\mathbf{V}^{(u)}_{\text{L},n}
\right)
(T_{\text{string}})^{|i-j|}T^{z}\right\}_{\alpha,\alpha;\beta,\beta} \\
& \quad 
\xrightarrow{|i-j|\nearrow \infty}
\sum_{\alpha,\beta}\left\{
(T_{\text{string}}^{z} \mathbf{V}^{(u)}_{\text{R},1})
(\mathbf{V}^{(u)}_{\text{L},1}T^{z})
\right\}_{\alpha,\alpha;\beta,\beta}  
= \sum_{\alpha,\beta}\left\{
(T_{\text{string}}^{z} \left\{ \mathbf{1}{\otimes} U^{\dagger}_{z}\right\}\mathbf{1})
(\mathbf{1}\left\{ \mathbf{1}{\otimes} U_{z}\right\}T^{z})
\right\}_{\alpha,\alpha;\beta,\beta}  \; ,
\end{split}
\label{eqn:strnig-OP2boundary}
\end{equation}
where $\mathbf{V}^{(u)}_{\text{L},1}$ ($\mathbf{V}^{(u)}_{\text{R},1}$) denotes the left (right) 
dominant eigenvector of $T_{\text{string}}$ and the tensor notations are defined as
\begin{equation}
\begin{split}
& \left[ \mathbf{1}{\otimes}B\right]_{\bar{\alpha},\alpha;\bar{\beta},\beta}
= \delta_{\bar{\alpha},\bar{\beta}}\left[B\right]_{\alpha\beta}\; , \\
& \left[ \mathbf{1}\left\{ \mathbf{1}{\otimes}B\right\}\right]_{\bar{\beta},\beta}
= \sum_{\bar{\alpha},\alpha} \delta_{\bar{\alpha},\alpha}
\delta_{\bar{\alpha},\bar{\beta}}\left[B\right]_{\alpha\beta} 
= [B]_{\bar{\beta}\beta} \; , \;\;
\left[\left\{ \mathbf{1}{\otimes}B\right\} \mathbf{1}\right]_{\bar{\alpha},\alpha}
= \sum_{\bar{\beta},\beta} 
\delta_{\bar{\alpha},\bar{\beta}}\left[B\right]_{\alpha\beta} \delta_{\bar{\beta},\beta}
= [B^{\text{t}}]_{\bar{\alpha}\alpha} \; .
\end{split}
\end{equation}
In fact, the eigenvalues of $T_{\text{string}}$ precisely coincides with 
those of $T$ for all the spin-$S$ VBS states and the different behaviors in the string order 
for the odd-$S$ and the even-$S$ chains comes {\em only} from 
these boundary factors. 

\begin{figure}[h]
\begin{center}
\includegraphics[scale=0.6]{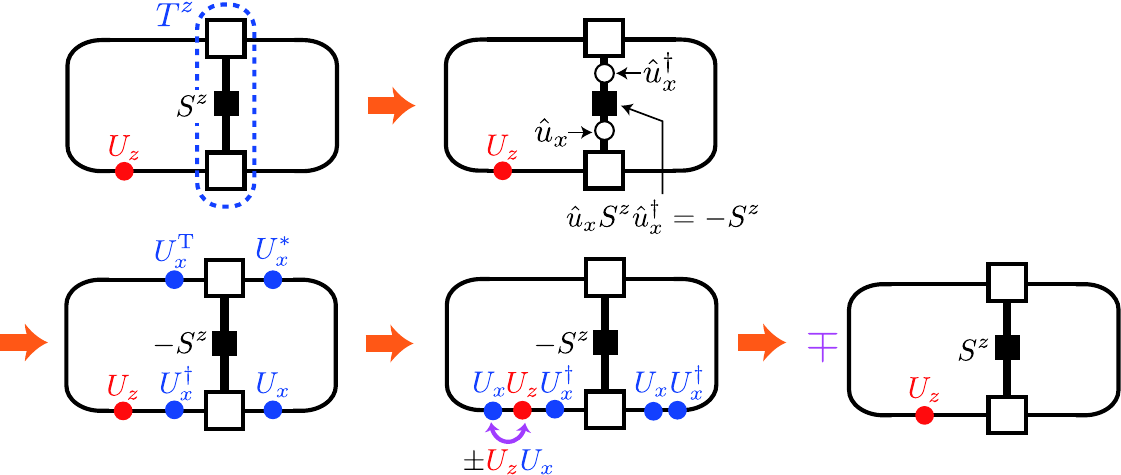}
\end{center}
\caption{(Color online) Rewriting the boundary factor 
$\mathbf{1}\left\{ \mathbf{1}{\otimes} U_{z}\right\}T^{z}$  
(for $a=z$) using $\hat{u}_{x}$ [see eq.\eqref{eqn:rewrite-boundary-factor}]. 
When $U_{x}$ and $U_{z}$ are anti-commuting, the minus sign coming from 
$\hat{u}_{x}S^{z}\hat{u}^{\dagger}_{x}=-S^{z}$ is cancelled and an overall 
plus sign is recovered. (Figure has been adapted from Ref.~\cite{Hasebe-Totsuka2012}.)
\label{fig:boundary-factor}}
\end{figure}

To see whether the boundary factors are non-vanishing or not, 
we consider the right-boundary factor 
$(\mathbf{1}\left\{ \mathbf{1}{\otimes} U_{z}\right\}T^{z})$ 
of $\mathcal{O}^{z}_{\text{string}}$.  
First we rewrite it by using (see the second figure of Fig.\ref{fig:boundary-factor}):
\begin{equation}
S^{z} = (\hat{u}^{\dagger}_{x}\hat{u}_{x})S^{z} (\hat{u}_{x}^{\dagger}\hat{u}_{x})
= \hat{u}^{\dagger}_{x}(-S^{z})\hat{u}_{x}  \quad 
(\hat{u}_{x}=\be^{-i\pi S^{x}})  \; .
\end{equation}
The unitary operators $\hat{u}^{\dagger}_{x}$ and $\hat{u}_{x}$ 
appearing on both sides of $-S^{z}$ can be absorbed into the MPS matrices 
by using eq.\eqref{eqn:U-Gamma-U-rotation} (the third figure of Fig.\ref{fig:boundary-factor}). 
By re-arranging the unitary matrices $U_{x}U_{z}$ with the help of 
eq.\eqref{eqn:UxUz-exchange} (the fourth figure of 
Fig.\ref{fig:boundary-factor}), we arrive at the expression:
\begin{equation}
\begin{split}
\mathbf{1}\left\{ \mathbf{1}{\otimes} U_{z}\right\}T^{z}
& = \mathbf{1}\left\{ \mathbf{1}{\otimes} (U_{x}U_{z}U_{x}^{\dagger})\right\}(-T^{z}) \\
&= \mathbf{1}\left\{ \mathbf{1}{\otimes} (
\pm U_{z}U_{x}U_{x}^{\dagger})\right\}(-T^{z}) 
= \mp \mathbf{1}\left\{ \mathbf{1}{\otimes} U_{z}\right\}T^{z} \; .
\end{split}
\label{eqn:rewrite-boundary-factor}
\end{equation}
Therefore, we see that the boundary factor $(\mathbf{1}\left\{ \mathbf{1}{\otimes} U_{z}\right\}T^{z})$ 
vanishes (and so does the string order parameter ${\cal O}_{\text{string}}^{z,\infty}$) when 
$U_{x}$ and $U_{z}$ are commuting ({\em i.e.} when the minus sign in the last expression is chosen). 
By the explicit construction of $U_{x}$ and $U_{z}$, we can easily see that  
for the even-$S$ VBS state, $U_{x}U_{z}=+U_{z}U_{x}$ holds.  
Therefore, it immediately results that the string order parameters identically vanish 
for the even-$S$ VBS state {\em solely for a symmetry reason}. 

In short, the existence of non-vanishing string order parameters 
${\cal O}_{\text{string}}^{x,\infty}\neq 0$ and ${\cal O}_{\text{string}}^{z,\infty}\neq 0$ 
implies that the existence of the two {\em anti-commuting} unitary matrices $U_{x}$ and $U_{z}$,  
and thereby guarantees the even-fold degeneracy in the entanglement spectrum.  
To put it another way, the string order parameters work as the {\em sufficient} condition 
for the topological Haldane phase.  
It is crucial that both of the string order parameters are finite. 
For instance, if we ``deform'' the original $SU(2)$-invariant VBS state by using the quantum group 
$U_{q}(su(2))$, we obtain a VBS state with uniaxial anisotropy \cite{Klumperetal1992}, 
where one of the string order parameters (${\cal O}_{\text{string}}^{x,\infty}\neq 0$) vanishes 
while the other is still finite \cite{Totsuka-Suzuki1994}. 
In this case, the unitary $U_{x}$ does not exist and the existence of the degenerate structure 
in the entanglement spectrum is no longer guaranteed.  In fact, explicit calculation shows 
that the two-fold degenerate entanglement levels in the $S=1$ VBS state split into two 
non-degenerate levels in the deformed state.  

It is straightforward to generalize \cite{Hasebe-Totsuka2012} this to SUSY cases by taking into account 
the appearance of the $P$-matrix. 
Again by using 
$\hat{u}_{x}(\pi)S^{z} \hat{u}_{x}^{\dagger}(\pi)= -S^{z}$ and $U_{x}U_{z} = \pm PU_{z}U_{x}$, 
the second factor of the right-hand side of eq.\eqref{eqn:strnig-OP2boundary} 
can be recasted as:
\begin{equation}
\begin{split}
\mathbf{1}\left\{ \mathbf{1}{\otimes} U_{z}\right\}T^{z}
& = \mathbf{1}\left\{ \mathbf{1}{\otimes} (U_{x}U_{z}U_{x}^{\dagger})\right\}(-T^{z}) \\
&= \mathbf{1}\left\{ \mathbf{1}{\otimes} (
\pm P U_{z}U_{x}U_{x}^{\dagger})\right\}(-T^{z}) 
= \mp \mathbf{1}\left\{ \mathbf{1}{\otimes} P U_{z}\right\}T^{z} \; .
\end{split}
\end{equation}
Since this implies 
\begin{equation}
\begin{split}
\sum_{\beta}\left\{
\mathbf{1}\left\{ \mathbf{1}{\otimes} U_{z}\right\}T^{z}\right\}_{\beta,\beta}
&=
\sum_{\alpha,\beta}\left\{
\left\{ \mathbf{1}{\otimes} U_{z}\right\}T^{z}\right\}_{\alpha,\alpha;\beta,\beta} \\
&=
\sum_{\alpha\in\text{B}}\;
\raisebox{-4.0ex}{\includegraphics[scale=0.5]{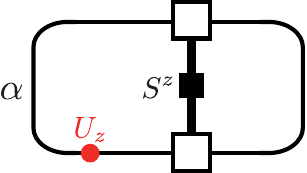}}
+ \sum_{\alpha\in\text{F}}\;
\raisebox{-4.0ex}{\includegraphics[scale=0.5]{boundary-SUSY-B}} \\
&=
\mp \sum_{\alpha\in\text{B}}\;
\raisebox{-4.0ex}{\includegraphics[scale=0.5]{boundary-SUSY-B}}
\pm \sum_{\alpha\in\text{F}}\;
\raisebox{-4.0ex}{\includegraphics[scale=0.5]{boundary-SUSY-B}}
\; ,
\end{split}
\end{equation}
we see that one of the two components (bosonic or fermionic) vanishes 
just by symmetry:
\begin{equation}
\raisebox{-4.0ex}{\includegraphics[scale=0.5]{boundary-SUSY-B}}
= 
\begin{cases}
\sum_{\alpha \in \text{F}} \;\; 
\raisebox{-4.0ex}{\includegraphics[scale=0.5]{boundary-SUSY-B}} & 
\text{when } \be^{i\Phi_{xz}}=+1 \\
\sum_{\alpha \in \text{B}} \;\; 
\raisebox{-4.0ex}{\includegraphics[scale=0.5]{boundary-SUSY-B}} &
\text{when } \be^{i\Phi_{xz}}=-1  \; .
\end{cases}
\end{equation}

In conclusion, we have established the connection between 
the string order parameters, which have been commonly used 
\cite{Hatsugai-K-91,Hatsugai1992,Alcaraz-H-92,Yamamoto-M-93,Nishiyama-T-H-S-95} 
to characterize the Haldane phase, and the entanglement spectrum \cite{LiHaldane2008} 
which is a modern tool to look at the topological properties in the bulk.  
As has been shown in Secs.~\ref{sec:inversion}--\ref{sec:Z2timesZ2}, SUSY 
guarantees, {\em regardless of} the parity of the bulk superspin $\mathcal{S}$, 
the existence of the even-fold degeneracy in the entanglement spectrum 
of either of the bosonic or the fermionic sector, 
provided at least one of the three symmetries (inversion, time-reversal and 
$\mathbb{Z}_2 \times \mathbb{Z}_2$) is present.   
Therefore, in contrast to the usual bosonic ({\em i.e.} $SU(2)$) VBS states, 
in which the Haldane phase is stable only for odd-integer spins,  
in the SVBS states, SUSY plays the crucial role in protecting the topological Haldane phase. 
The ``revival'' of the string order parameters upon doping the system (discussed in 
Sec.~\ref{sec:stringOP-in-SVBS}) may be naturally understood by the above argument. 

As has been seen above, that both of the two string order parameters are non-vanishing 
is the sufficient condition for the symmetry-protected Haldane phase.  
However, one may seek more faithful order parameters and, in fact, some results have been 
obtained in this direction.   
See, for instance, Refs. \cite{Haegeman-G-C-S-12} and \cite{Pollmann-T-12} 
for recent attempts at finding better order parameters.  
\section{Higher Symmetric Generalizations}\label{SecthigherDfuzzy}

In this section, we extend the previous SUSY formulation to the higher symmetric $UOSp(1|4)$ case.  
We begin with the construction of fuzzy four-supersphere based on the $UOSp(1|4)$ algebra \cite{Hasebe2011}. 
As the coordinates of fuzzy four-sphere correspond to the $SO(5)$ gamma matrices \cite{hep-th/9602115},  the coordinates of the fuzzy four-supersphere are constructed from ``super gamma matrices'' of the $UOSp(1|4)$ algebra. 
Next, based on the $UOSp(1|4)$ structure, we derive the  $UOSp(1|4)$ SVBS state \cite{Hasebe-Totsuka2012} whose bosonic counterpart  is the $SO(5)$ VBS state \cite{Tuetal2008I,Tuetal2008II,Tuetal2009}\footnote{ There still exist the correspondences among fuzzy geometry, QHE, and VBS in the higher dimension case.   
The $SO(5)$ VBS corresponds to the Laughlin-Haldane wave function in 4D QHE \cite{ZhangHu2001,ZhangPRL2002}, and 4D QHE realizes the fuzzy geometry of fuzzy four-sphere. }.   
We also develop the SMPS formalism for the $UOSp(1|4)$ SVBS state and investigate the topological properties.

\subsection{Fuzzy Four-supersphere}\label{subsecn1fuzzyfoursphere}

Here, we utilize $UOSp(1|4)$ algebra to construct fuzzy four-supersphere with $\mathcal{N}=1$ SUSY. 
It may be worthwhile to first point out a nice correspondence between algebras and fuzzy spheres.  For fuzzy two-sphere, we have  
\begin{equation}
SU(2)\simeq USp(2)~~\rightarrow ~~S^2_{\mathrm{f}},~~~~~~~UOSp(1|2)~~\rightarrow ~~S^{2|2}_{\mathrm{f}},
\end{equation}
and for fuzzy four-sphere,  
\begin{equation}
SO(5)\simeq USp(4)~~\rightarrow ~~S^{4}_{\mathrm{f}},~~~~~~~UOSp(1|4)~~\rightarrow ~~S^{4|2}_{\mathrm{f}}. 
\end{equation}
The $UOSp(1|4)$ algebra is constituted of fourteen generators: 
\begin{equation}
\dim[uosp(1|4)]=10|4=14,  
\end{equation}
ten of which are the bosonic generators  $\Gamma_{ab}=-\Gamma_{ba}$ $(a,b=1,2,\cdots,5)$ (the $SO(5)$ generators) and the remaining four are the fermionic ones $\Gamma_{\alpha}$ $(\alpha=1,2,3,4)$ (the $SO(5)$  spinor).   
They amount to satisfy the following algebra:  
\begin{align}
&[\Gamma_{ab},\Gamma_{cd}]=i(\delta_{ac}\Gamma_{bd}-\delta_{ad}\Gamma_{bc}-\delta_{bc}\Gamma_{ad}+\delta_{bd}\Gamma_{ac}),\nonumber\\
&[\Gamma_{ab},\Gamma_{\alpha}]=({\gamma}_{ab})_{\beta\alpha}\Gamma_{\beta},\nonumber\\
&\{\Gamma_{\alpha},\Gamma_{\beta}\}=\sum_{a<b}(C{\gamma}_{ab})_{\alpha\beta}\Gamma_{ab},\label{3rdalgebra}
\end{align}
where  $C$ is the $SO(5)(\simeq USp(4))$ charge conjugation matrix 
\begin{equation}
C=\mathcal{R}_{4}=
\begin{pmatrix}
 i\sigma_2 & 0\\
 0 & i\sigma_2\\
\end{pmatrix}, 
\label{so5chargeconjmat}
\end{equation}
and $\gamma_{ab}$ are the $SO(5)$ matrices.  
In the following discussion, we take the $SO(5)$ matrices as 
\begin{align}
&\gamma_{12}=
\frac{1}{2}
\begin{pmatrix}
\sigma_3 & 0\\
 0 & \sigma_3
\end{pmatrix}
,~~~~~~~~~
\gamma_{13} =
\frac{1}{2}
\begin{pmatrix}
-\sigma_2  & 0 \\
0  &-\sigma_2 
\end{pmatrix}
,~~~~\gamma_{14}=
\frac{1}{2}
\begin{pmatrix}
\sigma_1 & 0\\
0 & -\sigma_1
\end{pmatrix}
,\nonumber\\
&\gamma_{15} =
\frac{1}{2}
\begin{pmatrix}
0 & -\sigma_1\\
-\sigma_1 & 0 
\end{pmatrix}
,~~~~~\gamma_{23} =
\frac{1}{2}
\begin{pmatrix}
\sigma_1 & 0\\
0 & \sigma_1
\end{pmatrix},~~~~~~~~
\gamma_{24} =
\frac{1}{2}
\begin{pmatrix}
\sigma_2 & 0 \\
0 & -\sigma_2
\end{pmatrix}
,\nonumber\\
&\gamma_{25} =
\frac{1}{2}
\begin{pmatrix}
0 & -\sigma_2\\
-\sigma_2 & 0
\end{pmatrix}
,~~~~~
\gamma_{34} =
\frac{1}{2}
\begin{pmatrix}
\sigma_3  & 0\\
 0 & -\sigma_3
\end{pmatrix}
,~~~~~~\gamma_{35} =
\frac{1}{2}
\begin{pmatrix}
 0 & -\sigma_3\\
 -\sigma_3&0 
\end{pmatrix}
,\nonumber\\
&\gamma_{45} =
\frac{1}{2}
\begin{pmatrix}
 0 & i \mathbf{1}_2\\
 -i \mathbf{1}_2& 0
\end{pmatrix}
.\label{so5geneI1}
\end{align}
The $UOSp(1|4)$ quadratic Casimir is given by  
\begin{equation}
\mathcal{K}=\sum_{a<b}\Gamma_{ab}\Gamma_{ab}+C_{\alpha\beta}\Gamma_{\alpha}\Gamma_{\beta}.  
\label{osp14quadra}
\end{equation}
The fundamental 5-dimensional representation matrices of $uosp(1|4)$ are constructed as follows. 
From the (bosonic)  ``gamma matrices'' of $UOSp(1|4)$ algebra 
\begin{equation}
\Gamma_a=
\begin{pmatrix}
\gamma_a & 0 \\
 0 & 0 
\end{pmatrix} ,    
\label{gammauosp14}
\end{equation}
with 
\begin{align}
&\gamma_1=
\begin{pmatrix}
0 & i\sigma_1\\
-i\sigma_1 & 0
\end{pmatrix}
,~~\gamma_2=
\begin{pmatrix}
0 & i\sigma_2\\
-i\sigma_2 & 0
\end{pmatrix}
,~~\gamma_3=
\begin{pmatrix}
0 & i\sigma_3\\
-i\sigma_3 & 0
\end{pmatrix},\nonumber\\
&\gamma_4=
\begin{pmatrix}
0 & 1_2\\
1_2 & 0
\end{pmatrix}
,~~~~~~~\gamma_5=
\begin{pmatrix}
1_2 & 0\\
0 & -1_2
\end{pmatrix}, 
\label{so5gammaI1}
\end{align}
we can derive the $SO(5)$ generators 
\begin{equation}
\Gamma_{ab}=-i\frac{1}{4}[\Gamma_a,\Gamma_b]=
\begin{pmatrix}
 {\gamma}_{ab} & 0\\
 0 & 0
\end{pmatrix}. 
\label{gammablarge}
\end{equation}
The fermionic ``gamma matrices'' are also constructed as  
\begin{equation}
\Gamma_{\alpha}=\frac{1}{\sqrt{2}} 
\begin{pmatrix}
0_4 & \tau_{\alpha} \\
-(C\tau_{\alpha})^t & 0 
\end{pmatrix}, \label{fermimat14}
\end{equation}
where 
\begin{align}
\tau_1=
\begin{pmatrix}
1\\
0\\
0\\
0
\end{pmatrix},~~
\tau_2=
\begin{pmatrix}
0\\
1\\
0\\
0
\end{pmatrix},~~
\tau_3=
\begin{pmatrix}
0\\
0\\
1\\
0
\end{pmatrix},~~
\tau_4=
\begin{pmatrix}
0\\
0\\
0\\
1
\end{pmatrix}. 
\label{taucomponents}
\end{align}
They satisfy the ``hermiticity'' condition 
\begin{equation}
\Gamma_a^{\ddagger}=\Gamma_a,\quad\quad \Gamma_{ab}^{\ddagger}=\Gamma_{ab},\quad\quad \Gamma_{\alpha}^{\ddagger}=C_{\alpha\beta}\Gamma_{\beta}, 
\end{equation}
where the super-adjoint $\ddagger$ is defined in (\ref{adjointofosp12generators}).  
It is straightforward to check that (\ref{gammablarge}) and (\ref{fermimat14}) satisfy the $UOSp(1|4)$ algebra (\ref{3rdalgebra}). 
In the similar manner to the case of $S_{\mathrm{f}}^{2|2}$ (\ref{Schwingerconstrfuzzytwosphere}),  we introduce the coordinates of $S_{\mathrm{f}}^{4|2}$, $X_a$ and $\Theta_{\alpha}$, as 
\begin{equation}
X_a=\frac{R}{d}
\Psi^{\dagger}
\Gamma_a 
\Psi,~~~~ 
\Theta_{\alpha}= \frac{R}{d}\Psi^{\dagger} 
\Gamma_{\alpha} \Psi,
\label{bosonicscwingern1four}
\end{equation}
where  $\Psi$ is the $UOSp(1|4)$ Schwinger operator:  
\begin{equation}
\Psi=
(b^1, b^2, b^3, b^4, f)^t.   
\label{uosp14spinorschwinger}
\end{equation}
Here, $b^{\alpha}$ $(\alpha=1,2,3,4)$ and $f$ respectively denote four bosonic  and one fermionic operators that satisfy 
\begin{equation}
[b^{\alpha},{b^{\beta}}^{\dagger}]=\delta_{\alpha\beta},~~~~~\{f,f^{\dagger}\}=1,~~~~~[b^{\alpha},b^{\beta}]=[b^{\alpha},f]=\{f,f\}=0. 
\end{equation}
Square of the radius of  fuzzy four-supersphere is also readily derived as  
\begin{equation}
 X_aX_a+2C_{\alpha\beta}\Theta_{\alpha}\Theta_{\beta}=\biggl(\frac{R}{d}\biggr)^2(\Psi^{\dagger}\Psi)(\Psi^{\dagger}\Psi+3).  \label{n1fuzzy4dradius}
\end{equation}
In the Schwinger formalism, the Casimir (\ref{osp14quadra}) is represented as 
\begin{equation}
\mathcal{K}=\sum_{a<b}X_{ab}X_{ab}+C_{\alpha\beta}\Theta_{\alpha}\Theta_{\beta}=\frac{1}{2}\biggl(\frac{R}{d}\biggr)^2(\Psi^{\dagger}\Psi)(\Psi^{\dagger}\Psi+3),  
\label{casimirosp14fulre}
\end{equation}
where 
\begin{equation}
{X}_{ab}= \frac{R}{d}\Psi^{\dagger}  
\Gamma_{ab}  \Psi. 
\end{equation}
Notice that  in the Schwinger operator formalism the Casimir (\ref{casimirosp14fulre}) is equivalent to (\ref{n1fuzzy4dradius}) up to $1/2$ coefficient on the r.h.s. The basis states on fuzzy four-supersphere are given by the graded fully symmetric representation: 
\begin{subequations}
\begin{align}
&|n_1,n_2,n_3,n_4\rangle =\frac{1}{\sqrt{n_1!
~n_2!~n_3!~n_4!}}
 {{b^1}^{\dagger}}^{n_1} {{b^2}^{\dagger}}^{n_2} {{b^3}^{\dagger}}^{n_3} {{b^4}^{\dagger}}^{n_4}\vac,\label{uosp14fullysymbosonstates}\\ 
&|m_1,m_2,m_3,m_4)=\frac{1}{\sqrt{m_1!~m_2!~m_3!~m_4!}} 
{{b^1}^{\dagger}}^{m_1} {{b^2}^{\dagger}}^{m_2} {{b^3}^{\dagger}}^{m_3} {{b^4}^{\dagger}}^{m_4}f^{\dagger}\vac,
\label{uosp14fullysymfermistates}
\end{align}\label{uosp14fullysymstates}
\end{subequations}
where $n_1, n_2, \cdots, m_4$ are all non-negative integers satisfying the constraint,     $n_1+n_2+n_3+n_4=m_1+m_2+m_3+m_4+1=n \equiv \Psi^{\dagger}\Psi$.   
Therefore, the dimensions of bosonic states (\ref{uosp14fullysymbosonstates}) and fermionic states (\ref{uosp14fullysymfermistates}) are respectively given by 
\begin{subequations}
\begin{align}
&D_{\mathrm{B}}=D(n)\equiv  \frac{1}{3!}(n+1)(n+2)(n+3),\label{osp14bosonicdof}\\
&D_{\mathrm{F}}=D(n-1)= \frac{1}{3!}n(n+1)(n+2), \label{osp14fermidof} 
\end{align}\label{osp14totaldeg}
\end{subequations}
and the total dimension is
\begin{equation}
D_{\mathrm{T}}=D_{\mathrm{B}}+D_{\mathrm{F}}=\frac{1}{6}(n+1)(n+2)(2n+3). 
\label{totaldimn=1fuzzyfour}
\end{equation}
The bosonic and fermionic degrees of freedom 
(\ref{osp14totaldeg}) are respectively accounted for by the $SO(5)$ symmetric basis states with the Casimir indices, $n$ and $n-1$, and hence     
 the $\mathcal{N}=1$ fuzzy four-supersphere can be regarded as a ``compound'' of two fuzzy four-spheres with different radii $n$ and $n-1$: 
\begin{equation}
S_{\mathrm{f}}^{4|2}(n)~\simeq~ S_{\mathrm{f}}^{4}(n)\oplus S_{\mathrm{f}}^{4}(n-1).  
\end{equation}
The eigenvalues of $X_5$ take the following values,    
\begin{equation}
X_5=\frac{R}{d}(n-k),
\label{x5latitude}
\end{equation}
with $k=0,1,2,\cdots,2n$. These eigenvalues are equal to those of  $X_3$ of fuzzy two-supersphere 
(\ref{X3fuzzysupersphere}) except for the degeneracy of the basis states at each latitude.  
The degeneracies at the latitude (\ref{x5latitude}) for even $k=2l$ and for odd $k=2l+1$ are  respectively given by   
\begin{subequations}
\begin{align}
&D_{k=2l}(n)=(n-l+1)(l+1), \\
&D_{k=2l+1}(n)=(n-l)(l+1),  
\end{align}
\end{subequations}
which reproduce the total dimensions of bosonic and fermionic basis states (\ref{osp14totaldeg}) as  
\begin{equation}
\sum_{l=0}^{n}D_{k=2l}(n)=D_{\mathrm{B}},~~~~~~\sum_{l=0}^{n-1}D_{k=2l+1}(n)=D_{\mathrm{F}}. 
\end{equation}

From (\ref{n1fuzzy4dradius}), one may find that the condition for fuzzy four-supersphere  is invariant under the $SU(4|1)$ rotation of the Schwinger operator $\Psi$, which is larger than the original $UOSp(1|4)$ symmetry.    
It may be pedagogical to demonstrate how such ``hidden'' $SU(4|1)$ structure is embedded in the algebra of  fuzzy four-supersphere.  
First notice that the fuzzy four-supersphere coordinates, $X_a$ and $\Theta_{\alpha}$, do not satisfy a closed algebra by themselves, 
\begin{equation}
[X_a,X_b]=i\frac{4R}{d} X_{ab},~~~~[X_a,\Theta_{\alpha}]=\frac{R}{d}(\gamma_a)_{\beta\alpha}\varTheta_{\beta},~~~~
\{\Theta_{\alpha},\Theta_{\beta}\}=\frac{R}{d}\sum_{a<b}  (C\gamma_{ab})_{\alpha\beta}X_{ab}.   
\label{unclosedrelsu41first}
\end{equation}
On the right-hand sides of (\ref{unclosedrelsu41first}), there appear ``new'' operators:  
\begin{equation}
X_{ab}=\frac{R}{d} \Psi^{\dagger}\Gamma_{ab}\Psi,~~~~\varTheta_{\alpha}=\frac{R}{d}\Psi^{\dagger}D_{\alpha}\Psi,   
\end{equation}
where $\Gamma_{ab}$ and $D_{\alpha}$ are respectively given by (\ref{gammablarge}) and   
\begin{equation}
D_{\alpha}=\frac{1}{\sqrt{2}} 
\begin{pmatrix}
0_4 & \tau_{\alpha} \\
(C\tau_{\alpha})^t & 0
\end{pmatrix}. 
\label{defdalphasu41}
\end{equation}
 $X_{ab}$ and $\varTheta_{\alpha}$  respectively act as $SO(5)$  antisymmetric 2-rank tensor and spinor.  
Commutation relations for these new operators can be derived as  
\begin{align}
&[X_a,\Theta_{\alpha}]=\frac{R}{d}(\gamma_a)_{\beta\alpha} \varTheta_{\beta},~~~~~~~~~~~~~~[X_a,\varTheta_{\alpha}]=\frac{R}{d}(\gamma_a)_{\beta\alpha}\Theta_{\beta},\nonumber\\
&[X_{ab},\Theta_{\alpha}] = \frac{R}{d} (\gamma_{ab})_{\beta\alpha} \Theta_{\beta},~~~~~~~~~~~~[X_{ab},\varTheta_{\alpha}]=\frac{R}{d}(\gamma_{ab})_{\beta\alpha} \varTheta_{\beta},\nonumber\\ 
&\{\Theta_{\alpha},\Theta_{\beta}\}=\frac{R}{d}\sum_{a<b}(C\gamma_{ab})_{\alpha\beta}X_{ab},~~~~\{\varTheta_{\alpha},\varTheta_{\beta}\}=-\frac{R}{d} \sum_{a<b}(C\gamma_{ab})_{\alpha\beta}X_{ab},\nonumber\\
&\{\Theta_{\alpha},\varTheta_{\beta}\}=\frac{R}{4d}(C\gamma_a)_{\alpha\beta}X_a+\frac{R}{4d}C_{\alpha\beta}Z,  \label{lalphalalpharelations}
\end{align}
where the last equation further yields another new operator 
\begin{equation}
Z=\frac{R}{d} \Psi^{\dagger}H\Psi, 
\end{equation}
with 
\begin{equation}
H=
\begin{pmatrix}
1_4  & 0 \\
 0 & 4 
\end{pmatrix}. 
\label{gamma5times5}
\end{equation}
Commutation relations with $Z$ are obtained as  
\begin{equation}
[Z,X_a]=[Z,X_{ab}]=0,~~~~[Z,\Theta_{\alpha}] =-\frac{3R}{d}\varTheta_{\alpha},~~~~[Z,\varTheta_{\alpha}]=-\frac{3R}{d}\Theta_{\alpha}. 
\label{concernwithz}
\end{equation}
Consequently, for the closure of the algebra of the fuzzy coordinates $X_a$ and $\Theta_{\alpha}$, we have  introduced the  fifteen new  ``coordinates'', $X_{ab}$, $\varTheta_{\alpha}$ and $Z$.   
In total, with the original coordinates they amount to  twenty four operators that satisfy  the $SU(4|1)$ algebra.  
The basic concept of the non-commutative geometry is the algebraic construction of geometry, and  the $SU(4|1)$ structure, the symmetry of the basis states on fuzzy four-supersphere, has indeed appeared as the fundamental algebra of the fuzzy four-supersphere.

\subsection{$UOSp(1|4)$ SVBS States}
\label{sec:SO5-SVBS}

Similar to the $UOSp(1|2)$ SVBS case, 
the $UOSp(1|4)$ SVBS states \footnote{To be precise, there exist two types of $UOSp(1|4)$ VBS states, one of which is the tensor type (\ref{s05svbswavefun}) and the other is the vector type \cite{Hasebe-Totsuka2012}. Here, we focus on the tensor type.} can be constructed as   
\begin{align}
|\text{SVBS}\rangle& =\prod_{\langle i,j\rangle } (\Psi^{\dagger}_i(r) ~\mathcal{R}_{1|4}~ \Psi^{*}_{j}(r))^M\vac\nonumber\\
&=\prod_{i} ({b^1_i}^{\dagger} {b^2_{j}}^{\dagger}-{b^2_i}^{\dagger} {b^1_{j}}^{\dagger}+{b^3_i}^{\dagger} {b^4_{j}}^{\dagger}-{b^4_i}^{\dagger} {b^3_{j}}^{\dagger}-r f^{\dagger}_i f^{\dagger}_{j}   )^M\vac,   
\label{s05svbswavefun}
\end{align}
where $\Psi(r)$ denotes  the parameter-dependent $UOSp(1|4)$ Schwinger operator 
\begin{equation}
\Psi(r)\equiv (b^1, b^2, b^3, b^4,\sqrt{r}f)^t,    
\end{equation}
and $\mathcal{R}_{1|4}$ signifies the $UOSp(1|4)$ invariant matrix 
\begin{equation}
\mathcal{R}_{1|4}=\begin{pmatrix}
0 & 1  & 0  & 0   & 0 \\
-1 & 0  & 0   & 0  & 0 \\
0 & 0  & 0  & 1   & 0 \\
0 & 0  & -1  &  0  & 0 \\
0 & 0  & 0  & 0   & -1
\end{pmatrix}. 
\end{equation}
The $SO(5)$ spin magnitude on site $i$ reads as 
 \begin{equation}
 S_i=\frac{1}{2}{(b^1_i}^{\dagger}b^1_i+{b^2_i}^{\dagger}b^2_i+{b^3_i}^{\dagger}b^3_i+{b^4_i}^{\dagger}b^4_i)=M,~~M-\frac{1}{2}. 
 \end{equation}
In the following, we focus on the $M=1$ $UOSp(1|4)$ SVBS chain and its corresponding SMPS representation: 
\begin{equation}
|\text{SVBS}\rangle_{a_L, a_R}=\prod_{i=1}^L (\Psi^{\dagger}_i  \mathcal{R}_{1|4} \Psi_{i+1}^{*})_{a_L, a_R} |\text{vac}\rangle=(\mathcal{A}_1\mathcal{A}_2\cdots\mathcal{A}_L)_{a_L, a_R},   
\end{equation}
where $\mathcal{A}$ is a supermatrix given by 
\begin{equation}
\mathcal{A}_i=\mathcal{R}_{1|4}  \Psi^*_i(r)\Psi^{\dagger}_i(r) \nonumber\\
=\begin{pmatrix}
|1,2\rangle_i & \sqrt{2}|2,2\rangle_i  & |2,3\rangle_i & |2,4\rangle_i & \sqrt{r} |2\rangle_i \\
-\sqrt{2}|1,1\rangle & -|1,2\rangle_i & -|1,3\rangle & -|1,4\rangle_i & -\sqrt{r}  |1\rangle_i  \\
|1,4\rangle &|2,4\rangle & |3,4\rangle_i & \sqrt{2}|4,4\rangle_i &\sqrt{r} |4\rangle_i \\
-|1,3\rangle & -|2,3\rangle_i & -\sqrt{2}|3,3\rangle & -|3,4\rangle_i & -\sqrt{r} |3\rangle_i  \\
-\sqrt{r}|1\rangle_i  &\sqrt{r}|2\rangle_i  & -\sqrt{r}|3\rangle_i  & -\sqrt{r}|4\rangle_i  & 0
\end{pmatrix},  
\end{equation}
with the matrix elements  
\begin{align}
&|\alpha,\alpha\rangle =\frac{1}{\sqrt{2}}({b^{\alpha}}^{\dagger})^2|\text{vac}\rangle~~\text{(no~sum~for~$\alpha$)},\nonumber\\
&|\alpha,\beta\rangle_{\alpha\neq \beta} ={b^{\alpha}}^{\dagger}{b^{\beta}}^{\dagger}|\text{vac}\rangle,\nonumber\\
&|\alpha\rangle ={b^{\alpha}}^{\dagger}f^{\dagger}|\text{vac}\rangle.  
\end{align}
The basis states $|\alpha,\alpha\rangle$ and $|\alpha,\beta\rangle_{\alpha\neq \beta}$ are  $SO(5)$ $\bold{10}$-dimensional adjoint representation  while $|\alpha\rangle$ are $SO(5)$ $\bold{4}$-dimensional spinor. In total, 
the components of $\mathcal{A}$ consist of $UOSp(1|4)$ $\bold{14}$-dimensional representation of the graded fully  symmetric representation (\ref{uosp14fullysymstates}) for $n=2$.

\subsection{Entanglement spectrum and $(\mathbb{Z}_2\times \mathbb{Z}_2)^2$ symmetry}

\begin{figure}[!t]
\centering 
~~~~
\includegraphics[width=7.3cm]{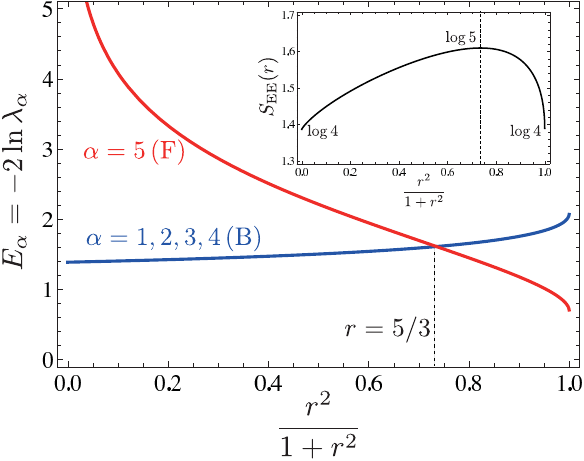}
\caption{The behaviors of the Schmidt coefficients and entanglement entropy (inset) of the $\mathcal{S}=1$ $UOSp(1|4)$ VBS chain.  [Figure and Caption are taken from Ref.\cite{Hasebe-Totsuka2012}.] 
\label{entropyS5I.fig} }
\end{figure}

For the $UOSp(1|4)$ SVBS infinite chain,  the Schmidt coefficients are computed as 
\begin{subequations}
\begin{align}
&{\lambda_{\mathrm{B}}}^2\equiv{\lambda_1}^2={\lambda_2}^2={\lambda_3}^2={\lambda_4}^2=\frac{1}{8}+\frac{5}{8\sqrt{25+16r^2}},\\
&{\lambda_{\mathrm{F}}}^2\equiv{\lambda_5}^2=\frac{1}{2}-\frac{5}{2\sqrt{25+16r^2}}. 
\end{align} 
\end{subequations}\label{s=1osp(14)Schmidt}
The bosonic Schmidt coefficients are quadratically degenerate while the fermionic one is non-degenerate. 
The behaviors of the Schmidt coefficients and the entanglement entropy,  
$S_{E.E.}=-4\lambda_{\mathrm{B}}^2 \log_2 \lambda^2_{\mathrm{B}}- \lambda_{\mathrm{F}}^2 \log_2 \lambda^2_{\mathrm{F}}$, 
are plotted in Fig.~\ref{entropyS5I.fig}.  
The qualitative behaviors of the entanglement spectra of the $UOSp(1|4)$ VBS chain are quite similar to those of  the type-I  SVBS chain [see Fig. \ref{entropySL11.fig}] except for the quadratical degeneracy in the blue curve.  

The degeneracy of the entanglement spectra of the $UOSp(1|4)$ SVBS chain can  be understood based on the arguments of the symmetry protected topological order. Before proceeding to the case of $UOSp(1|4)$, we introduce the original arguments 
for its bosonic counterpart, the $SO(5)$ VBS state \cite{Tuetal2008I,Tuetal2008II,Tuetal2009}.   
Each of inversion symmetry  and  time reversal symmetry guarantees at least double degeneracy of the entanglement spectra of the $SO(5)$ VBS states as proven by similar manner to the $SU(2)$ VBS states. This is simply because each of the inversion symmetry and time reversal symmetry is a realization of $\mathbb{Z}_2 $ symmetry.   
The crucial difference to the $SU(2)$ case is the existence of $(\mathbb{Z}_2 \times \mathbb{Z}_2 )^2$ symmetry of the $SO(5)$ VBS states originating from the $SO(5)$ spin degrees of freedom at edge \cite{Tuetal2011}. 
The $SO(5)$ rotation represents a rotation in five-dimensional space, and we ``divide'' the five-dimensional coordinates (1, 2, 3, 4, 5) to two three-dimensional subsectors, (1, 2, 5) and (3, 4, 5).  The  $\pi$-rotational symmetry around $x$ and $z$ axes in each sector generates the $\mathbb{Z}_2\times \mathbb{Z}_2$ symemtry and in total two independent sets of such discrete rotations give rise to $(\mathbb{Z}_2\times \mathbb{Z}_2)^2$ symmetry in the five-dimensional space.    
The group elements of $(\mathbb{Z}_2\times \mathbb{Z}_2)^2$ consist of 16 bases: 
\begin{align}
\overbrace{(1,u_{12})\times (1,u_{15})}^{\mathbb{Z}_2 \times \mathbb{Z}_2 }\times \overbrace{(1,u_{34})\times (1,u_{35})}^{\mathbb{Z}_2 \times \mathbb{Z}_2 }
\end{align}
where $u_{ab}$ are the $SO(5)$ group elements of $\pi$ rotation generated by the $SO(5)$ generator $\sigma_{ab}$:  
\begin{equation}
u_{ab}(\pi)=e^{i\pi \sigma_{ab}}. 
\end{equation}
As the $\mathbb{Z}_2\times \mathbb{Z}_2$ symmetry generates at least doubly degeneracy in the entanglement spectra, 
the $(\mathbb{Z}_2 \times \mathbb{Z}_2 )^2$ symmetry guarantees the four-fold degeneracy of the entanglement spectra of the $SO(5)$ VBS chain. 
For the SUSY case, 
just as we have discussed in the $UOSp(1|2)$ SVBS case, symmetry transformations are attributed to those of the bosonic and fermionic sectors, 
\begin{equation}
U_{ab}=
\begin{pmatrix}
u_{ab}^{(B)} & 0 \\
0 & u_{ab}^{(F)}
\end{pmatrix}. 
\end{equation}
Similar to the discussions about double degeneracy in the $UOSp(1|2)$ case,  
either of the bosonic and fermionic sectors brings the quadruple degeneracy to the entanglement spectra  of the $UOSp(1|4)$ SVBS state in the presence of $(\mathbb{Z}_2\times \mathbb{Z}_2)^2$ symmetry.

\section{Summary and Discussions}\label{SectSummary}

We reviewed the constructions and basic properties of the fuzzy superspheres and SVBS models.     
Particularly, fuzzy superspheres and SVBS models with $UOSp(1|2)$, $UOSp(2|2)$ and $UOSp(1|4)$ symmetries were discussed in detail. We clarified the mutual relations among the fuzzy spheres, QHE and VBS states were also emphasized based on the Schwinger operator formalism.  It was illustrated that, though the SVBS states incorporate fermionic degrees of freedom, they ``inherit''  all the  nice properties of the VBS:   
\begin{itemize} 
\item Solvable parent Hamiltonian   
\item Gapped bulk and gapless edge  excitations   
\item Generalized hidden order.  
\end{itemize}
We explicitly derived the spectra for gapped  excitations (magnon and hole excitations) on 1D SVBS chain within SMA. 
Physical properties of the SVBS models are qualitatively different from those of the other generalized VBS models based on bosonic Lie groups, in the sense that the SVBS states accommodate the charge sector in addition to the spin sector. In each sector, the SVBS states exhibit the following properties:       
\begin{itemize}
\item In charge sector, the SVBS states have the superconducting property (SSB).  
\item In spin sector,  the SVBS states show non-trivial topological order of QAFM (no SSB).  
\end{itemize}
We also established the SMPS formalism for the SVBS states, which naturally incorporates the edge degrees of freedom to provide a powerful tool in investigating topological property.  For practical use, the (S)MPS formalism greatly simplifies the calculations of physical quantities, such as the string order and entanglement spectrum.   
The SVBS states bear a finite string order regardless of the parity of the bulk superspin unlike the original VBS states.     
From the explicit calculations of the entanglement spectra of the SVBS chains, we demonstrated that there exists a hallmark of topological phase, the double degeneracy in SUSY entanglement spectrum.        
The degeneracy is naturally understood by invoking particular edge states in SUSY chain:    
Since  SUSY relates the integer and half-integer edge-spin states, the SVBS chains always accommodate  half-integer edge spin  (as well as integer edge spin) that guarantees at least double degeneracy of the entanglement spectra.  
Consequently, the topological order is stabilized regardless of the parity of bulk-spin in the presence of SUSY.   

Though we focused on the SVBS states, the arguments of the present SUSY protected topological order  are applicable to general boson-fermion systems. By reformulating boson-fermion system, such as  
 boson-fermion mixture cold atom system \cite{YuYang2008} with SUSY, we may apply the present results to discuss the stability for their topological phases.   
  The idea of topological insulator has begun to be applied to particle theory model \cite{KaplanSun2012}. It may also be interesting to apply the present arguments to other SUSY models that are not directly related to condensed matter physics.

\section*{Acknowledgment}

K.H. would like to thank Daniel P. Arovas, Xiaoliang Qi, and Shoucheng Zhang 
for bringing him the idea of SVBS states and the collaboration from which the sequent works were initiated.  
The authors also thank H. Katsura and F. Pollmann for very useful discussions. 
This work was supported in
part by Grants-in-Aid for Scientific Research (B) 23740212
(K.H.), (C) 20540375, (C) 24540402 (K.T.) 
and by the global COE (GCOE) program, ``The next generation of physics, spun from university and emergent'' of Kyoto University.

\section*{Appendix}

\appendix

\section{$UOSp(\mathcal{N}|2K)$}\label{Appen:SecUOSp(MN)}

Here, we summarize the basic properties of $UOSp(\mathcal{N}|2K)$ algebra. 
We denote the generators of the orthosymplectic group $OSp(\mathcal{N}|2K)$ as $\Sigma_{AB}$, which satisfy the following relation\footnote{
 The minus sign in front of $1_{\mathcal{N}}$ in (\ref{defospmn}) is not important for the definition of $OSp(\mathcal{N}|2K)$, but added to be consistent with the notation of the present paper.}:  
\begin{equation}
\Sigma_{AB}^{st}
\begin{pmatrix}
J_{2K} & 0 \\
0 & -1_{\mathcal{N}}
\end{pmatrix} + \begin{pmatrix}
J_{2K} & 0 \\
0 & -1_{\mathcal{N}}
\end{pmatrix}\Sigma_{AB}=0,   
\label{defospmn}
\end{equation}
where  the supertranspose, $st$, is defined in (\ref{definitionsupertranspose}), $1_{\mathcal{N}}$ denotes $\mathcal{N}\times \mathcal{N}$ unit matrix, and 
 $J_{2K}$ represents the $Sp(2K, \mathbb{C})$ invariant matrix    
\begin{equation}
J_{2K}=
\begin{pmatrix}
0 & 1_{K} \\
-1_{K} & 0 
\end{pmatrix}.    
\label{sympecticinvmat} 
\end{equation}
$\Sigma_{AB}$ can be expressed by a linear combination of the following matrices:  
\begin{equation}
\Sigma_{\alpha\beta}=\begin{pmatrix}
\sigma_{\alpha\beta} & 0 \\
0 & 0 
\end{pmatrix},~~\Sigma_{lm}=\begin{pmatrix}
0 & 0 \\
0 & \sigma_{lm}
\end{pmatrix}
,~~\Sigma_{l\alpha}=
\begin{pmatrix}
0 & \sigma_{l\alpha} \\
(J_{2K}\sigma_{l\alpha})^t & 0 
\end{pmatrix}, \label{blockmatrixsigmas}
\end{equation}
where   $\alpha$ and $\beta$ are the indices of  $Sp(2K, \mathbb{C})$ $(\alpha,\beta=1,2,\cdots,2K)$ and $l$ and $m$ are those of $O(\mathcal{N})$ ($l,m=1,2,\cdots,\mathcal{N}$). 
$\sigma_{l\alpha}$ stand for arbitrary $2K\times 2K$ matrices, while  
$\sigma_{\alpha\beta}$ and $\sigma_{lm}$ respectively signify $2K\times 2K$ and $\mathcal{N}\times \mathcal{N}$ matrices that satisfy 
\begin{subequations}
\begin{align}
&{\sigma_{lm}}^t+\sigma_{lm}=0,\label{defortho}\\
&{\sigma_{\alpha\beta}}^t J_{2K} + J_{2K}\sigma_{\alpha\beta}=0\label{defsymp}.   
\end{align} \label{blocksatisfyequations}
\end{subequations}
The $OSp(\mathcal{N}|2K)$ algebra contains the maximal bosonic subalgebra, $sp(2K, \mathbb{C}) \oplus o(\mathcal{N}, \mathbb{C})$, whose generators are $\sigma_{\alpha\beta}$ and $\sigma_{lm}$.  The off-diagonal block matrices  $\Sigma_{l\alpha}$ are  fermionic generators that  transform as the fundamental representation under each of the transformations of  $Sp(2K, \mathbb{C})$ and $O(\mathcal{N}, \mathbb{C})$.   The $o(\mathcal{N}, \mathbb{C})$ matrices $\sigma_{lm}$ are antisymmetric matrices (\ref{defortho}), and then we can take the indices of $\sigma_{lm}$ to be antisymmetric,    
$\sigma_{lm}=-\sigma_{ml}$.    In the following, we consider the real antisymmetric matrices, the generators of $o(\mathcal{N})$, with  $\mathcal{N}(\mathcal{N}-1)/2$ real degrees of freedom.  
Meanwhile from the relation (\ref{defsymp}), the generators of $Sp(2K, \mathbb{C})$ $\sigma_{\alpha\beta}$ take the form of 
\begin{equation}
\sigma_{\alpha\beta}=
\begin{pmatrix}
k & s \\
s' & -k^t
\end{pmatrix},  
\end{equation}
where $k$ stands for an arbitrary $K \times K$ complex matrix, and similarly $s$ and $s'$ 
 are $K \times K$ symmetric complex matrices.  
If the Hermiticity condition is imposed, $\sigma_{\alpha\beta}$ are reduced to 
the generators of $USp(2K)$ that take the form of 
\begin{equation}
\sigma_{\alpha\beta}=
\begin{pmatrix}
h & s \\
s^{\dagger} & -h^* 
\end{pmatrix},  
\label{sp4generators}
\end{equation}
where $h$ represents an arbitrary Hermitian matrix and $s$ also signifies an arbitrary symmetric complex matrix. 
Consequently, $UOSp(\mathcal{N}|2K)$ generators consist of (\ref{blockmatrixsigmas}) whose blocks satisfy (\ref{blocksatisfyequations}) and (\ref{sp4generators}).  The real independent degrees of freedom of $\sigma_{\alpha\beta}$ are $K(2K+1)$.  
Then, for $usp(2K)$ matrices $\sigma_{\alpha\beta}$, we can take the indices to be  symmetric,   
$\sigma_{\alpha\beta}=\sigma_{\beta\alpha}.$ 
Meanwhile, the real degrees of freedom of the fermionic generators 
$\Sigma_{l\alpha}$ are $2KN$.  
Then in total, th real degrees of freedom of $uosp(\mathcal{N}|2K)$ are given by 
\begin{equation}
\dim[uosp(\mathcal{N}|2K)]=\frac{1}{2}(4K^2+{\mathcal{N}}^2+2K-\mathcal{N})|2K\mathcal{N}=\frac{1}{2}((2K+\mathcal{N})^2+2K-\mathcal{N}).
\end{equation}

In the present paper, instead of $J_{2K}$ (\ref{sympecticinvmat}), we adopt the following $2K\times 2K$ matrix (which is unitarily equivalent to $J_{2K}$):   
\begin{equation}
\mathcal{R}_{2K}=\begin{pmatrix}
i\sigma_2 &    0        & 0      &  0 \\
0         &   i\sigma_2 & 0      &  0 \\
0         &    0        & \ddots & 0       \\

0         &    0        &   0    & i\sigma_2 
\end{pmatrix}, 
\end{equation}
and the following $UOSp(\mathcal{N}|2K)$ invariant matrix,  
\begin{equation}
\mathcal{R}_{\mathcal{N}|2K}=
\begin{pmatrix}
\mathcal{R}_{2K} & 0 \\
0 & -1_{\mathcal{N}}
\end{pmatrix}.   
\end{equation}
For instance, for the $UOSp(1|2)$, $UOSp(2|2)$, and $UOSp(1|4)$,  $\mathcal{R}_{\mathcal{N}|2K}$ are respectively given by 
\begin{equation}
\mathcal{R}_{1|2}=\begin{pmatrix}
0 & 1 & 0 \\
-1 & 0    & 0 \\
0 & 0  &  -1 
\end{pmatrix},~~~
\mathcal{R}_{2|2}=\begin{pmatrix}
0 & 1 & 0 & 0 \\
-1 & 0 & 0 & 0 \\
0 & 0 & -1 & 0 \\
0 & 0 & 0 &   -1 
\end{pmatrix}, ~~~\mathcal{R}_{1|4}=\begin{pmatrix}
0 & 1  & 0  & 0   & 0 \\
-1 & 0  & 0   & 0  & 0 \\
0 & 0  & 0  & 1   & 0 \\
0 & 0  & -1  &  0  & 0 \\
0 & 0  & 0  & 0   & -1
\end{pmatrix}. 
\end{equation}

\section{Fuzzy Four-Supersphere with Higher Supersymmetries }\label{SectMoreSUSY}

By generalizing the construction of $\mathcal{N}=1$ fuzzy four-supersphere  based on 
$UOSp(1|4)$ [Sec.\ref{subsecn1fuzzyfoursphere}],  
we can construct  $\mathcal{N}$-SUSY fuzzy four-sphere with use of the $UOSp(\mathcal{N}|4)$ algebra \cite{Hasebe2011}.   
The dimension of the $UOSp(\mathcal{N}|4)$ algebra is given by 
\begin{equation}
\dim [uosp(\mathcal{N}|4)]= 10+\frac{1}{2}\mathcal{N}(\mathcal{N}-1)|4\mathcal{N} =10+\frac{1}{2}\mathcal{N}(\mathcal{N}+7).
\end{equation}
We denote the bosonic generators of $uosp(\mathcal{N}|4)$ as $\Gamma_{ab}=-\Gamma_{ba}$ $(a,b=1,2,3,4,5)$, $\tilde{\Gamma}_{lm}=-\tilde{\Gamma}_{ml}$ $(l,m=1,2,\cdots,\mathcal{N})$, and fermionic generators as $\Gamma_{l\alpha}$ $(\alpha=\theta_1,\theta_2,\theta_3,\theta_4)$.  In total, they satisfy 
\begin{align}
&[\Gamma_{ab},\Gamma_{cd}]= i
(\delta_{ac}{\Gamma}_{bd}-\delta_{ad}{\Gamma}_{bc}+\delta_{bc}{\Gamma}_{ad}-\delta_{bd}{\Gamma}_{ac}),\nonumber\\
&[\Gamma_{ab},\Gamma_{l\alpha}]=(\gamma_{ab})_{\beta\alpha} \Gamma_{l\beta},\nonumber\\
&[\Gamma_{ab},\tilde{\Gamma}_{lm}] = 0,  \nonumber\\
&\{\Gamma_{l\alpha},\Gamma_{m\beta}\}= \sum_{a<b} (C\gamma_{ab})_{\alpha\beta} \Gamma_{ab}\delta_{lm}+\frac{1}{4}C_{\alpha\beta}\tilde{\Gamma}_{lm}, \nonumber\\
&[\Gamma_{l\alpha},\tilde{\Gamma}_{mn}]=   (\gamma_{mn})_{pl}\Gamma_{p\alpha},\nonumber\\
&[\tilde{\Gamma}_{lm},\tilde{\Gamma}_{np}]= -
\delta_{ln}{\tilde{\Gamma}}_{mp}+\delta_{lp}\tilde{\Gamma}_{mn}-\delta_{mp}\tilde{\Gamma}_{ln}+\delta_{mn}\tilde{\Gamma}_{lp}, 
\label{algebraosp4k}
\end{align}
where $C$ is the $SO(5)$ charge conjugation matrix (\ref{so5chargeconjmat}), and 
$\gamma_{lm}=-\gamma_{ml}$ $(l<m)$ are $SO(\mathcal{N})$ generators given by 
\begin{equation}
(\gamma_{lm})_{np}=\delta_{ln}\delta_{mp}-\delta_{lp}\delta_{mn}.
\label{explicitspksmall}
\end{equation}
We introduce the coordinates of $S_{\mathrm{f}}^{4|2\mathcal{N}}$ as\footnote{ $X_a$, $\Theta_{\alpha}^{(l)}$, $Y_{lm}$ do not satisfy a closed algebra by themselves,  and the minimally extended closed algebra including these operators is $SU(4|\mathcal{N})$ \cite{Hasebe2011}.}         
\begin{equation}
X_a=\frac{R}{d} \Psi^{\dagger}\Gamma_a\Psi,~~~~\Theta_{\alpha}^{(l)}=\frac{R}{d} \Psi^{\dagger}\Gamma_{l\alpha}\Psi,~~~~{Y}_{lm}=\frac{R}{d} \Psi^{\dagger}\tilde{\Gamma}_{lm}\Psi, \label{Schwingerrepgeneral}
\end{equation}
where  $\Psi=(b_1, b_2, b_3, b_4, f_1, f_2,\cdots, f_\mathcal{N})^t$ denotes the $UOSp(\mathcal{N}|4)$ Schwinger operator whose first four components,  $b_{\alpha}$ $(\alpha=1,2,3,4)$, are the bosonic operators while the remaining $\mathcal{N}$ components, $f_{l}$ $(l=1,2,\cdots,\mathcal{N})$, are the fermionic operators.  
The sandwiched matrices in (\ref{Schwingerrepgeneral}) 
\begin{equation}
\Gamma_{a}=\begin{pmatrix}
\gamma_{a} & 0 \\
0 & 0_\mathcal{N} 
\end{pmatrix},~~~~\Gamma_{l\alpha}=\begin{pmatrix}
0_{3+l} & \tau_{\alpha} & 0 \\
 -(C\tau_{\alpha})^t & 0 & 0 \\
 0 & 0 & 0_{\mathcal{N}-l} 
\end{pmatrix},~~~~\tilde{\Gamma}_{lm}=\begin{pmatrix} 
0_4 & 0 \\
0 & \gamma_{lm}
\end{pmatrix},  
\label{fundrepgeneuspn4}
\end{equation}
are the fundamental representation matrices of $uosp(\mathcal{N}|4)$. 
Here, $0_k$ denotes $k\times k$ zero-matrix, and 
$\tau_{\alpha}$ are given by (\ref{taucomponents}).      
Square of  the radius of the $\mathcal{N}$-SUSY fuzzy four-supersphere is readily derived as  
\begin{equation}
X_a X_a +2\sum_{l=1}^{\mathcal{N}} C_{\alpha\beta}\Theta_{\alpha}^{(l)}\Theta_{\beta}^{(l)}+\sum_{l<m=1}^{\mathcal{N}} Y_{lm}Y_{lm}=\biggl(\frac{R}{d} \biggr)^2 {n}({n}+4-\mathcal{N}),  
\label{square42kfuzzy}
\end{equation}
with ${n}=\Psi^{\dagger}\Psi$.  
Notice that  the square of the radius of $\mathcal{N}$-SUSY fuzzy supersphere is proportional to $n(n+4-\mathcal{N})$ and takes  negative values for sufficiently small $n$ that satisfy $n < \mathcal{N}-4$. For the positive definiteness of the radius,  the SUSY number should be restricted to $\mathcal{N}\le 4$. 
The basis states on $\mathcal{N}$-SUSY fuzzy supersphere $S_{\mathrm{f}}^{4|\mathcal{N}}$ are given by the graded fully symmetric representation:  
\begin{align}
&\!\!\!\!\!\!|l_1,l_2,l_3,l_4\rangle =\frac{1}{\sqrt{l_1!~l_2!~l_3!~l_4!}} 
{b_1^{\dagger}}^{l_1} {b_2^{\dagger}}^{l_2} {b_{3}^{\dagger}}^{l_3} {b_4^{\dagger}}^{l_4} \vac, \nonumber\\
&\!\!\!\!\!\!|m_1,m_2,m_3,m_4)_{i_1} =\frac{1}{\sqrt{m_1!~m_2!~m_3!~m_4!}} 
{b_1^{\dagger}}^{m_1} {b_2^{\dagger}}^{m_2} {b_3^{\dagger}}^{m_3} {b_4^{\dagger}}^{m_4} f_{i_1}^{\dagger}\vac \nonumber\\
&\!\!\!\!\!\!|n_1,n_2,n_3,n_4\rangle_{i_1<i_2} =\frac{1}{\sqrt{n_1!~n_2!~n_3!~n_4!}} 
{b_1^{\dagger}}^{n_1} {b_2^{\dagger}}^{n_2} {b_3^{\dagger}}^{n_3} {b_4^{\dagger}}^{n_4} f_{i_1}^{\dagger}  f_{i_2}^{\dagger} \vac \nonumber\\
&\vdots\nonumber\\
&\!\!\!\!\!\!|q_1,q_2,q_3,q_4\rangle_{i_1<i_2< \cdots< i_{\mathcal{N}-1}}=
\frac{1}{\sqrt{q_1!q_2!q_3!q_4!}} 
{b_1^{\dagger}}^{q_1} {b_2^{\dagger}}^{q_2} {b_3^{\dagger}}^{q_3} {b_4^{\dagger}}^{q_4} ~f_{i_1}^{\dagger}  f_{i_2
}^{\dagger}f^{\dagger}_{i_3} \cdots f_{i_{\mathcal{N}-1}}^{\dagger}  \vac,\nonumber\\
&\!\!\!\!\!\!|r_1,r_2,r_3,r_4)=
\frac{1}{\sqrt{r_1!r_2!r_3!r_4!}} 
{b_1^{\dagger}}^{r_1} {b_2^{\dagger}}^{r_2} {b_3^{\dagger}}^{r_3} {b_4^{\dagger}}^{r_4} ~f_{1}^{\dagger}  f_{2
}^{\dagger}f_3 \cdots f_{\mathcal{N}-1}^{\dagger}  f_{\mathcal{N}}^{\dagger}\vac, \label{gradedfulrepgeneral}
\end{align}
where $l_1+l_2+l_3+l_4=m_1+m_2+m_3+m_4+1=n_1+n_2+n_3+n_4+2=\cdots=q_1+q_2+q_3+q_4+\mathcal{N}-1=r_1+r_2+r_3+r_4+\mathcal{N}=n$.  
Therefore with $D(n)$ (\ref{osp14bosonicdof}),  
the dimension of (\ref{gradedfulrepgeneral}) is given by 
\begin{equation}
D_{\mathrm{T}}=\sum_{l=0}^{\mathcal{N}} 
\begin{pmatrix}
\mathcal{N} \\
l 
\end{pmatrix} \cdot  D(n-l)=\frac{1}{3}  (2n+4-\mathcal{N})\biggl((2n+4-\mathcal{N})^2-4+3\mathcal{N}\biggr) 2^{\mathcal{N}-4},  
\label{totaldimingeneral}
\end{equation}
for $n\ge \mathcal{N}-3$. 
The degrees of freedom of the basis states (\ref{gradedfulrepgeneral}) imply that     
  $S_{\mathrm{f}}^{4|2\mathcal{N}}(n)$ can  be interpreted as a ``compound'' of lower SUSY fuzzy four-spheres with different radii: 
\begin{align}
S_{\mathrm{f}}^{4|2\mathcal{N}}(n)&~\simeq~ \sum_{m=0}^l ~ {}_{l}C_m\cdot S_{\mathrm{f}}^{4|2\mathcal{N}-2l}(n-m)\nonumber\\
&~\simeq~ S_{\mathrm{f}}^{4|2\mathcal{N}-2l}(n)\oplus \biggl[l\times S_{\mathrm{f}}^{4|2\mathcal{N}-2l}(n-1)\biggr] \oplus \biggl[\frac{l(l-1)}{2!} \times S_{\mathrm{f}}^{4|2\mathcal{N}-2l}(n-2)\biggr] 
\oplus \cdots \oplus S_{\mathrm{f}}^{4|2\mathcal{N}-2l}(n-l).
\end{align}
More explicitly, 
\begin{align}
S_{\mathrm{f}}^{4|2\mathcal{N}}(n)&~\simeq~ S^{4|2\mathcal{N}-2}_{\mathrm{F}}(n) \oplus S_{\mathrm{f}}^{4|2\mathcal{N}-2}(n-1)\nonumber\\
&~\simeq ~S_{\mathrm{f}}^{4|2\mathcal{N}-4}(n)\oplus \biggl[2 \times S_{\mathrm{f}}^{4|2\mathcal{N}-4}(n-1)\biggr]\oplus  S_{\mathrm{f}}^{4|2\mathcal{N}-4}(n-2)\nonumber\\
&~\simeq ~S_{\mathrm{f}}^{4|2\mathcal{N}-6}(n)\oplus \biggl[ 3 \times S_{\mathrm{f}}^{4|2\mathcal{N}-6}(n-1)\biggr]\oplus \biggl[ 3 \times S_{\mathrm{f}}^{4|2\mathcal{N}-6}(n-2)\biggr]\oplus S_{\mathrm{f}}^{4|2\mathcal{N}-6}(n-3)\nonumber\\
&~\simeq~ \cdots. 
\end{align}
For $\mathcal{N}=1$, this reproduces the relation for $S_{\mathrm{f}}^{4|2}(n)$ (\ref{compoundfuzzysupersph}).



\end{document}